\documentclass[aps,prd,longbibliography,reprint,superscriptaddress,preprintnumbers,floatfix,nofootinbib,notitlepage,showkeys]{revtex4-1}

\usepackage{amsmath,amssymb,amsfonts,mathtools} 
\usepackage{hepunits}
\usepackage[normalem]{ulem}
\usepackage[utf8]{inputenc} 
\usepackage[T1]{fontenc}
\usepackage{hyperref}
\usepackage{graphicx}
\usepackage{latexsym}
\usepackage{bbm}
\usepackage{epstopdf}
\usepackage{color}
\usepackage{braket}
\usepackage{hyperref}
\hypersetup{colorlinks=true,citecolor=OliveGreen,linkcolor=BrickRed,urlcolor=MidnightBlue}
\usepackage{lineno}
\usepackage{slashed}
\usepackage{placeins} 
\usepackage{diagbox}
\usepackage{dcolumn}
\usepackage{xfrac}
\usepackage[dvipsnames]{xcolor}
\usepackage{aligned-overset}

\newcommand{\Tc}{T_{\rm c}}

\newcommand{\Tr}{\textrm{Tr}\,}

\newcommand{\be}{\begin{equation}} 
\newcommand{\ee}{\end{equation}}
\def\ba#1\ea{\begin{linenomath}\begin{align}#1\end{align}}
\newcommand{\bea}{\begin{eqnarray}} 
\newcommand{\eea}{\end{eqnarray}}
\def\lsim{\mathrel{\raise.3ex\hbox{$<$\kern-.75em\lower1ex\hbox{$\sim$}}}}
\def\gsim{\mathrel{\raise.3ex\hbox{$>$\kern-.75em\lower1ex\hbox{$\sim$}}}}

\begin{document}
%\linenumbers

\title{Lattice study of correlators of chromoelectric fields for heavy quarkonium dynamics in the quark-gluon plasma}
\author{Nora Brambilla}
\email{nora.brambilla@ph.tum.de}
\affiliation{Technical University of Munich, TUM School of Natural Sciences, Physics Department, James-Franck-Strasse 1, 85748 Garching, Germany}
\affiliation{Institute for Advanced Study, Technische Universit\"at M\"unchen,
Lichtenbergstrasse 2a, 85748 Garching, Germany}
\affiliation{Munich Data Science Institute, Technische Universit\"at M\"unchen, \\
Walther-von-Dyck-Strasse 10, 85748 Garching, Germany}

\author{Saumen Datta}
\email{saumen@theory.tifr.res.in}
\affiliation{Tata Institute of Fundamental Research, Homi Bhabha Road, Mumbai 400005, India}

\author{Marc Janer}
\email{m.janer@tum.de}
\affiliation{Technical University of Munich, TUM School of Natural Sciences, Physics Department, James-Franck-Strasse 1, 85748 Garching, Germany}

\author{Viljami Leino}
\email{viljami.leino@uni-mainz.de}
\affiliation{Helmholtz Institut Mainz, Staudingerweg 18, 55128 Mainz, Germany}
\affiliation{Institut für Kernphysik, Johannes Gutenberg-Universität Mainz, 
             Johann-Joachim-Becher-Weg 48, 55128 Mainz, Germany}

\author{Julian Mayer-Steudte}
\email{julian.mayer-steudte@tum.de}
\affiliation{Technical University of Munich, TUM School of Natural Sciences, Physics Department, James-Franck-Strasse 1, 85748 Garching, Germany}
\affiliation{Munich Data Science Institute, Technische Universit\"at M\"unchen, \\
Walther-von-Dyck-Strasse 10, 85748 Garching, Germany}

\author{Peter Petreczky}
\email{petreczk@bnl.gov}
\affiliation{Physics Department, Brookhaven National Laboratory,
  Upton, New York 11973, USA}

\author{Antonio Vairo}
\email{antonio.vairo@ph.tum.de}
\affiliation{Technical University of Munich, TUM School of Natural Sciences, Physics Department, James-Franck-Strasse 1, 85748 Garching, Germany}

\collaboration{TUMQCD Collaboration}
\noaffiliation

\date{\today}
\preprint{TUM-EFT 189/24}
\preprint{MITP-24-077}

\begin{abstract}
We perform a lattice calculation of  the correlators of two chromoelectric fields in the adjoint representation
connected by adjoint Wilson lines at non-zero temperature.
These correlators arise in the study of  quarkonium  dynamics and of adjoint heavy quark diffusion in deconfined matter. 
We work 
in SU(3) gauge theory using either gradient flow or multi-level algorithms for noise reduction, and discuss the renormalization of the correlators on the lattice. 
We find that  a Casimir factor rescaling relates the adjoint correlators corresponding
to the diffusion of an adjoint heavy quark and the octet-octet quarkonium transitions  
to the chromoelectric correlator in the fundamental representation describing the diffusion of a heavy quark.
\end{abstract}

\maketitle

\section{Introduction}\label{sec:intro}
Quarkonia were originally proposed as a probe of the hot matter produced in heavy
ion collisions by Matsui and Satz ~\cite{Matsui:1986dk}. 
Since then, a large experimental program dedicated to studying quarkonia production in heavy ion experiments has emerged -- see Refs.~\cite{Aarts:2016hap,Zhao:2020jqu,Andronic:2024oxz,Brambilla:2014jmp} for recent reviews. 
However, interpreting the corresponding experimental findings remains a challenge for the theory.

There are three distinct energy scales relevant for quarkonia: the heavy
quark mass, the inverse of the size of the system, and the binding energy.
Because quarkonia are non-relativistic bound states, these scales are hierarchically ordered, 
which naturally calls for an effective field theory treatment, 
with the ultimate effective field theory being potential non-relativistic QCD (pNRQCD)~\cite{Brambilla:2004jw}.
In pNRQCD, only light degrees of freedom with energy and momentum smaller than the inverse of the size of the system are dynamic.
If quarkonium is a Coulombic, weakly coupled bound state, which applies to the lowest bottomonium and charmonium states, then pNRQCD may be constructed by multipole expanding the gluon fiels in terms of the relative distance of the heavy quark-antiquark pair.
In this situation, the non-perturbative quarkonium dynamics are encoded at leading order in the correlation function of two chromoelectric fields connected by a Wilson line in the adjoint representation.
In a thermal bath, if the temperature is smaller than the inverse of the size of the system, 
the thermal correlator of two chromoelectric fields connected by a Wilson line in the adjoint representation encodes the leading order thermal contributions to the quarkonium mass shift and width induced by the medium~\cite{Brambilla:2008cx,Brambilla:2010vq,Escobedo:2010tu,Eller:2019spw}.
Furthermore, in the pNRQCD setting one can treat the open quantum system dynamics
of quarkonium formation and dissociation in a hot medium via Lindblad equations~\cite{Brambilla:2016wgg,Brambilla:2017zei,Brambilla:2019tpt}. 
The transport coefficients entering these equations can be calculated from correlators
of chromoelectric fields connected by Wilson lines in the adjoint representation.
We indicate them in short as \emph{adjoint chromoelectric correlators}. 
These correlators and the associated transport coefficients provide the necessary QCD input for
the study of quarkonium dynamics in heavy-ion collisions.

Despite their importance for the study of the nonequilibrium evolution 
of quarkonium in the quark-gluon plasma~\cite{Brambilla:2016wgg,Brambilla:2017zei,Yao:2020eqy,Scheihing-Hitschfeld:2023tuz,Binder:2021otw,Brambilla:2024tqg,Brambilla:2022ynh,Sharma:2021vvu}, 
the adjoint chromoelectric correlators  at finite temperature have not been calculated on the lattice so far. 
The goal of this study is to calculate, for the first time, the adjoint chromoelectric correlators in lattice
SU(3) gauge theory (quenched QCD). 
This requires noise reduction techniques. 
We use gradient flow~\cite{Luscher:2010iy} and the multi-level algorithm~\cite {Luscher:2001up} for noise reduction. 
At zero temperature, the adjoint correlators are also related to the mass spectrum of gluelumps~\cite{Foster:1998wu}.

A somewhat similar correlator, the thermal chromoelectric correlator with  fundamental Wilson lines  
wrapping around the periodic Euclidean time direction, has already been  studied in lattice QCD~\cite{Banerjee:2011ra,Francis:2015daa,Brambilla:2020siz,Altenkort:2020fgs,Brambilla:2022xbd,Altenkort:2023oms}.
This correlation function can be connected to the diffusion of a heavy quark in  a hot medium~\cite{Casalderrey-Solana:2006fio,Caron-Huot:2009ncn}. 
We  shall discuss the relation of these correlators to the adjoint chromoelectric
correlators calculated here. 
We also study a related correlator, the chromoelectric correlator with an adjoint Wilson line wrapping around the Euclidean time direction. The rest of the paper is organized as follows: 
The precise definitions of the various correlators and the methods on noise reductions in the calculations of these correlators
are given in the next section. In Section \ref{sec:lattice_setup} we discuss the details of our lattice setup. Our main numerical results
are given in Section \ref{sec:results}. Finally, Section \ref{sec:concl} contains our conclusions.

\section{Lattice methodology}

\subsection{Adjoint correlators on the lattice}

Within pNRQCD, the propagation and the interaction of  quarkonium in a hot medium are described by transport coefficients, which are encoded at leading order in the adjoint correlators of two chromoelectric fields. 
The correlators can be derived from QCD employing pNRQCD and the open quantum system approach detailed in Refs.~\cite{Brambilla:2016wgg,Brambilla:2017zei,Brambilla:QuarkoniumTransportCoefficients}.
In the following, we consider the adjoint Euclidean correlators 
\begin{linenomath}\begin{align}
    G_{E}(\tau) &= -\frac{1}{3}\sum_{i=1}^3 \langle E_{i,a}(\tau) U^{\mathrm{adj}}_{ab}(\tau,0) E_{i,b}(0)\rangle, \label{NonSymCorr} \\
    G_{E}^{\mathrm{oct}}(\tau) &= -\frac{1}{3\langle l_8\rangle}\sum_{i=1}^3 \langle U^{\mathrm{adj}}_{ea}(1/T,\tau)d_{abc}E_{i,c}(\tau)\nonumber\\
    &\ \hspace{2cm} \times U^{\mathrm{adj}}_{bd}(\tau,0)d_{def}E_{i,f}(0)\rangle,\label{octCorr}
\end{align}\end{linenomath}
where $T$ is the temperature, $U^{\mathrm{adj}}(\tau_1,\tau_2)$ denotes the adjoint temporal Wilson line connecting the time slices $\tau_1$ and $\tau_2$ at the same spatial coordinate, $E_{i,a}(0)\equiv E_{i,a}(\mathbf{x},t)$ is  the chromoelectric field component at space-time coordinate $x=(\mathbf{x},t)$, $E_{i,a}(\tau)\equiv E_{i,a}(\mathbf{x},t+\tau)$ is the chromoelectric  field component at the same spatial position but shifted along the temporal axis by $\tau$, and $d_{abc}$ are the symmetric structure constants of SU(3). 

The first index of the chromoelectric field component refers to the spatial component, 
while the second  is  the color index. 
The adjoint Polyakov loop $l_8$ is defined as
\begin{linenomath}\begin{align}
    l_8 = \Tr~U^{\mathrm{adj}}(1/T, 0) = 
    U_{aa}^{\mathrm{adj}}(1/T, 0)\,,
\end{align}\end{linenomath}
where repeated color indices are summed over.
When evaluating the expectation values $\langle ... \rangle$ we perform averaging over all lattice sites, i.e., over all space-time coordinates to increase the signal before averaging over the field configurations. This is possible because of the translational symmetry 
on the lattice with periodic boundary conditions.

The correlators describe transitions between  
quark-antiquark pairs in different color configurations. 
Specifically, $G_E(\tau)$ describes singlet-octet transitions, while $G_E^{\mathrm{oct}}(\tau)$ octet-octet transitions. 
As noted in Ref~\cite{Brambilla:QuarkoniumTransportCoefficients}, a third correlator exists for octet-singlet transitions:
\begin{linenomath}\begin{align}
    \Tilde{G}_E(\tau) = -\frac{1}{3}\sum_{i=3}^3\langle U_{ba}^\mathrm{adj}(\tau,1/T)E_{i,b}(\tau)E_{i,a}(0)\rangle.
\end{align}\end{linenomath}
Due to the periodicity of $G_E(\tau)$ in Euclidean time, 
it holds that
\begin{linenomath}\begin{align}
    \Tilde{G}_E(\tau) = G_E(1/T-\tau).
\end{align}\end{linenomath}
The diffusion of a single heavy quark is described by the following correlation function~\cite{Casalderrey-Solana:2006fio,Caron-Huot:2009ncn}
\begin{equation}
    G_{E}^{\mathrm{fund}}(\tau) = -\sum_{i=1}^3 \frac{\langle \mathrm{Re}\Tr [U(1/T,\tau)E_{i}(\tau)U(\tau,0)E_{i}(0)]\rangle}{3\langle l_3\rangle}, \label{fundCorr}
\end{equation}
where the Polyakov loops $l_3$, the Wilson lines $U(\tau_1,\tau_2)$ and the chromoelectric fields are in the fundamental representation. 
The chromoelectric fields in the fundamental representation are defined as
\begin{linenomath}\begin{align}
    E_i = E_{i,a}  \frac{\lambda _a}{2}
\end{align}\end{linenomath}
with $\lambda_a$ the Gell-Mann matrices.
We define the adjoint counterpart of  the correlator~\eqref{fundCorr} as
\begin{linenomath}\begin{align}
    G_{E}^{\mathrm{sym}}(\tau) &= \frac{1}{3\langle l_8\rangle} \sum_{i=1}^3 \langle U^{\mathrm{adj}}_{ea}(1/T,\tau)f_{abc}E_{i,c}(\tau)\nonumber \\
    &\ \hspace{1.58cm}\times U^{\mathrm{adj}}_{bd}(\tau,0)f_{def}E_{i,f}(0)\rangle,\label{symCorr}
\end{align}\end{linenomath}
which describes the diffusion of a heavy adjoint source of color. 
Here $f_{abc}$ are the antisymmetric structure constants of SU(3).

The adjoint Wilson line and the chromoelectric fields are related to their fundamental counterparts by
\begin{linenomath}\begin{align}
    U^{\mathrm{adj}}_{ab}(\tau_1,\tau_2) &= \frac{1}{2}\Tr[U(\tau_1,\tau_2)\lambda_aU(\tau_2,\tau_1)\lambda_b]\label{eq:adjoint_wilson_line_definition},\\
    E_{i,a} &= \Tr[\lambda_aE_i]\label{eq:adjoint_field_definition}.
\end{align}\end{linenomath}
Using these relations and the Fierz  identity, we obtain the correlators in terms of fundamental links and   chromoelectric fields in the fundamental representation as\\
\begin{equation}
    G_E(\tau) = -\frac{2}{3}\sum_{i=1}^3 \langle\Tr[E_i(\tau)U(\tau, 0)E_i(0)U(0,\tau)]\rangle\label{eq:non_symmetric_Gee_fundamental_representation},
\end{equation}
\begin{linenomath}\begin{align}
    G_E^{\mathrm{oct}}(\tau) =& -\frac{1}{3\langle l_8\rangle} \sum_{i=1}^3 \langle\Tr[E_i(\tau)P(\tau)^\dagger]\Tr[E_i(0)P(0)] \nonumber\\
    +&\Tr[E_i(0)P(0)^\dagger]\Tr[E_i(\tau)P(\tau)] \nonumber\\
    +&\Tr[E_i(\tau)U(\tau, 1/T)E_i(0)U(0,\tau)]\Tr[P(0)] \nonumber\\
    +&\Tr[E_i(\tau)U(\tau, 0)E_i(0)U(1/T,\tau)]\Tr[P(0)^\dagger] \nonumber\\
    -&\frac{4}{3} \Tr[E_i(\tau)U(\tau,0)E_i(0)U(0,\tau)] \nonumber \\
    -&\frac{4}{3} \Tr[E_i(\tau)U(\tau, 1/T)E_i(0)U(1/T,\tau)]\rangle\label{eq:octet_Gee_fundamental_representation},
\end{align}\end{linenomath}
\begin{linenomath}\begin{align}
    G_E^{\mathrm{sym}}(\tau) =& \frac{1}{3\langle l_8\rangle} \sum_{i=1}^3 \langle\Tr[E_i(\tau)P(\tau)^\dagger]\Tr[E_i(0)P(0)] \nonumber\\
    +&\Tr[E_i(0)P(0)^\dagger]\Tr[E_i(\tau)P(\tau)] \nonumber\\
    -&\Tr[E_i(\tau)U(\tau, 1/T)E_i(0)U(0,\tau)]\Tr[P(0)] \nonumber\\
    -&\Tr[E_i(\tau)U(\tau, 0)E_i(0)U(1/T,\tau)]\Tr[P(0)^\dagger] \rangle\label{eq:symmetric_Gee_fundamental_representation},
\end{align}\end{linenomath}
where $P(\tau)=U(\tau,0)U(1/T,\tau)$ is the temporal Wilson line wrapping around the whole temporal lattice axis starting at $\tau$. 
The adjoint Polyakov loop can be expressed in terms of the fundamental one   as 
\begin{linenomath}\begin{align}
    l_8 &= |l_3|^2-1, \label{polyakov_representations}\\
    l_3 &= \mathrm{Tr}\ U(1/T,0), 
\end{align}\end{linenomath}
with     
\begin{linenomath}\begin{align}
    \langle l_3\rangle &= \frac{1}{N_\mathrm{s}^3N_\tau} \sum_{\mathbf{x},\tau}\mathrm{Tr}\ P(\tau) =\frac{1}{N_\mathrm{s}^3}\sum_{\mathbf{x}}\mathrm{Tr}\ U(1/T,0),
\end{align}\end{linenomath}
and  $N_\mathrm{s}$ and $N_\tau$ the number of spatial and temporal lattice sites,  respectively. 
We denote the trace normalized versions of the Polyakov loops with as
\begin{linenomath}\begin{align}
    L_3 &\equiv \frac{1}{3}l_3,\\
    L_8 &\equiv \frac{1}{8}l_8.
\end{align}\end{linenomath}
Details on the Fierz relation and on the derivation of the above relations can be found  
in Appendix \ref{app:fierz_stuff}.

We use the clover and a 2-plaquette discretization scheme to implement the field strength components $F_{\mu\nu}$ at a certain space-time coordinate $x$ on the lattice. 
The clover discretization, which we label as CLO, is given  by 
\begin{linenomath}\begin{align}
    a^2F_{\mu\nu} &= -\frac{i}{8}(Q_{\mu\nu}-Q_{\nu\mu}), \\
    Q_{\mu\nu} &= U_{\mu,\nu} + U_{\nu,-\mu} + U_{-\mu,-\nu} + U_{-\nu,\mu} = Q_{\nu\mu}^\dagger, 
\end{align}\end{linenomath}
 i.e., it consists of the sum over all four plaquettes $U_{\mu\nu}$ over both orientations starting and ending at $x$; $a$ is the lattice spacing. 
The chromoelectric field components  are the temporal components of the field strength tensor:
\begin{linenomath}\begin{align}
    E_i = -F_{i,4}.
\end{align}\end{linenomath}
The 2-plaquette discretization, which we label 2PL, consists of a partially summed discretization where the field components contain only two plaquettes. For $G_E(\tau)$, we choose the plaquettes without overlapping with the connecting Wilson lines. In terms of plaquettes, the 2PL discretization is explicitly given by
\begin{linenomath}\begin{align}
    E_i(\tau) &= \frac{i}{4}(U_{i,4}+U_{4,-i}-U_{4,i}-U_{-i,4}),\\
    E_i(0) &= \frac{i}{4}(U_{-i,-4}+U_{-4,i}-U_{-4,-i}-U_{i,-4}).
\end{align}\end{linenomath}
Note that both field descriptions are interchangeable for the octet and symmetric correlator as long as the source is set   at  $0-a/2$ and  the   sink at $\tau+a/2$. 
In Appendix~\ref{app:cont_extra}, we compare both discretizations, and we find that the 2PL is the one that gives better continuum extrapolations.

Additionally, we modify the definition of the chromoelectric field to ensure that it is traceless on the lattice
\begin{equation}
    E_i(\tau) \rightarrow E_i(\tau) - \frac{1}{3} \Tr[E_i(\tau)],
\end{equation}
which corresponds to an $a^2$-improvement~\cite{Bilson-Thompson:2002xlt}.

The fundamental correlator introduced in Eq.~\eqref{fundCorr} at leading order (LO) in perturbation theory reads~\cite{Caron-Huot:2009ncn}
\begin{equation}
    \frac{G^{\mathrm{fund}}_E(\tau)\vert_{\mathrm{LO}}}{g^2C_F} \equiv f(\tau) = \pi^2T^4\left[\frac{\cos^2(\pi\tau T)}{\sin^4(\pi\tau T)} + \frac{1}{3\sin^2(\pi\tau T)}\right]
\end{equation}
where $g$ is the bare coupling. On the lattice, the LO term $f(\tau)$ takes the form
\begin{linenomath}\begin{align}
    f(\tau)\vert_{\mathrm{lat}} &= \frac{1}{3a^4} \int_{-\pi}^\pi \frac{d^3q}{(2\pi)^3}\frac{\cosh[\bar{q} N_\tau (\frac{1}{2}-\tau T)]}{\sinh (\bar{q} N_\tau/2)} \frac{1}{\sinh (\bar{q})} \nonumber \\
    &\times \begin{cases} \;
    \left( 1 + \frac{\tilde{q}^2}{4} \right) \, \left(\tilde{q}^2 - 
    \frac{(\tilde{q}^2)^2}{8} + \frac{\tilde{q}^4}{8} \right) & (\mathrm{CLO}) \\
    \left( \tilde{q}^2 + \frac{\tilde{q}^4 - \left( \tilde{q}^2 \right)^2}{8}
    \right) & (\mathrm{2PL})\label{eq:f_of_tau}
\end{cases}
\end{align}\end{linenomath}
with
\begin{linenomath}\begin{align}
    \bar{q} &= 2\arcsin\left(\frac{\sqrt{\tilde{q}^2}}{2}\right),\\
    \tilde{q}^n &= \sum_{i=1}^3 2^n\sin^n\left(\frac{q_i}{2}\right).
\end{align}\end{linenomath}
For the adjoint correlators introduced in Eqs.~\eqref{NonSymCorr},~\eqref{octCorr},~\eqref{symCorr}, at LO, we obtain that they are proportional to the fundamental correlator: 
\begin{linenomath}\begin{align}
    \frac{G_E(\tau)\vert_{\mathrm{LO}}}{g^2 C_F} &= 2Nf(\tau), \\
    \frac{G^{\mathrm{oct}}_E(\tau)\vert_{\mathrm{LO}}}{g^2 C_F} &= 2\frac{N^2-4}{N^2-1}f(\tau), \label{G_oct_constant} \\
    \frac{G^{\mathrm{sym}}_E(\tau)\vert_{\mathrm{LO}}}{g^2 C_F} &= \frac{C_A}{C_F}f(\tau). \label{G_sym_constant}
\end{align}\end{linenomath}
with $N=3$ for SU(3).
To further reduce discretization errors, we define an improved correlator that includes tree-level improvement as
\begin{equation}
    G^{\mathrm{imp}}_E(\tau_F, \tau T) = \frac{G_E(0,\tau T)\vert_{\mathrm{LO}}}{G_E(0,\tau T)\vert^{\mathrm{lat}}_{\mathrm{LO}}}G_E^{\mathrm{measured}}(\tau_F, \tau T).
\end{equation}
We use the tree-level correlators at zero flow time to improve the correlators because the gradient flow effects are below \SI{1}{\percent} within the considered flow time range~\cite{Altenkort:2020fgs}; hence, gradient flow effects are negligible at tree-level.
Since at leading order Eqs.~\eqref{NonSymCorr},~\eqref{octCorr}, and~\eqref{symCorr} are proportional to $f(\tau)$, we use it directly as normalization for the tree-level improvement.

\subsection{Noise reduction for the lattice calculations of the adjoint chromo-electric correlators}
The aim of this study is to perform lattice calculations of the adjoint chromoelectric correlators in quenched QCD, i.e.
pure SU(3) gauge theory. The lattice calculations of these correlators require some techniques to improve the signal-to-noise 
ratio.
We utilize the gradient flow method~\cite{Narayanan:2006rf,Luscher:2009eq,Luscher:2010iy} to measure the correlators along the flow time axis. 
It is known that gradient flow improves the signal-to-noise ratio and also renormalizes  chromo-electric field insertions~\cite{Brambilla:2023fsi}. 
The Yang--Mills gradient flow evolves the gauge fields $A_\mu$ towards the minimum of the Yang--Mills gauge action along an auxiliary dimension referred to as flow time $\tau_F$. 
The evolution is governed by the differential equation
\begin{equation}
    \partial _{\tau_F} B_\mu = \dot{B}_\mu = D_\nu G_{\nu\mu}, \qquad B_\mu\vert_{\tau_F = 0} = A_\mu, \\
\end{equation}
\begin{equation}
    G_{\mu\nu} = \partial_\mu B_\nu - \partial_\nu B_\mu + [B_\mu, B_\nu], \; D_\mu = \partial_\mu + [B_\mu,.].
\end{equation}
These equations are an explicit representation of
\begin{equation}
    \partial_{\tau_F}B_\mu(\tau_F,x) = -g^2\frac{\delta S_{\mathrm{YM}}[B]}{\delta B_\mu(\tau_F,x)},
\end{equation}
where $S_{\mathrm{YM}}[B]$ represents the Yang--Mills action evaluated with the flowed gauge fields $B(\tau_F)$.

In the context of pure-gauge lattice theory, these equations are adapted as
\begin{linenomath}\begin{align}
    \dot{V}_{\tau_F}(x,\mu) &= -g^2\{\partial_{x,\mu}S_G(V_{\tau_F})\}V_{\tau_F}(x,\mu), \\
    V_{\tau_F}(x,\mu)\vert_{\tau_F = 0} &= U_\mu(x),
\end{align}\end{linenomath}
where $V_{\tau_F}$ are the flowed link variables and $U_\mu(x)$ the unflowed ones. 
$S_G(V)$ is some pure-gauge action evaluated with the gauge field $V$.

The gradient flow evolves the unflowed gauge fields $A_\mu$ into the flowed fields $B_\mu$, which corresponds at leading order to smeared fields with a Gaussian kernel of flow radius $\sqrt{8\tau_F}$~\cite{Luscher:2010iy}. 
This smearing systematically cools off the UV physics and automatically renormalizes the gauge-invariant observables~\cite{Luscher:2011bx}.

The flow radius $\sqrt{8\tau_F}$ is the characteristic length scale of the gradient flow, which must not interfere with the physical length scale of interest, $\tau$, defined as the separation between the inserted chromoelectric fields.
To ensure that the flow smooths the UV fluctuations sufficiently while preserving the physics at the scale $\tau$, we consider the range
\begin{equation}
    a \leq \sqrt{8\tau_F} \leq \frac{\tau}{3}.
    \label{flowtimecondition}
\end{equation}
To obtain the renormalized continuum correlator, we first take the continuum limit ($a \to 0$) and subsequently extrapolate to zero flow time. 
This is discussed in detail in the next section. 

To double-check  the systematics of the lattice calculations using the gradient flow we compare the continuum correlators  $G_{E}(\tau)$ with those obtained from lattice calculations using the multi-level algorithm \cite{Luscher:2001up}. Unfortunately, in
this approach we do not have the non-perturbative renormalization of the chromo-electric fields. Therefore, we use tree-level tadpole
renormalization for the chromo-electric fields.
The details of the calculations using the multi-level algorithm can be found in Appendix~\ref{app:multilevel}.

\section{Lattice setup}
\label{sec:lattice_setup}

As mentioned in the previous section, we calculate the adjoint chromoelectric correlators in SU(3) gauge theory.
We use the publicly available MILC Code~\cite{MILC} to generate a set of pure-gauge SU(3) configurations using the standard Wilson gauge action~\cite{Wilson:1974sk}. 
At least 120 sweeps separate successive configurations, with each sweep consisting of 15-20 over-relaxation steps and 5-15 heat-bath steps. 
We use the same configurations used in \cite{Brambilla:2022xbd}, where we considered two temperatures: a low temperature, $1.5~T_c$, and a high temperature, $10^4~T_c$, with $T_c$ being the deconfinement phase transition temperature. 
These temperatures are achieved by relating them to the lattice coupling $\beta=6/g^2$, which determines the lattice spacing $a$ via the scale setting \cite{Francis:2015lha}. 
This scale setting relates $\beta$ to a gradient flow parameter $t_0$ via a renormalization-group-inspired fit form, which is then further related to the temperature with $T_c\sqrt{t_0} = 0.2489(14)$ \cite{Francis:2015lha}.
We use lattices with varying numbers of temporal sites, $N_\tau = 16,~20,~24,~28,~\mathrm{and}~34$, and with corresponding spatial extents of $N_\mathrm{s} = 48,~48,~48,~56,~\mathrm{and}~68$ sites. 
Based on our previous study \cite{Brambilla:2020siz}, we do not expect a notable dependence on the spatial size of the lattice. 
Table \ref{tab:simulation_parameters} gives the full list of lattice parameters.
\begin{table}
\caption{The simulation parameters for the gradient flow lattices.}
\label{tab:simulation_parameters}
    \centering
    \begin{ruledtabular}
    \begin{tabular}{cccdc}
        $T/\Tc$ & $N_\tau$ & $N_\mathrm{s}$ & \multicolumn{1}{c}{$\beta$} & $N_\mathrm{conf}$\\\hline
         1.5 & 16 & 48 & 6.872 & 1000 \\
             & 20 & 48 & 7.044 & 2002 \\
            & 24 & 48 & 7.192 & 2060 \\
            & 28 & 56 & 7.321 & 1882 \\
            & 34 & 68 & 7.483 & 1170 \\
       10000 & 16 & 48 & 14.443 & 1000 \\
            & 20 & 48 & 14.635 & 1178 \\
            & 24 & 48 & 14.792 & 998 \\
            & 34 & 68 & 15.093 & 599
    \end{tabular}
    \end{ruledtabular}
\end{table}

 We  select $S_G$ as the Symanzik action to solve the gradient flow equation because it shows a more stable behavior~\cite{Altenkort:2020fgs,Brambilla:2022xbd}. 
 
 To solve the differential equation numerically, we can use either a fixed step-size or an adaptive step-size solver~\cite{Fritzsch:2013je, Bazavov:2021pik}, where the step size is varied individually for each gauge field. 
 The adaptive step size solver increases the step sizes when the changes of the flowed fields are small and decreases them when they are larger, which speeds up gradient flow computations as it optimizes the required number of steps. 
 While for a given gauge configuration we obtain data along the same discrete flow time axis  with fixed step sizes, we would have to interpolate the data evaluated on the  different lattice configurations to the same flow time axis.  
 It turns out that the step sizes with the adaptive step-size solver hardly vary between different configurations within a given ensemble.
 Therefore, we use the same step sizes for every gauge configuration for a given ensemble to avoid additional systematics originating from the interpolation. 
 To ensure that the chosen step size covers all lattice configurations properly within a given ensemble, we keep the stepsize slightly smaller than the smallest step sizes from the trial run. 
 Although this procedure requires some additional computations to estimate the proper stepsize  the added computational
 costs are still negligible because computing the correlation functions is still quite expensive.

\section{Numerical results on the adjoint chromoelectric correators}
\label{sec:results}

\subsection{Symmetric correlators}
\begin{figure}
    \centering
    \includegraphics[width=0.9\linewidth]{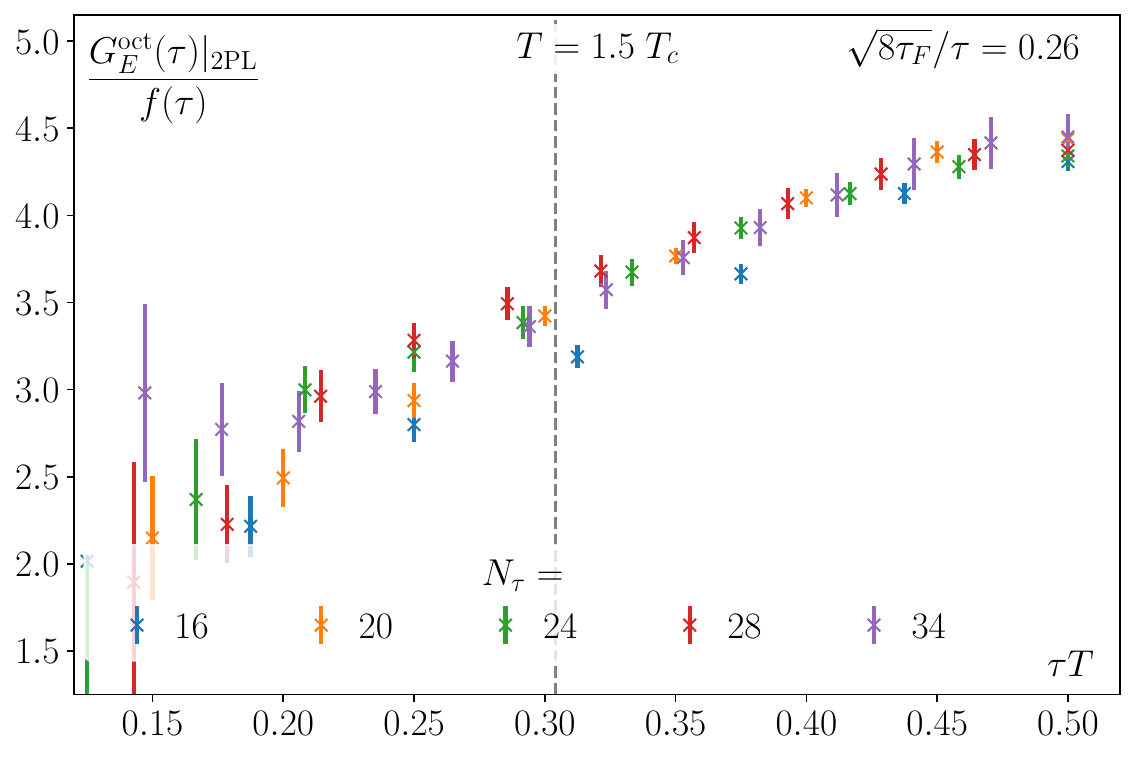} \\
    \includegraphics[width=0.9\linewidth]{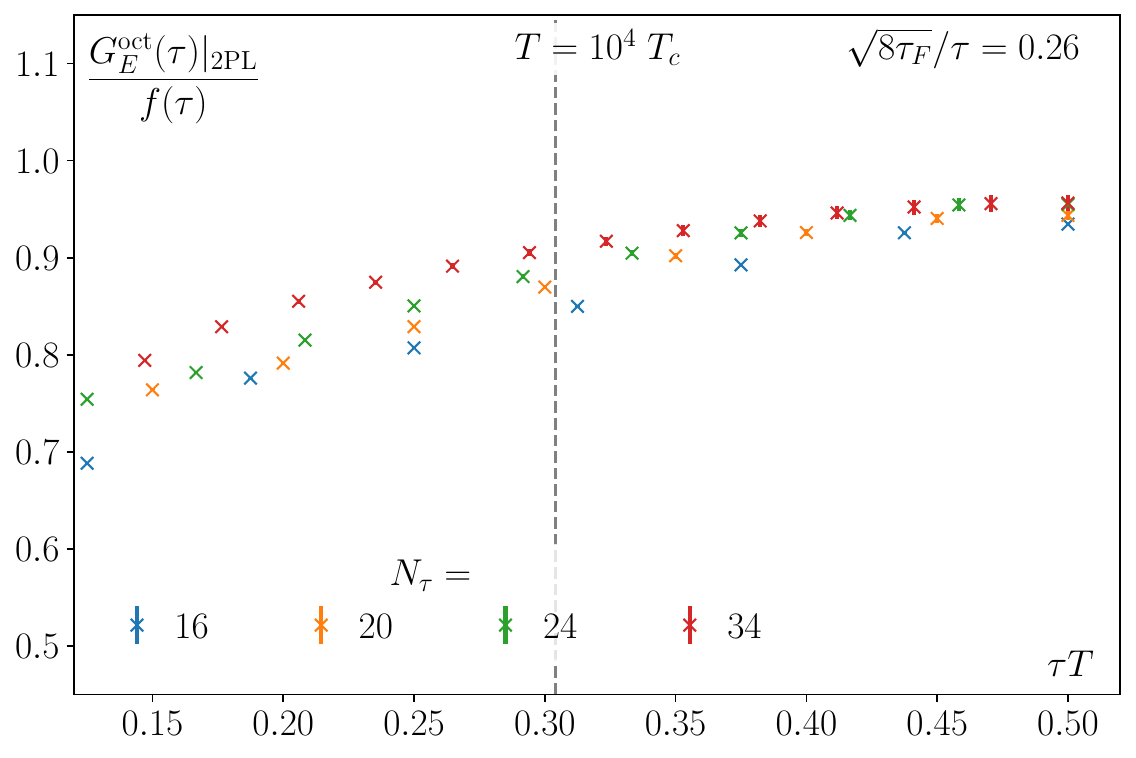}
    \caption{$G_E^{\mathrm{oct}}$ normalized with $f(\tau)$ for different lattice sizes with the ratio between the flow time radius and the time separation fixed to $\sqrt{8\tau_F}/\tau = 0.26$. The upper plot shows the low-temperature case, and the lower one shows the high-temperature case. The vertical line indicates the validity threshold according to Eq.~\eqref{flowtimecondition}, where the points on the right-hand side are the valid points.}
    \label{G_E_oct_ratio}
\end{figure}

In Fig.~\ref{G_E_oct_ratio}, we present the lattice data obtained using the gradient flow for $G_E^\mathrm{oct}$ for both temperatures studied for a representative value of the flow time and normalized with $f(\tau)\vert_{\rm lat}$ given by Eq.~\eqref{eq:f_of_tau}. 
Note that we have fixed the ratio $\sqrt{8\tau_F}/\tau$; 
Thus the statistical errors have the same order of magnitude for each ratio and decrease when increasing the ratio. 
The values to the right of the vertical line indicate the lower condition of Eq.~\eqref{flowtimecondition} for $N_\tau = 16$, since it is the most restrictive condition.

\begin{figure}
    \centering
    \includegraphics[width=0.9\linewidth]{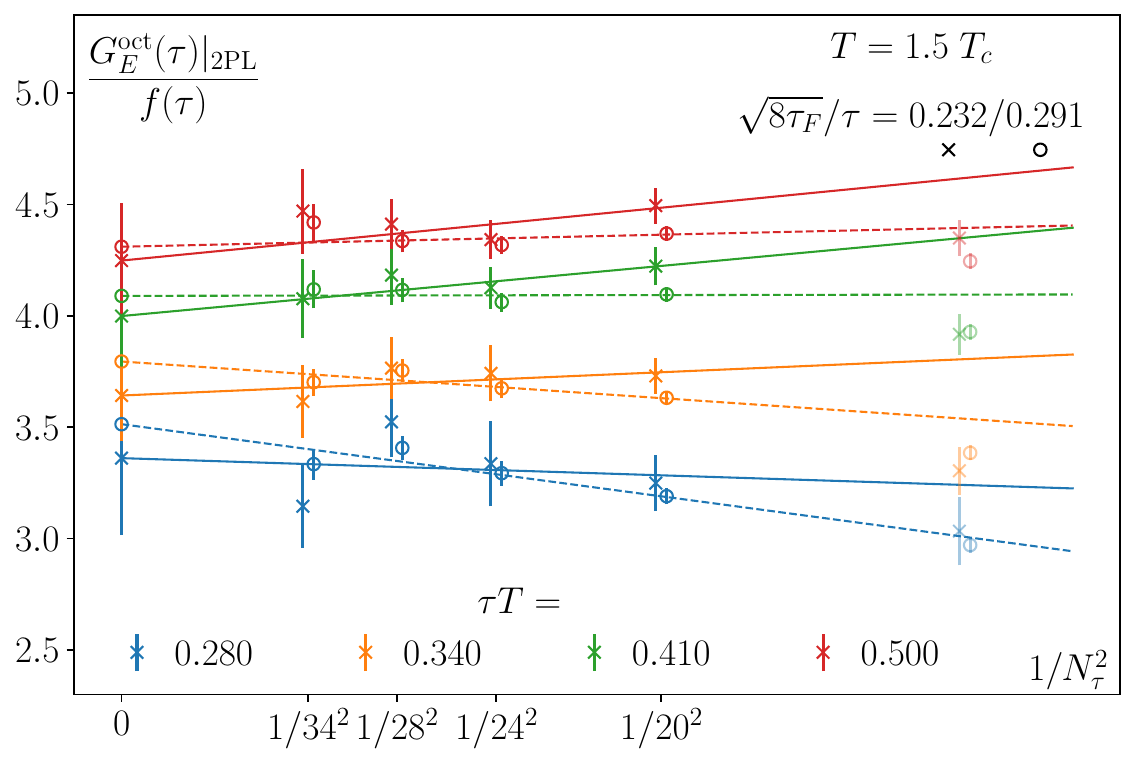} \\
    \includegraphics[width=0.9\linewidth]{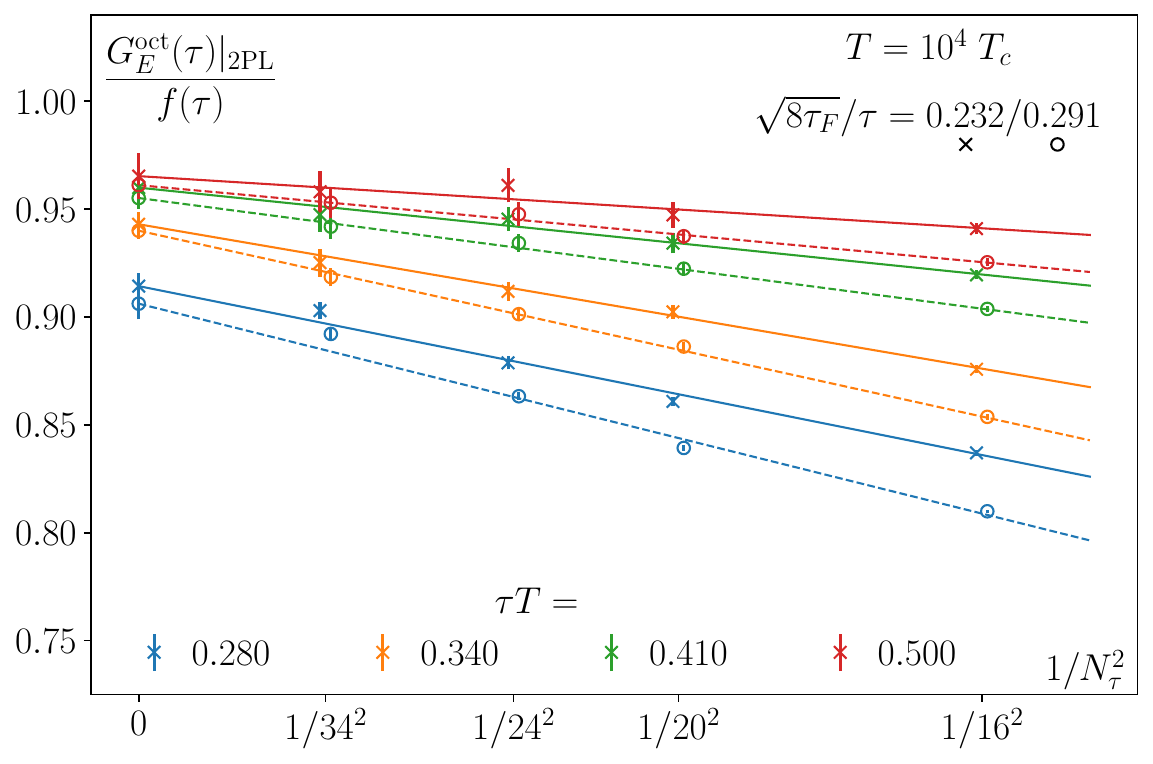} 
    \caption{Examples of the continuum extrapolations for the normalized $G_E^{\mathrm{oct}}$ at two different fixed ratios $\sqrt{8\tau_F/}\tau$ for different values of $\tau T$. 
    The solid lines and crosses indicate the ratio $\sqrt{8\tau_F/}\tau = 0.232$, while the dashed lines and circles $\sqrt{8\tau_F/}\tau = 0.291$. 
    The upper plot shows the low-temperature case, and the lower one shows the high-temperature case. The dimmed points are the ones not included in the linear fit.}
    \label{G_E_oct_cont_limit}
\end{figure}

To perform the continuum extrapolation, we first interpolate the data of each lattice size in $\tau_F$ and then in $\tau T$ with cubic splines. Since this correlator is symmetric by construction around $\tau T = 0.5$, we impose the condition that the first derivative of the cubic spline is zero at $\tau T = 0.5$. 
Since we generated the configurations with the standard Wilson gauge action, we perform a linear continuum extrapolation in $1/N_\tau^2 = (aT)^2$ at a fixed flow-time ratio. In Fig.~\ref{G_E_oct_cont_limit}, we show an example of the linear extrapolation at different $\tau T$ and $\sqrt{8\tau_F}/\tau$. 
We   vary  the number of included lattice sizes in the continuum extrapolation to estimate the systematic error. For $T=1.5\; T_c$, the inclusion of the $N_\tau=16$ lattice requires an additional $1/N_\tau^4 = (aT)^4$ term to keep $\chi^2/d.o.f$ at around 1 or below, whereas for $T=10^4\; T_c$, the linear ansatz describes the data well. 
In Appendix~\ref{app:cont_extra}, we present the values of $\chi^2$ and explain how the systematics are estimated.

\begin{figure}
    \centering
    \includegraphics[width=0.9\linewidth]{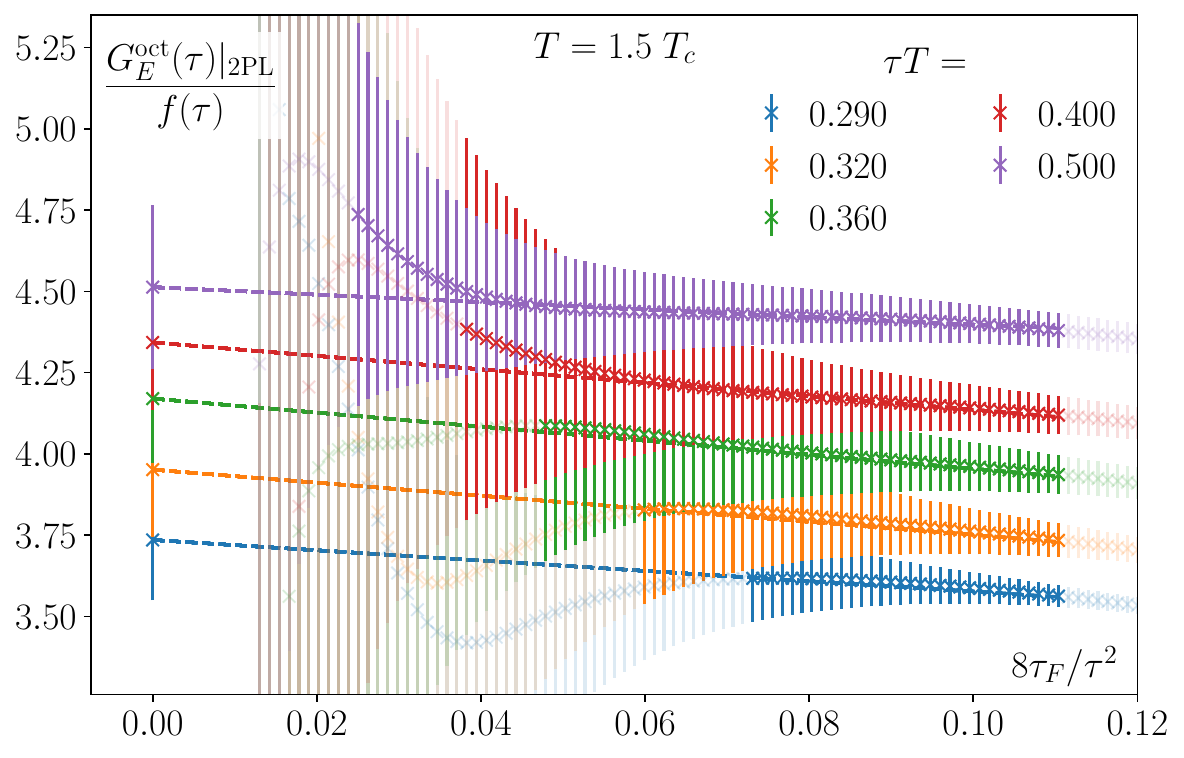} \\
    \includegraphics[width=0.9\linewidth]{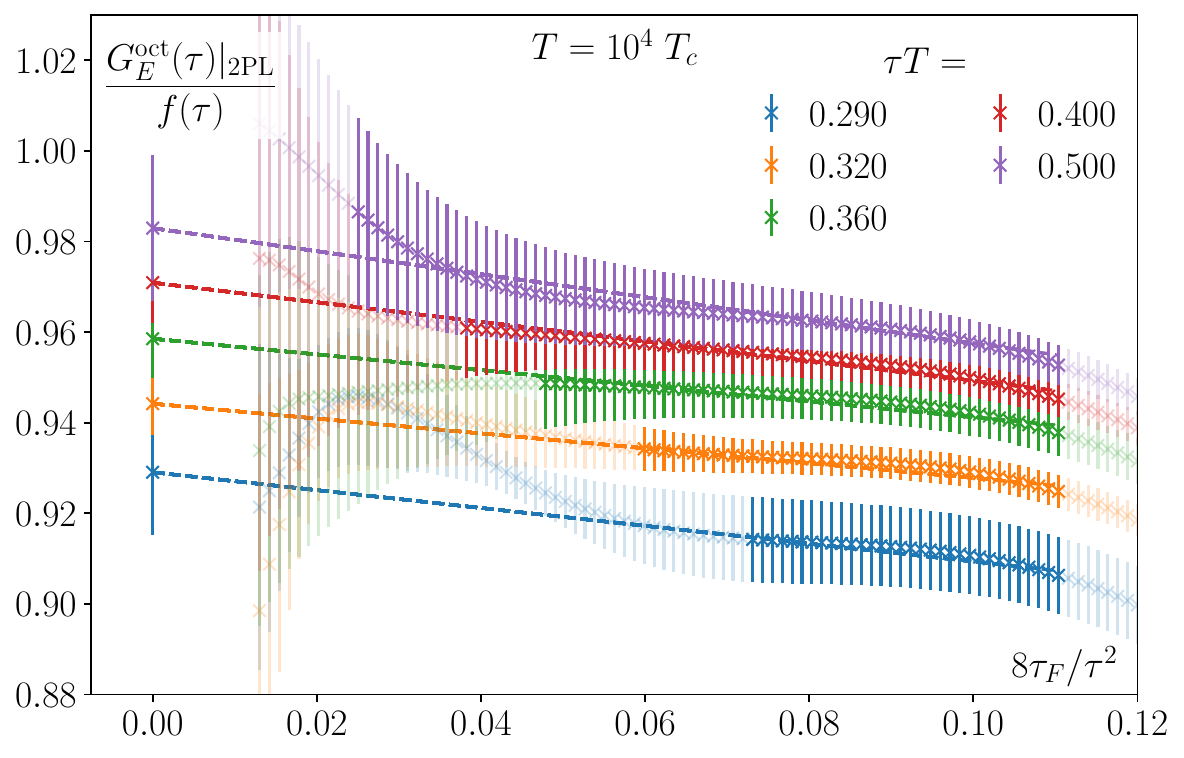} 
    \caption{Examples of the zero-flow-time extrapolation for the normalized $G_E^{\mathrm{oct}}$ at different $\tau T$ values. 
    { Only} the highlighted points fulfill the condition given in Eq.~\eqref{flowtimecondition}. The upper plot shows the low-temperature case, and the lower one the high-temperature case.}
    \label{G_E_oct_zft_limit}
\end{figure}

Next, we perform the zero-flow-time extrapolation by assuming a linear ansatz in $\tau_F$. The fundamental correlator $G_E^{\mathrm{fund}}$ behaves linearly with $\tau_F$ at NLO~\cite{delaCruz:2024cix}.%\cite{Eller:2021qpp}. 
We do not have similar calculations for the adjoint correlators in this study, but it is plausible to assume a similar behavior in $\tau_F$. Examples of the zero-flow-time extrapolation at different $\tau T$ are shown in Fig.~\ref{G_E_oct_zft_limit}.

\begin{figure}
    \centering
    \includegraphics[width=0.9\linewidth]{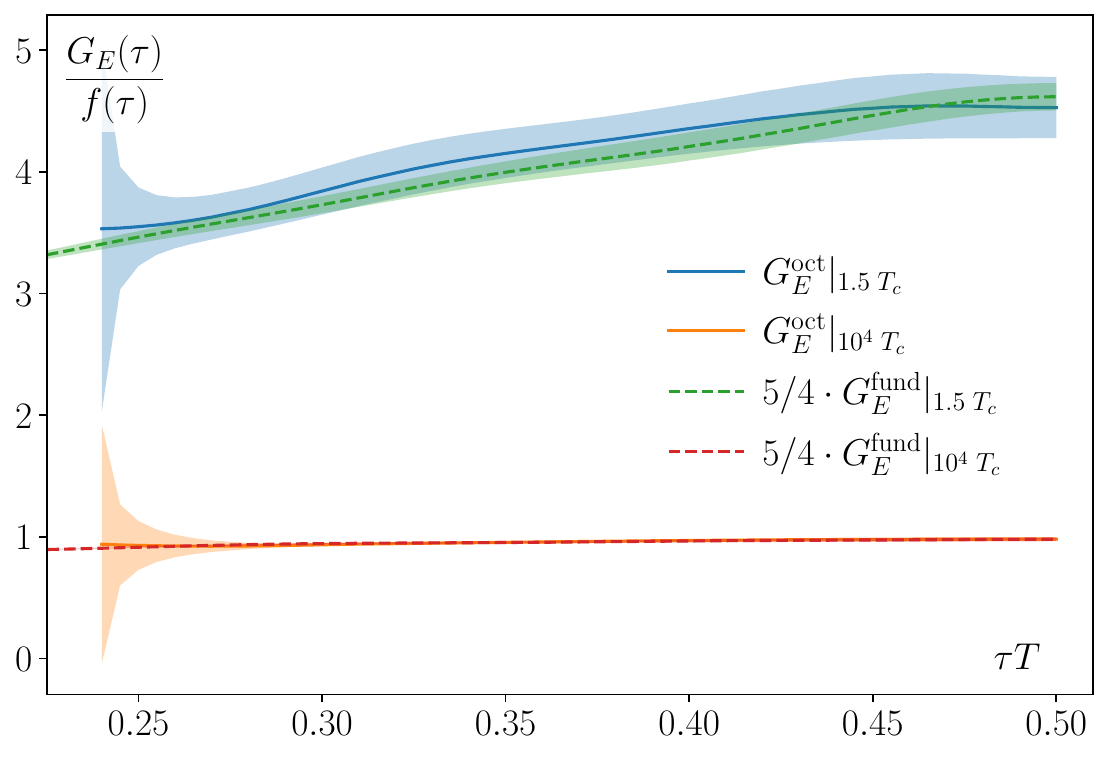} \\
    \includegraphics[width=0.9\linewidth]{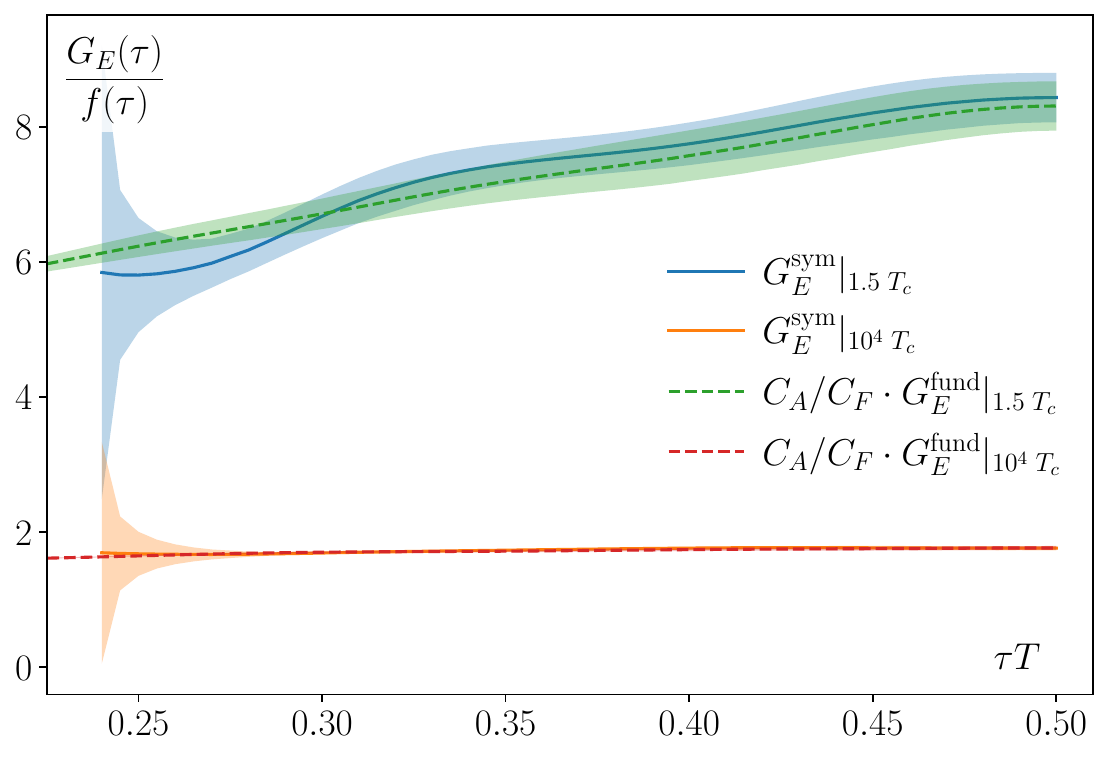} 
    \caption{Final continuum results of the normalized $G_E^{\mathrm{oct}}$ (top) and $G_E^{\mathrm{sym}}$ (bottom) after the zero-flow-time extrapolation. The dashed lines are the results of $G_E^{\mathrm{fund}}$ obtained in \cite{Brambilla:2022xbd} multiplied by the  constants given in Eq.~\eqref{G_oct_constant} and~\eqref{G_sym_constant}.
    }
    \label{G_E_oct_final}
\end{figure}

The final continuum and zero-flow-time extrapolated results for $G_E^{\mathrm{oct}}$ are shown together with $G_E^{\mathrm{fund}}$ from \cite{Brambilla:2022xbd} in the top of Fig.~\ref{G_E_oct_final}. 
We observe how both correlators agree within errors up to an overall constant. 
This constant is the same as the one we obtained when doing the LO calculation in Eq.~\eqref{G_oct_constant}, meaning that the ratio between correlators  { that}  we found at LO also seems to hold  in the non-perturbative regime. 

For $G_E^\mathrm{sym}$, where the only difference is the multiplicative constant, we used the linear ansatz in the continuum extrapolation even when including the $N_\tau = 16$ lattice. 
The final results are shown at the bottom of Fig.~\ref{G_E_oct_final}, and we observe the same behavior as 
for $G_E^\mathrm{oct}$ but with the constant given by Eq.~\eqref{G_sym_constant}. 
All three correlators have the same shape at both temperatures, and the LO result gives the ratio between them even non-perturbatively.

\subsection{Non-symmetric correlator}
The correlator $G_E$ is not symmetric by construction because we only have a single Wilson line of length $\tau$ between the fields. 
This Wilson line introduces a divergence $e^{\delta m \tau}$, where $\delta m \propto 1/\sqrt{8\tau_F}$ which makes it impossible to perform the zero-flow-time extrapolation without removing the divergence explicitly. 
This problem is absent for $G_E^{\mathrm{oct}}$ since the divergence is $e^{\delta m / T}$, 
which cancels when normalizing with the Polyakov loop $l_8$.

To renormalize $G_E$, we match our measured Polyakov loop $L_8(\tau_F)$ with the values of the renormalized Polyakov loops $L_8^{\mathrm{r}}$ from \cite{Gupta:2007ax} using the relation
\begin{equation}
    L^{\mathrm{r}}_8 = e^{\delta m(\tau_F)/T}L_8(\tau_F),
\end{equation}
%where $L_8 = l_8/8$ is the Polyakov loop normalized to one, 
and we use Eq.~(14) in~\cite{Gupta:2007ax} to obtain the Polyakov loop in the adjoint representation. With this relation, we can extract $\delta m(\tau_F)$ and use it to obtain the renormalized correlator
\begin{linenomath}\begin{align}
    G_E^{\mathrm{r}}(\tau, \tau_F) &= e^{\delta m(\tau_F) \tau} G_E(\tau, \tau_F) \nonumber \\
    &= \left(\frac{L^{\mathrm{r}}_8}{L_8(\tau_F)}\right)^{\tau T}G_E(\tau,\tau_F). \label{renormalized_G_E}
\end{align}\end{linenomath}
For the $T = 10^4~T_c$ case, we approximate $L_8\approx 1$, which is valid at this high temperature.\\
%{\color{blue}\bf [Is this the case or do we use the NNLO expression of 1512.08443?] JMS: We tested the NNLO expression, but it had only a minor/negligible effect.}

\begin{figure}
    \centering
    \includegraphics[width=0.9\linewidth]{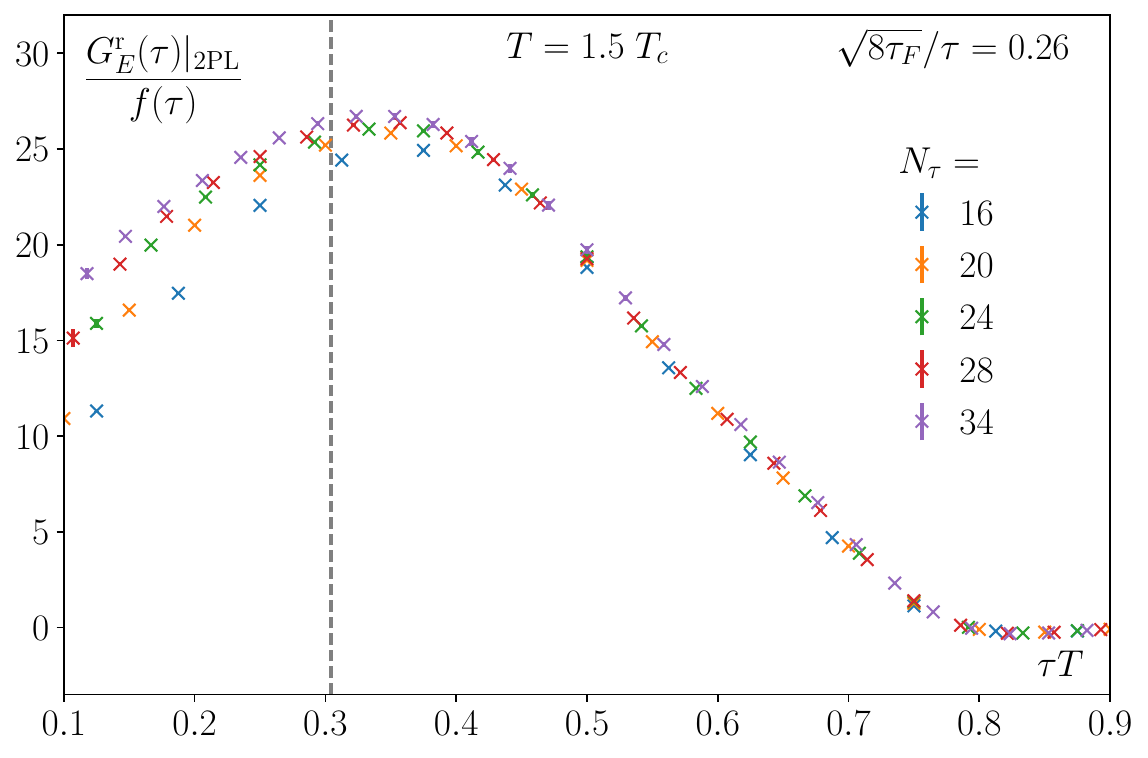} \\
    \includegraphics[width=0.9\linewidth]{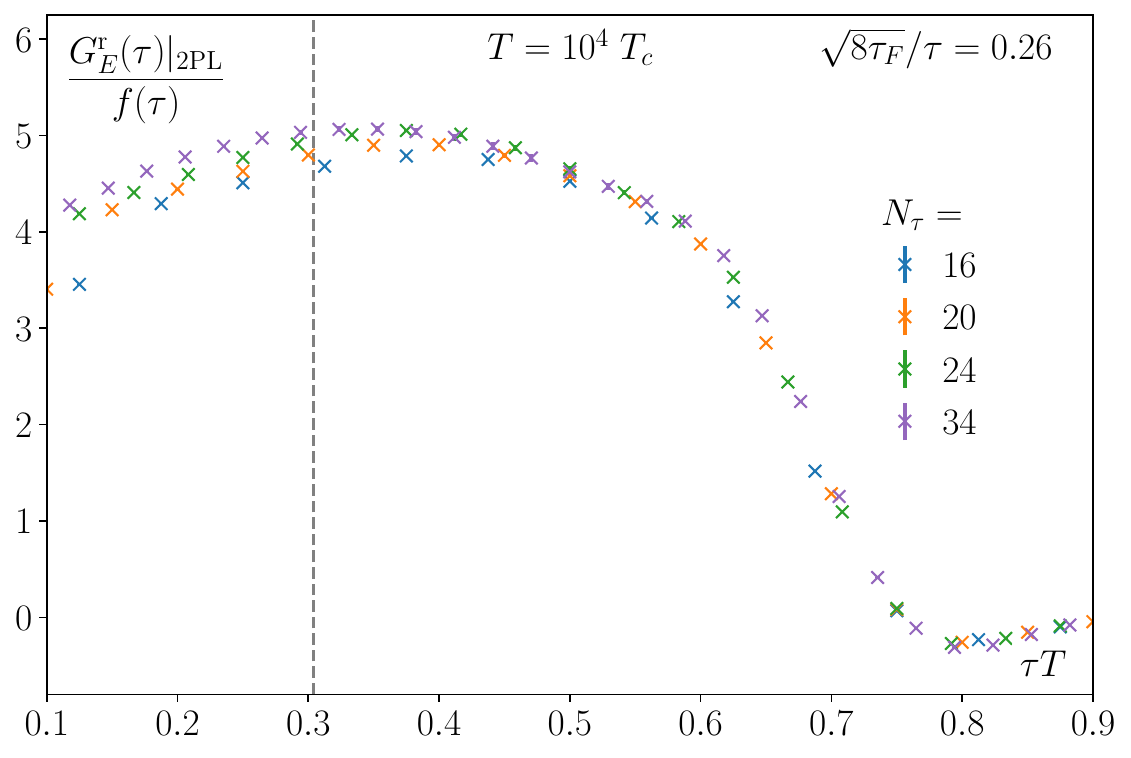}
    \caption{$G_E^{\mathrm{r}}$ normalized with $f(\tau)$ for different lattice sizes with the ratio between the flow time radius and the time separation fixed to $\sqrt{8\tau_F}/\tau = 0.26$. 
    The upper plot shows the low-temperature case, and the lower one shows the high-temperature case. 
    The vertical line indicates the validity threshold according to Eq.~\eqref{flowtimecondition}, where the points on the right-hand side are the valid points.}
    \label{G_E_r_ratio}
\end{figure}

In Fig.~\ref{G_E_r_ratio}, we show the lattice data for $G_E^{\mathrm{r}}$ for different lattice spacings
at representative values of the flow time at both temperatures, normalized with $f(\tau)$. For the flow time we again use a fixed ratio $\sqrt{8\tau_F}/\tau=0.26$. The vertical line represents the lower condition of Eq.~\eqref{flowtimecondition}, and the data on the right side of the line fulfills this condition.
We use the data in the valid regime for the continuum and zero-flow-time extrapolation, where we follow the same procedure as with $G_E^{\mathrm{oct}}$.

\begin{figure}
    \centering
    \includegraphics[width=0.9\linewidth]{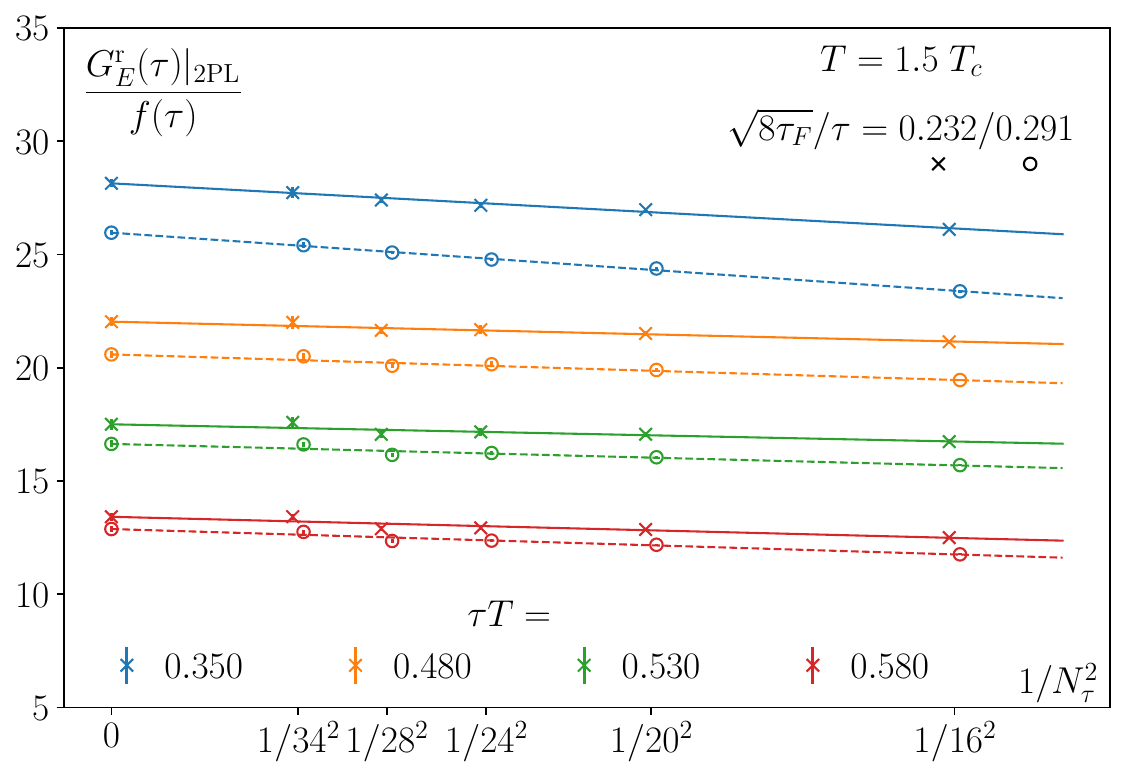} \\
    \includegraphics[width=0.9\linewidth]{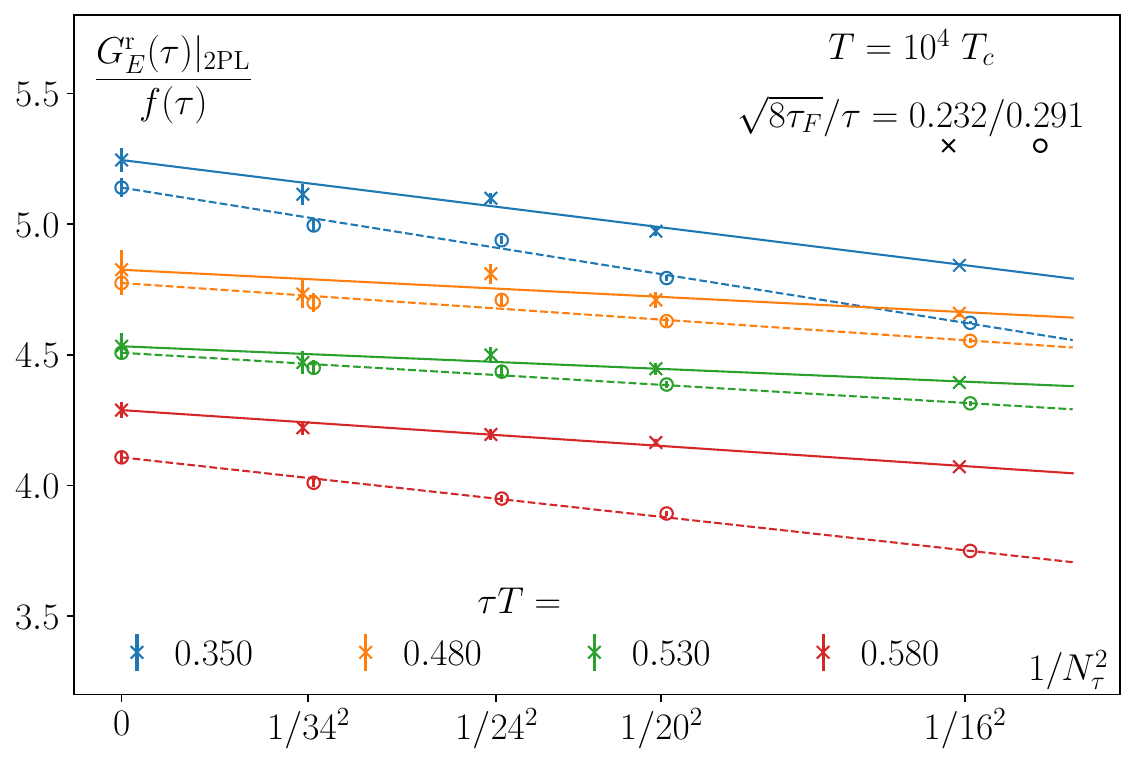}
    \caption{Examples of the continuum extrapolations for the normalized $G_E^{\mathrm{r}}$ at two different fixed ratios $\sqrt{8\tau_F/}\tau$ for different values of $\tau T$. 
    The solid lines and crosses indicate the ratio $\sqrt{8\tau_F/}\tau = 0.232$, while the dashed lines and circles $\sqrt{8\tau_F/}\tau = 0.291$. 
    The upper plot shows the low-temperature case, and the lower one shows the high-temperature case.}
    \label{G_E_cont_limit}
\end{figure}

\begin{figure}
    \centering
    \includegraphics[width=0.9\linewidth]{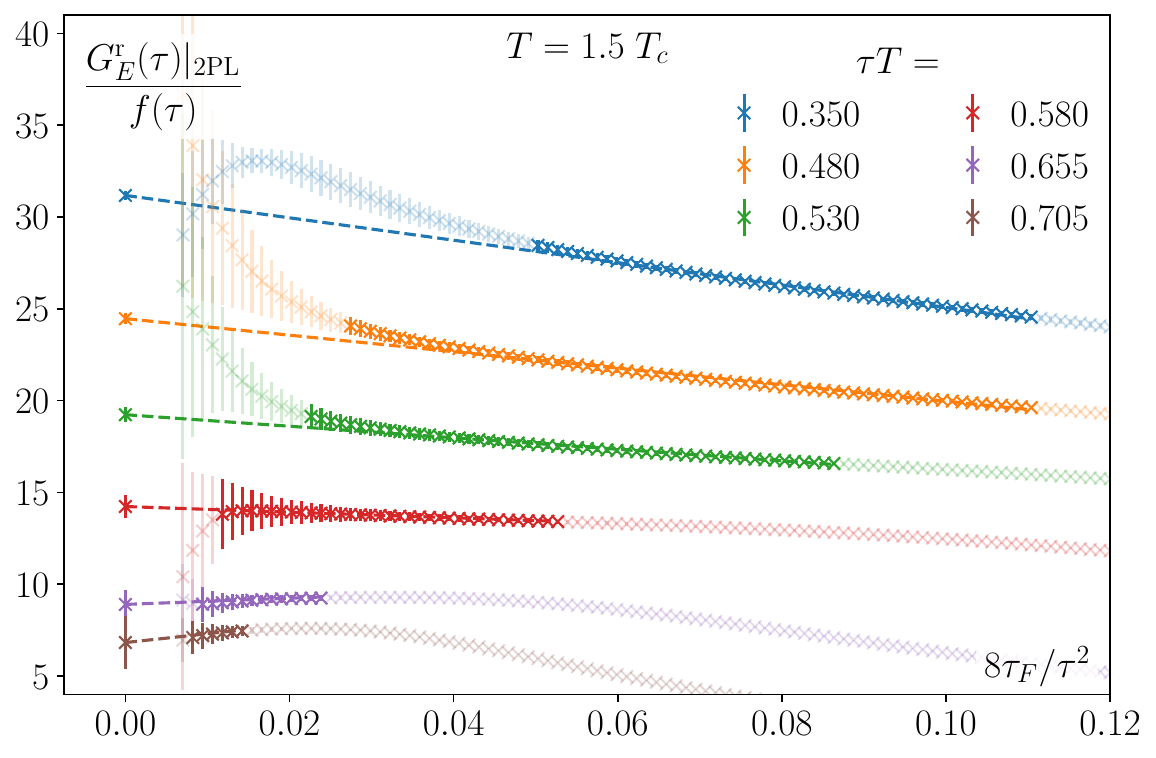} \\
    \includegraphics[width=0.9\linewidth]{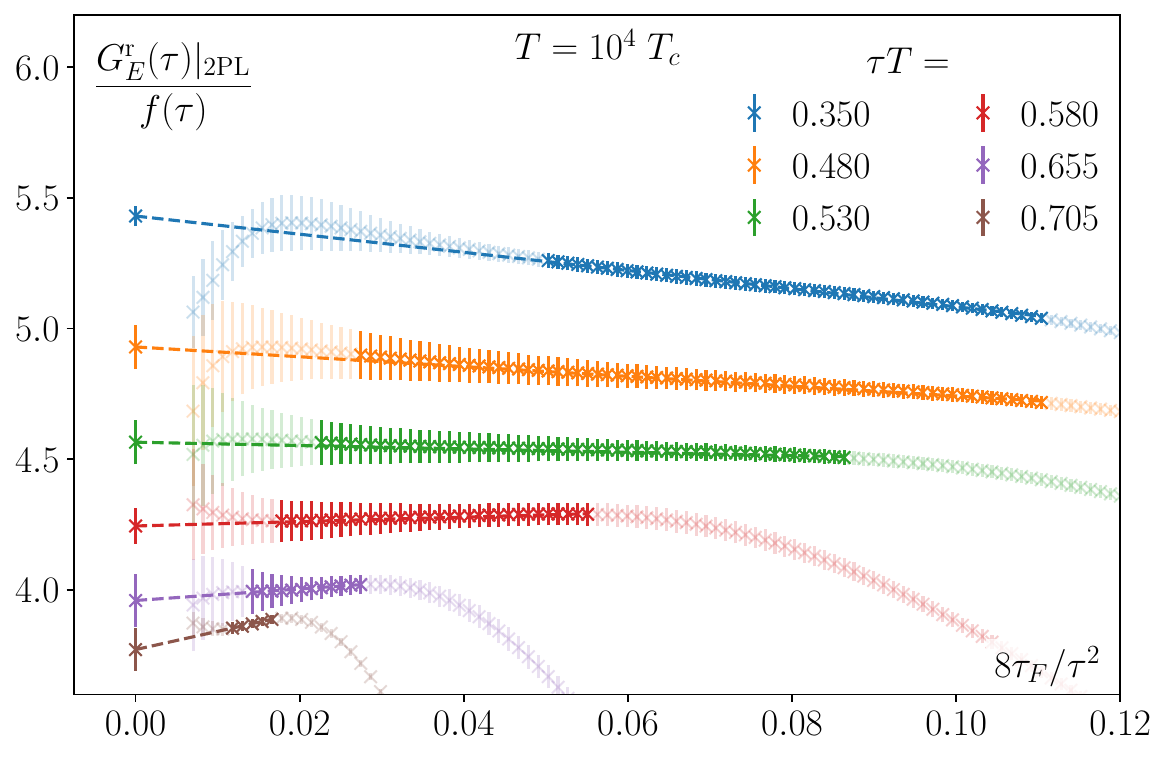}
    \caption{Examples of the zero-flow-time extrapolation for the normalized $G_E^{\mathrm{r}}$ at different $\tau T$ values. 
    { Only} the highlighted points fulfill the condition given in Eq.~\eqref{flowtimecondition}. The upper plot shows the low-temperature case, and the lower one shows the high-temperature case.}
    \label{G_E_zft_limit}
\end{figure}

The continuum extrapolation at fixed $\tau T$ is shown in Fig.~\ref{G_E_cont_limit}. 
As explained previously, we vary the number of included lattice sizes to estimate the systematic error. 
In this case, including a $1/N_\tau^4$ term does not improve the $\chi^2/d.o.f$ even when using the $N_\tau = 16$ lattice in the analysis. 
The $\chi^2/d.o.f$ values are not as good as for the symmetric correlators, where they exceed 1 in some ranges. Neglecting the $N_\tau = 24$ lattice significantly improves the continuum extrapolation for these problematic points. 
However, the extrapolated results remain consistent within the 1$\sigma$ band. 
The details  of the continuum extrapolations are given  in Appendix~\ref{app:cont_extra}. 
The zero-flow-time extrapolation is shown in Fig.~\ref{G_E_zft_limit}. 
We observe a clear linear behavior in $\tau_F$, meaning that Eq.~\eqref{renormalized_G_E} effectively removes the divergence.

\begin{figure}
    \centering
    \includegraphics[width=0.9\linewidth]{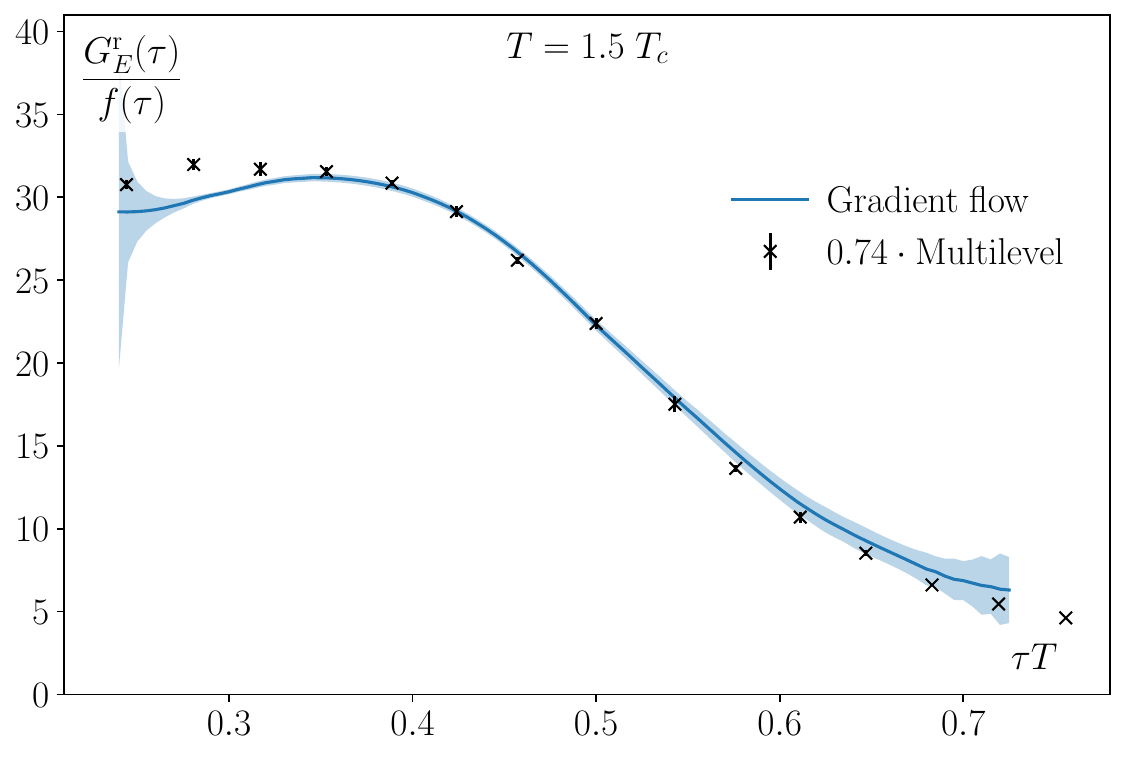} \\
    \includegraphics[width=0.9\linewidth]{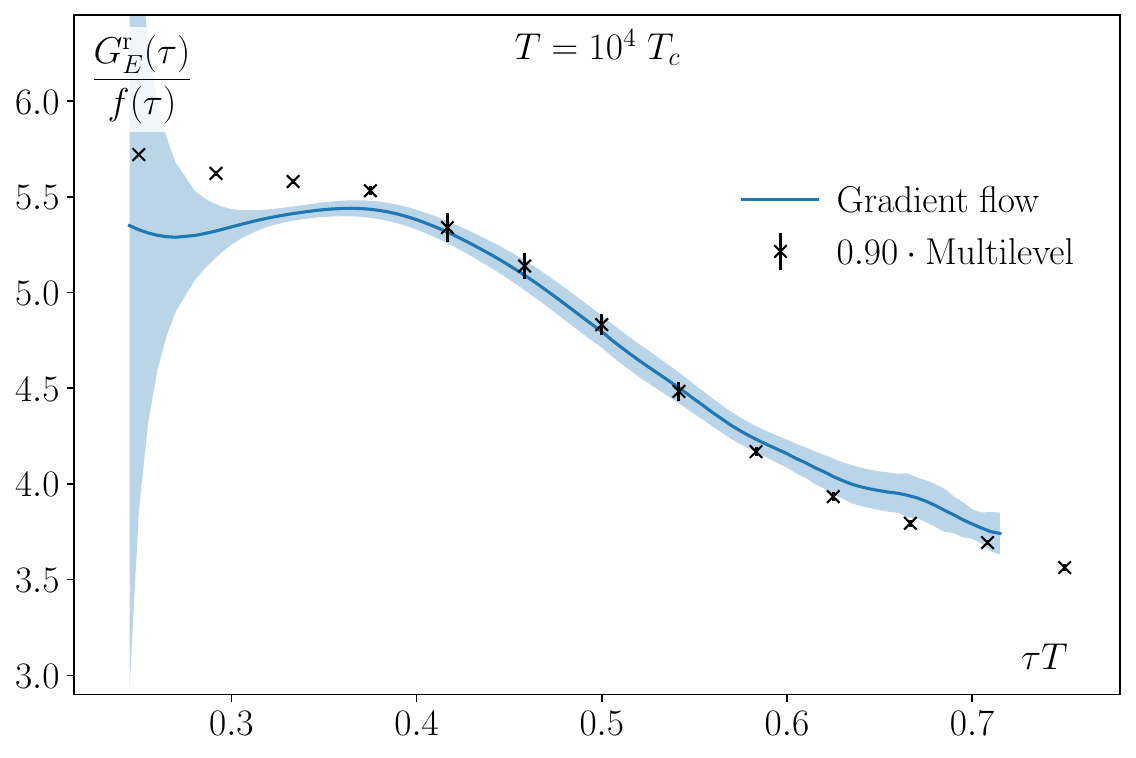}
    \caption{Final results of the normalized $G_E^{\mathrm{r}}$ with gradient flow compared with the multilevel algorithm, both obtained with the 2-plaquette discretization. The upper plot shows the low-temperature case, and the lower one shows the high-temperature case.}
    \label{G_E_r_final}
\end{figure}

In Fig.~\ref{G_E_r_final}, we show the gradient flow correlator $G_E^{\mathrm{r}}$  after the continuum and zero-flow-time extrapolations.  We compare the gradient flow results to the corresponding
computed with the multilevel algorithm after the continuum extrapolation. 
The details of the multilevel calculation are given in Appendix~\ref{app:multilevel}. 
As mentioned before the non-perturbative renormalization of the chromoelectric field insertions on the lattice is not known
when the calculations are performed using the multi-level algorithm. Therefore, the calculation of the adjoint chromoelectric 
correlators using gradient flow and multi-level algorithm can only agree up to a constant. 
We  see how the results from the different methods seem to agree up to an overall constant. 
Since this constant approaches one at high temperatures,  and the $E$ fields in the multilevel calculation are only tadpole-improved, 
we  suspect that the constant originates  from the (finite) renormalization of the chromoelectric field in the multilevel algorithm.

\subsection{Comparison with NLO results}
\begin{figure}
    \centering
    \includegraphics[width=0.9\linewidth]{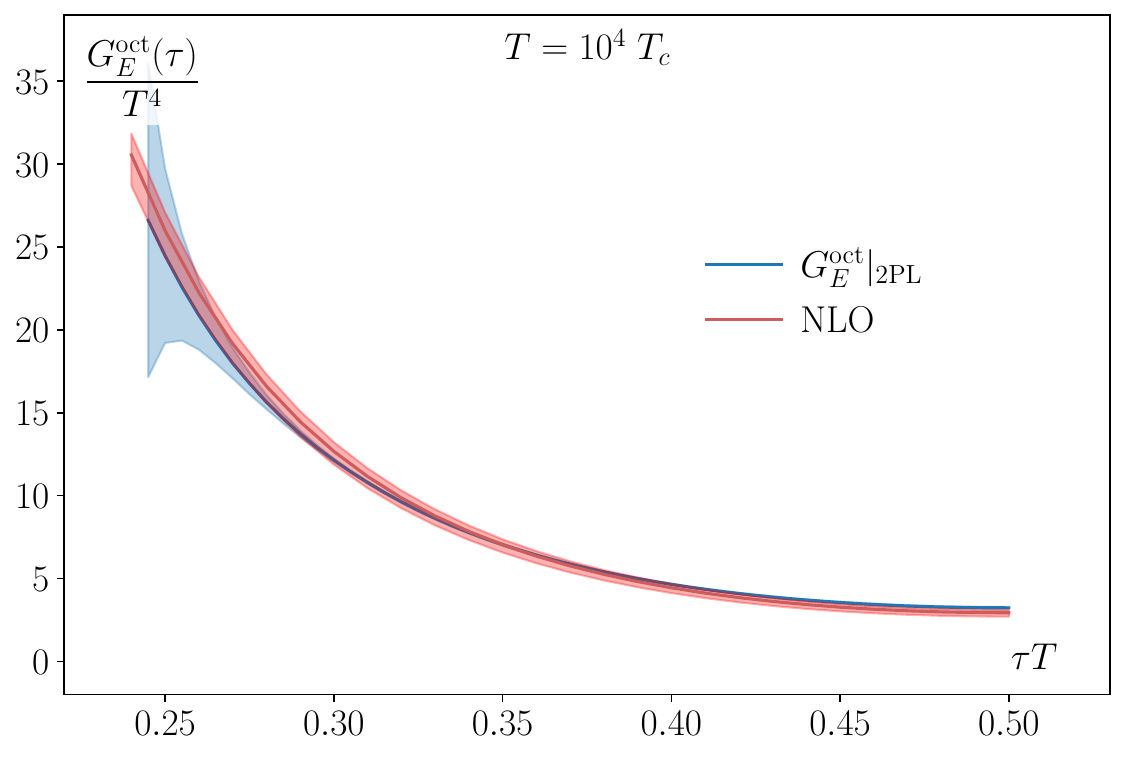} \\
    \includegraphics[width=0.9\linewidth]{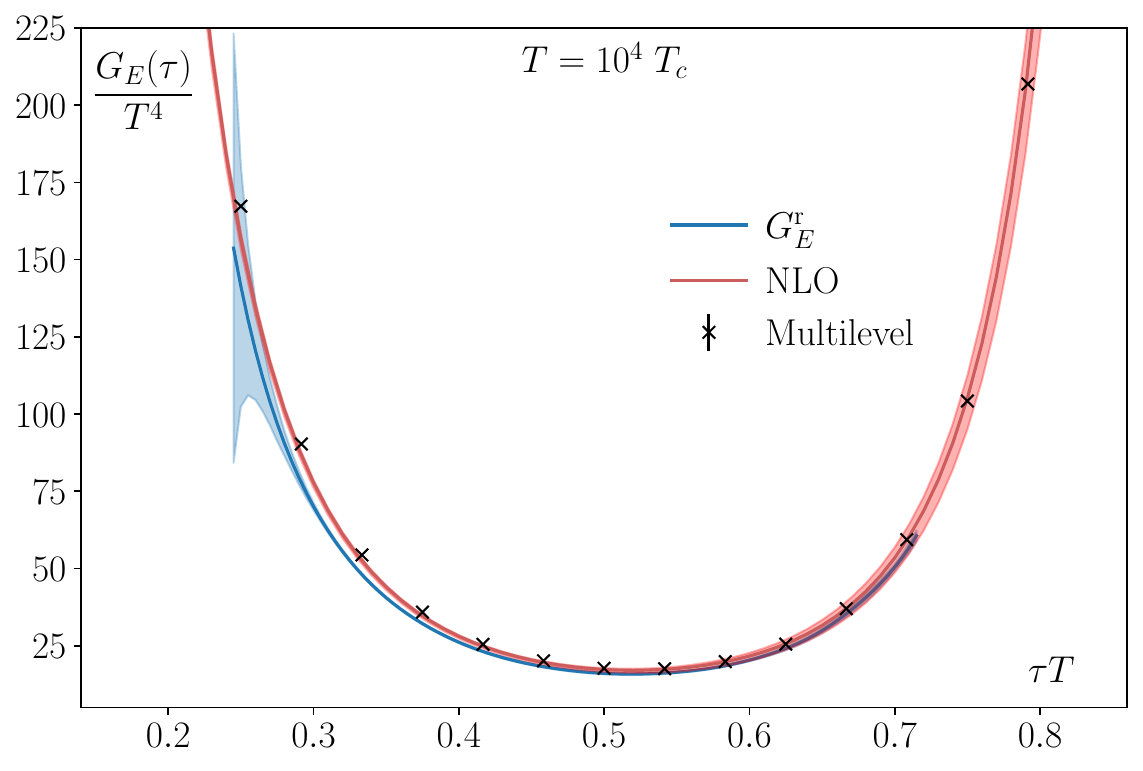}
    \caption{Gradient flow correlator $G_E^{\mathrm{oct}}$ (top) and $G_E^{\mathrm{r}}$ (bottom) after the continuum and zero-flow-time extrapolation at $T=10^4~T_c$ of the non-symmetric correlator compared to the NLO perturbative results. $G_E^{\mathrm{r}}$ is also compared with the multilevel algorithm.
    }
    \label{comparison_NLO}
\end{figure}

The next-to-leading order (NLO)  expressions, supplemented with the $g^5$ contribution, for the correlators $G_{E}(\tau)$ and $G_{E}^{\mathrm{oct}}(\tau)$ have been recently obtained in~\cite{Brambilla:AdjointCorrelatorsNLO}. 
In Fig.~\ref{comparison_NLO}, we compare our results of $G_E^{\mathrm{oct}}$ and $G_E^{\mathrm{r}}$ for $T=10^4~T_c$ with the perturbative results. 
We do not show the comparison for $G_E^{\mathrm{sym}}$, as it is similar to that of $G_E^{\mathrm{oct}}$. 
At NLO these two correlators differ by the same multiplicative factor as at LO.   Furthermore, Fig.~\ref{G_E_oct_final} shows that the same multiplicative factor also works for the nonperturbative correlators, within our error band.

The perturbative bands shown in Fig.~\ref{comparison_NLO} are obtained by varying the ${\rm \overline{MS}}$ renormalization scale in the range $[\Lambda/2, 2 \Lambda]$, where $\Lambda \, = \, 4 \pi T \, \exp(- \gamma_E - 1/22) \, \approx 6.74 \,T$. 
For $G_E^{\mathrm{oct}}$, the lattice (gradient flow) and the NLO results agree within the error bands. 
The comparison for $G_E^{\mathrm{r}}$ is shown in the lower figure, where we display
the lattice results obtained from both the gradient flow calculation and the multilevel calculation. 
The lattice results are close to the perturbative band, and show similar trends of asymmetry: 
in particular, the minimum of $G_E^{\mathrm{r}}$ is shifted to $tT > 0.5$.
The agreement with the NLO curves observed for both correlators at $T=10^4~T_c$ confirms that we are in the perturbative regime at these high temperatures.

\section{Conclusions}
\label{sec:concl}

In this paper, we have performed the first nonperturbative calculation of chromoelectric field correlators related to quarkonium transport in quenched QCD. 
We used gradient flow for noise reduction and its renormalization property for all correlators and multilevel for the non-symmetric correlator $G_E$. 

For the symmetric correlators $G_E^{\mathrm{oct}}$ and $G_E^{\mathrm{sym}}$, we find that non-perturbatively they have the same shape as the fundamental correlator $G_E^{\mathrm{fund}}$ studied previously in~\cite{Brambilla:2022xbd}. 
The ratio between the correlators fulfills the same relation as at LO, which is helpful for the extraction of the momentum diffusion coefficients $\kappa_{\mathrm{}}$ coming from them, as they turn out to be 
just a scaled version of the fundamental one: $\kappa_{\mathrm{oct}} = (5/4)\kappa_{\mathrm{fund}}$ and $\kappa_{\mathrm{sym}} = (C_A/C_F)\kappa_{\mathrm{fund}}$. 

For the non-symmetric  correlator $G_E$, we successfully extracted the divergence from the Polyakov loop  and renormalized according to Eq.~\eqref{renormalized_G_E}. 
We obtained similar results for gradient flow and multilevel. 
Furthermore, these results also agree with NLO perturbation theory at $T = 10^4~T_c$, confirming that Eq.~\eqref{renormalized_G_E} provides an effective renormalization method.

The extraction of the diffusion coefficient for the non-symmetric correlator is more involved. 
On one hand, we need the NLO perturbative result of the non-symmetric spectral function associated with that correlator. 
On the other hand, the spectral function for a non-symmetric correlator  needs a different kernel in the convolution integral that represents the corresponding chromoelectric correlator.
Once we obtain the spectral function, we plan to extract the corresponding diffusion coefficient of quarkonium. 
We also intend to explore different means of extracting the Wilson line divergence that do not depend on the Polyakov loop. 
In this way, we can extend the study to zero temperature, which is necessary for exploring other applications of the adjoint correlators, such as the extraction of gluelump masses or the calculation of the chromoelectric correlators entering quarkonium P-wave decays~\cite{Brambilla:2001xy,Brambilla:2002nu,Brambilla:2020xod}.

\acknowledgements
The simulations were carried out on the computing facilities of the Computational Center for Particle and Astrophysics (C2PAP) in the project 
\emph{Calculation of finite $T$ QCD correlators} (pr83pu) and of the SuperMUC cluster at the Leibniz-Rechenzentrum (LRZ) in the project 
\emph{Static force and other operators with field strength tensor insertions} (pn49lo), 
both located in Munich (Germany), and on the computing facilities of the Department of Theoretical Physics, TIFR.
The authors gratefully acknowledge the Gauss Centre for Supercomputing e.V. 
(\href{www.gauss-centre.eu}{www.gauss-centre.eu})
for funding this project by providing computing time on the GCS Supercomputer SuperMUC-NG 
at Leibniz Supercomputing Centre (\href{www.lrz.de}{www.lrz.de}). 
This research was funded by the Deutsche Forschungsgemeinschaft (DFG, German Research Foundation) cluster of excellence "ORIGINS" (\href{https://www.origins-cluster.de}{www.origins-cluster.de}) under Germany's Excellence Strategy EXC-2094-390783311.
N.B. acknowledges the European Research Council advanced grant  ERC-2023-ADG-Project EFT-XYZ.  
N.B. and J.M.-S. acknowledge the DFG Grant No. BR 4058/2-2. 
S.D. acknowledges support of the Department of Atomic Energy, Government of India, under Project
Identification No.\ RTI 4002.
J.M.-S. acknowledges support by the Munich Data Science Institute (MDSI) at the Technical University of Munich (TUM) via the Linde/MDSI Doctoral Fellowship program.
This work was also supported by  U.S. Department of Energy, Office of Science, Office of Nuclear Physics through Contract No.~DE-SC0012704
and through "Heavy Flavor Probes of QCD Matter (HEFTY)" topical collaboration in nuclear theory. 

\appendix

\section{Deriving the fundamental representations of the adjoint correlators}\label{app:fierz_stuff}
The Fierz completeness relation for Gell-Mann matrices~\cite{Gell-Mann:1962yej} reads
\begin{linenomath}\begin{align}
    \lambda_a^{\alpha\beta}\lambda_a^{\gamma\delta} = 2\delta_{\alpha\delta}\delta_{\gamma\beta} - \frac{2}{3}\delta_{\alpha\beta}\delta_{\gamma\delta}\,, \label{eq:fierz_identity}
\end{align}\end{linenomath}
where $\delta_{\alpha\beta}$ is the Kronecker delta and color indices are summed over. 
The structure constants can be expressed in terms of Gell-Mann matrices:
\begin{linenomath}\begin{align}
    f_{abc} &= -\frac{1}{4}i\mathrm{Tr}\left( \lambda_a[\lambda_b,\lambda_c] \right),\\
    d_{abc} &= \frac{1}{4}\mathrm{Tr}\left( \lambda_a\{ \lambda_b,\lambda_c\} \right),
\end{align}\end{linenomath}
with $[.,.]$ the commutator and $\{ .,.\}$ the anticommutator operator.

We start by introducing two identities for frequently used contractions (see e.g., Ref.~\cite{Luscher:1995np}). In the following, $X$ and $Y$ are general two $3\times 3$ matrices, and the Einstein sum convention applies. 
The two identities read
\begin{linenomath}\begin{align}
    \mathrm{Tr}\left[\lambda_a X\right]\mathrm{Tr}\left[ \lambda_a Y\right] &= \mathrm{Tr}\left[ XY\right] - \frac{2}{3}\mathrm{Tr}\left[ X\right]\mathrm{Tr}\left[ Y\right] \label{eq:trxtry_fierz_identity}\\
    \mathrm{Tr}\left[ \lambda_a X \lambda_a Y\right] &= 2\mathrm{Tr}\left[ X\right] \mathrm{Tr}\left[ Y\right] - \frac{2}{3}\mathrm{Tr}\left[ XY\right]\label{eq:trxy_fierz_identity}\,.
\end{align}\end{linenomath}
Both identities can be derived by expanding the left-hand side and inserting the Fierz relation iteratively.

We can now prove Eq.~\eqref{eq:non_symmetric_Gee_fundamental_representation}. 
For some adjoint Wilson line $U^{\mathrm{adj}}$, according to Eq.~\eqref{eq:adjoint_wilson_line_definition}, and two fields $O$ and $\hat{O}$, we obtain~\cite{Foster:1998wu}
\begin{align*}
    O_aU^{\mathrm{adj}}_{ab}\hat{O}_b &= \mathrm{Tr}\left[\lambda_aO\right]\frac{1}{2}\mathrm{Tr}\left[U^\dagger\lambda_aU\lambda_b\right]\mathrm{Tr}\left[\lambda_b\hat{O}\right]\\
    &= 2\mathrm{Tr}[OU\hat{O}U^\dagger] - \frac{2}{3}\mathrm{Tr}[O]\mathrm{Tr}[\hat{O}],
\end{align*}
where we have used Eq.~\eqref{eq:trxtry_fierz_identity}, the cyclicity of the trace, that the Gell-Mann matrices are traceless, and that $U$ are SU(3) matrices. 
Eq.~\eqref{eq:non_symmetric_Gee_fundamental_representation} follows from identifying $O=E_i(\tau)$, $\hat{O}=E_i(0)$, $U=U(\tau,0)$, and $U^\dagger=U(0,\tau)$. 
By taking traceless fields, only the first term remains.

For the octet correlator in the left-hand side of  Eq.~\eqref{eq:octet_Gee_fundamental_representation}, we have an additional adjoint temporal Wilson line $\hat{U}^{\mathrm{adj}}$, and the general structure of the operator in the correlator reads
\begin{align*}
    U^{\mathrm{adj}}_{ea}d_{abc}O_c\hat{U}^{\mathrm{adj}}_{bd}d_{def}\hat{O}_f\,.
\end{align*}
We go through the following steps to find the representation in terms of fundamental Wilson lines. 
First, we establish the contraction of the structure constant with the field component
\begin{align*}
    &d_{abc}O_c = \frac{1}{4}\mathrm{Tr}\left[ \lambda_a\{\lambda_b,\lambda_c\}\right] \mathrm{Tr}\left[\lambda_cO\right] \\
    &\quad = \frac{1}{2}\left(\mathrm{Tr}\left[\lambda_a\lambda_bO\right] + \mathrm{Tr}\left[\lambda_b\lambda_aO\right] - \frac{2}{3}\mathrm{Tr}\left[\lambda_a\lambda_b\right] \mathrm{Tr}\left[O\right]\right),
\end{align*}
and second, the contraction with an adjoint Wilson line
\begin{align*}
    U^{\mathrm{adj}}_{ea}d_{abc}O_c &= \frac{1}{2}\mathrm{Tr}\left[ U^\dagger\lambda_eU\lambda_a\right] \frac{1}{4}\mathrm{Tr}\left[ \lambda_a\{\lambda_b,\lambda_c\}\right] \mathrm{Tr}\left[\lambda_cO\right]\\
    &= \frac{1}{2}\Big( \mathrm{Tr}\left[U^\dagger\lambda_eU\lambda_bO\right] +\mathrm{Tr}\left[U^\dagger\lambda_eUO\lambda_b\right] \\
    &\ \ \ - \frac{2}{3}\mathrm{Tr}\left[U^\dagger\lambda_eU\lambda_b\right]\mathrm{Tr}[O]\Big)\,.
\end{align*}
This leads to
\begin{align*}
    &U^{\mathrm{adj}}_{ea}d_{abc}O_c\hat{U}^{\mathrm{adj}}_{bd}d_{def}\hat{O}_f \\ &= \mathrm{Tr}[U\hat{U}]\mathrm{Tr}[OU^\dagger\hat{O} \hat{U}^\dagger] + \mathrm{Tr}[O\hat{U}\hat{O}U]\mathrm{Tr}[\hat{U}^\dagger U^\dagger]\\
    &\ \ \ \ +\mathrm{Tr}[OU^\dagger\hat{U}^\dagger ]\mathrm{Tr}[U\hat{U}\hat{O}] + \mathrm{Tr}[O\hat{U}U]\mathrm{Tr}[\hat{U}^\dagger U^\dagger \hat{O}]\\
    &\ \ \ \ -\frac{4}{3}\mathrm{Tr}[O\hat{U}\hat{O}\hat{U}^\dagger] - \frac{4}{3}\mathrm{Tr}[OU^\dagger \hat{O}U]\\
    &\ \ \ \ -\frac{2}{3}\mathrm{Tr}[OU^\dagger\hat{U}^\dagger]\mathrm{Tr}[U\hat{U}]\mathrm{Tr}[\hat{O}] - \frac{2}{3}\mathrm{Tr}[O\hat{U}U]\mathrm{Tr}[\hat{U}^\dagger U^\dagger] \mathrm{Tr}[\hat{O}] \\
    &\ \ \ \ -\frac{2}{3}\mathrm{Tr}[O]\mathrm{Tr}[U\hat{U}]\mathrm{Tr}[\hat{U}^\dagger U^\dagger\hat{O}] - \frac{2}{3}\mathrm{Tr}[O]\mathrm{Tr}[\hat{U}^\dagger U^\dagger]\mathrm{Tr}[U\hat{U}O] \\
    &\ \ \ \ + \frac{4}{9} \mathrm{Tr}[O]\mathrm{Tr}[U\hat{U}]\mathrm{Tr}[\hat{U}^\dagger U^\dagger]\mathrm{Tr}[\hat{O}] + \frac{8}{9}\mathrm{Tr}[O]\mathrm{Tr}[\hat{O}]\,.
\end{align*}
Eq.~\eqref{eq:octet_Gee_fundamental_representation} follows from identifying
$U = U(1/T, \tau)$, $\hat{U} = U(\tau, 0)$, $O = E(\tau)$, $\hat{O} = E(0)$, and $\mathrm{Tr}[O] = \mathrm{Tr}[E] = 0$.

For the symmetric adjoint correlator in the left-hand side of Eq.~\eqref{eq:symmetric_Gee_fundamental_representation}, 
we go through similar steps as for the octet correlator: first,
\begin{align*}
    U^{\mathrm{adj}}_{ea}f_{abc}O_c &= -\frac{1}{2}\mathrm{Tr}[U^\dagger \lambda_eU\lambda_a]\frac{i}{4}\mathrm{Tr}[\lambda_a[\lambda_b,\lambda_c]]\mathrm{Tr}[\lambda_cO]\\
    &= -\frac{i}{2}\left( \mathrm{Tr}\left[U^\dagger\lambda_e U \lambda_b O\right] - \mathrm{Tr}\left[ U^\dagger \lambda_e U O\lambda_b\right] \right),
\end{align*}
and then
\begin{align*}
    &U^{\mathrm{adj}}_{ea}f_{abc}O_c\hat{U}^{\mathrm{adj}}_{bd}f_{def}\hat{O}_f \\
    &= \mathrm{Tr}[OU^\dagger \hat{U}^\dagger]\mathrm{Tr}[U\hat{U}\hat{O}] + \mathrm{Tr}[O\hat{U}U]\mathrm{Tr}[\hat{U}U\hat{E}] \\
    &\ \ \ \ -\mathrm{Tr}[OU^\dagger \hat{O}\hat{U}^\dagger] - \mathrm{Tr}[O\hat{U}\hat{O}U]\mathrm{Tr}[\hat{U}^\dagger U^\dagger]\,.
\end{align*}
Eq.~\eqref{eq:symmetric_Gee_fundamental_representation} follows then from the same identifications as before.

\section{Multilevel calculation of $G_{E}(\tau)$}\label{app:multilevel}
Here, we give the details of the calculation of the non-flowed correlators
using the multilevel algorithm \cite{Luscher:2001up}. After thermalization,
the lattice is split into $N_{\rm sub}$ sublattices in the time direction. For the sublattice averaging, $N_{\rm avg}$ sweeps of each sublattice were conducted, each sweep consisting of 1 heatbath and 3 overrelaxation sweeps. 
The sublattice averaged parts were then combined to give a measurement of the correlator. $N_{\rm meas}$ gives the number of such measurements used for the calculation of each correlator.
The various
parameters for the different lattices are shown in
Table~\ref{tab:multilevel}. 

\begin{table}
\caption{Simulation details for the multilevel runs.}
\label{tab:multilevel}
\centering
\begin{ruledtabular}
\begin{tabular}{cccccc}
$T/\Tc$ & $N_\tau$ & $N_s$ & $N_{\rm sub}$  & $N_{\rm avg}$ & $N_{\rm meas}$ \\
\hline
1.5 & 16 & 48 & 4 & 500 & 5850 \\
& 20 & 64 & 5 & 500 & 4668 \\
& 24 & 72 & 6 & 500 & 5624 \\
& 28 & 64 & 7 & 500 & 5850 \\
10000 & 16 & 48 & 4 & 200 & 11700 \\
& 20 & 48 & 5 & 200 & 9100 \\
& 24 & 48 & 4 & 250 & 12100 \\
\end{tabular} \end{ruledtabular} 
\end{table}

The mass renormalization of the adjoint Wilson line in the correlator is performed in the same way as in Eq.~\eqref{renormalized_G_E}. 
We also need to renormalize the electric field operators. 
Unfortunately, there are no independent estimates of the nonperturbative renormalization constant
for the electric field operators. 
We have used the tadpole renormalization, where the tadpole factor $Z_{\rm tad}$ is calculated from the plaquette
operator. The renormalized multilevel correlator is then
\begin{equation}
  G_E^{\mathrm{r}}(\tau, T; a=1/N_\tau T) \; = \; Z_{\rm tad}^2 \;
  \left(\frac{L^{\mathrm{r}}_8(T)}{L_8(T;a)}\right)^{\tau T} G_E(\tau, T; a).
  \label{Gmlnorm}
\end{equation}

\begin{figure*}
    \centering
    \includegraphics[width=0.45\linewidth]{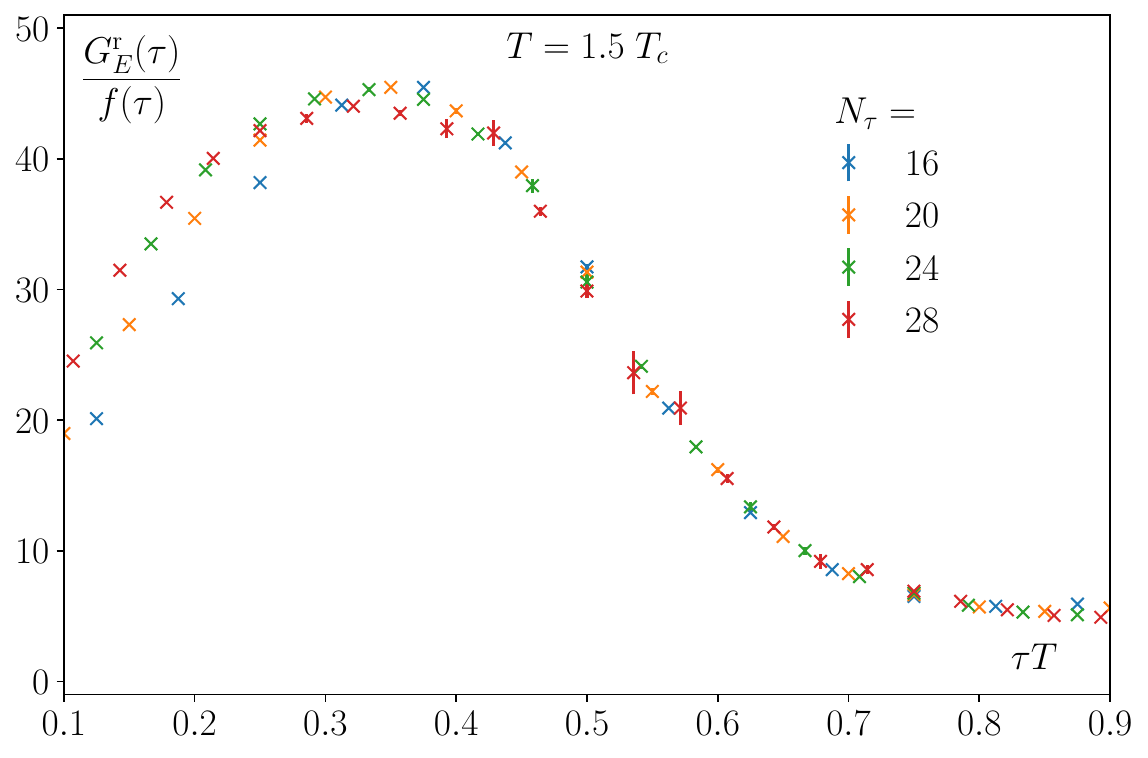} \includegraphics[width=0.45\linewidth]{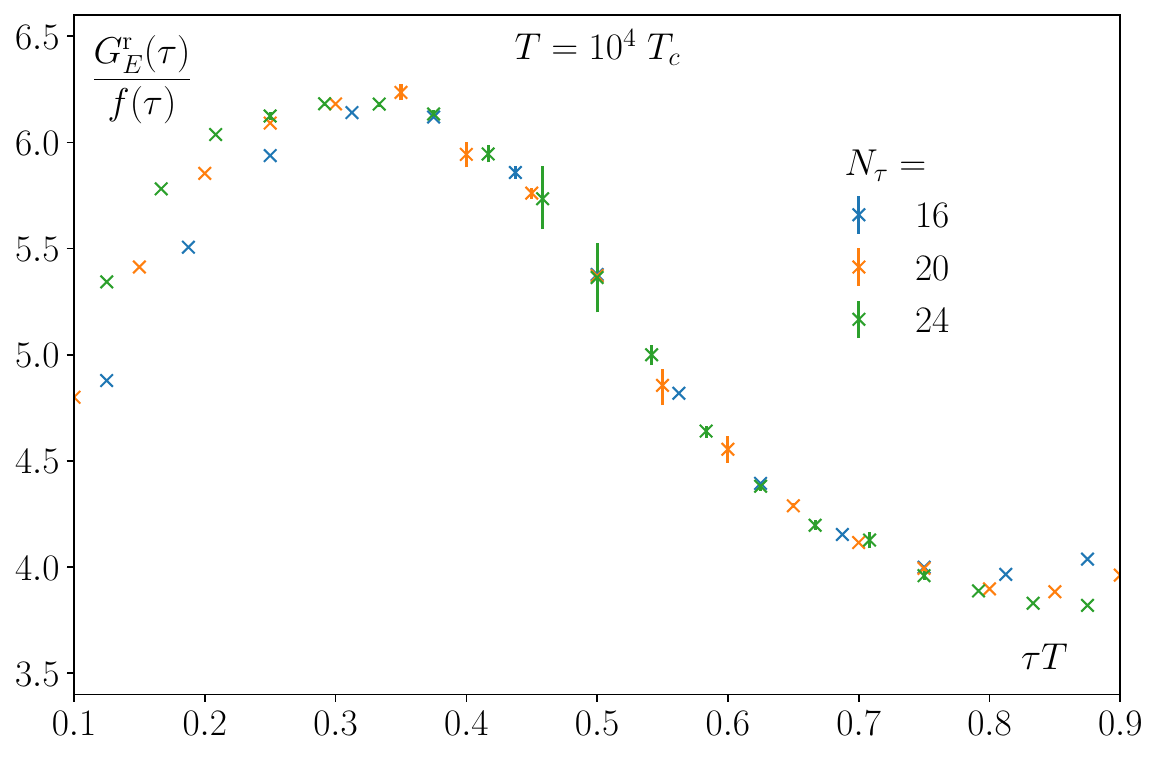} 
    \caption{Results for the multilevel calculation of $G_E^{\mathrm{r}}(\tau;a)$
    at 1.5 $T_c$ (left) and at red $10^4\; T_c$ (right) with the 2PL discretization.}
    \label{ML-2PL}
\end{figure*}    

The results for the correlator with the 2PL discretization are shown in Fig.~\ref{ML-2PL}. 
The cutoff effect is small for $\tau T \gtrsim 0.3$, and a reliable
continuum extrapolation is possible from these lattices. 
This continuum extrapolated result is shown in Fig.~\ref{G_E_r_final}. 
As mentioned there, an overall normalization $\sim 0.74$ and $0.9$ are needed to have agreement
between the results from the multilevel and the flow calculations. 
We expect this to be due to the use of the simple $Z_{\rm tad}$ factor in Eq.~\eqref{Gmlnorm}.

\begin{figure*}
    \centering
    \includegraphics[width=0.45\linewidth]{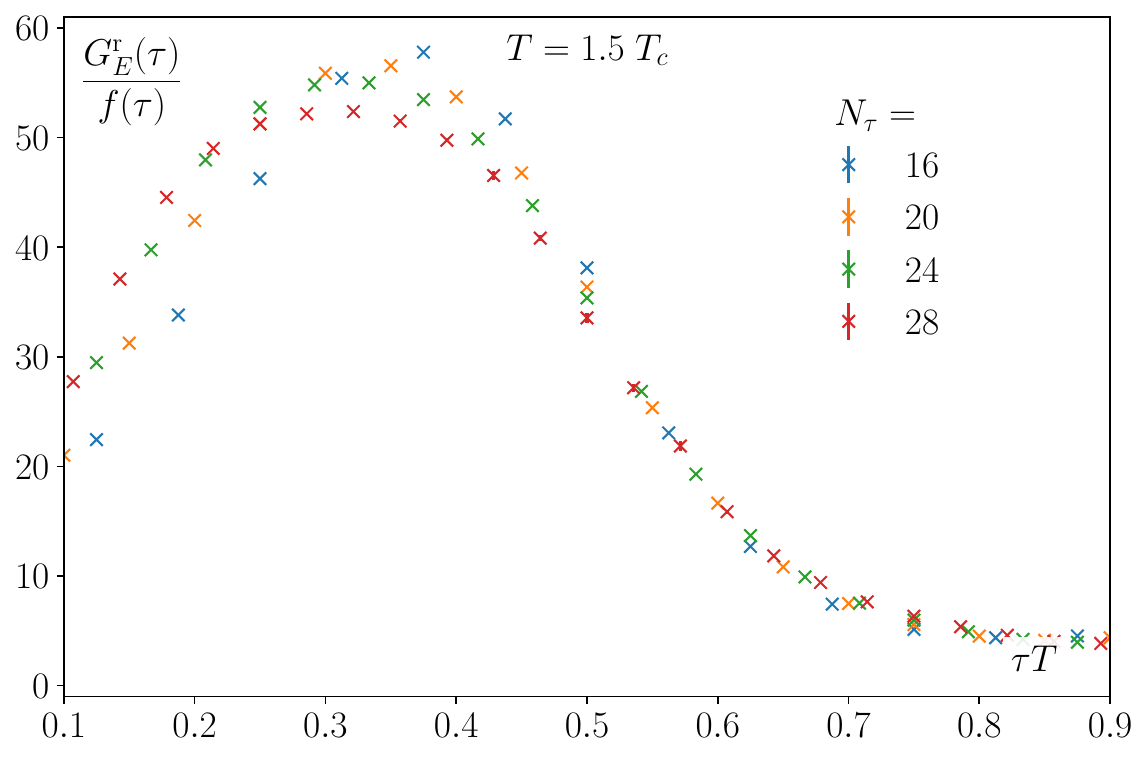} \includegraphics[width=0.45\linewidth]{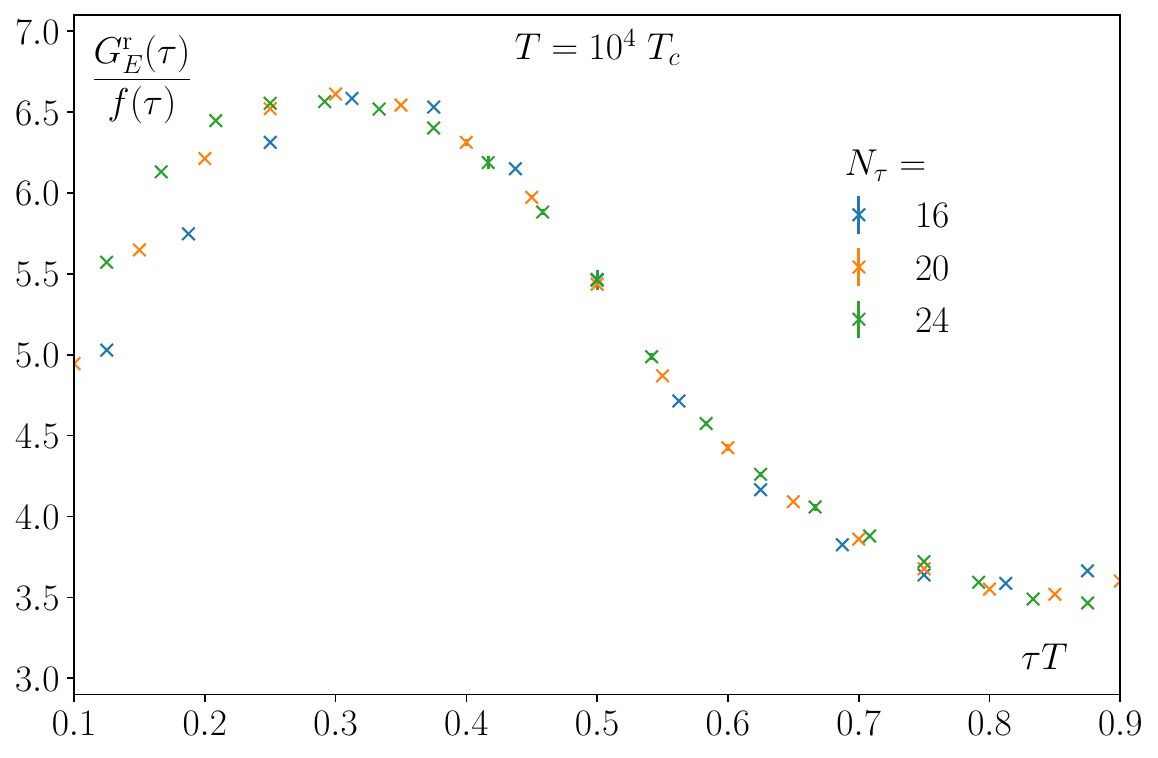} 
    \caption{Results for the multilevel calculation of $G_E^{\mathrm{r}}(\tau;a)$
    at 1.5 $T_c$ (left) and at $10^4 \;T_c$ (right) with the CLO discretization.}
    \label{ML-clov}
\end{figure*}

The clover discretized correlators are shown in Fig.~\ref{ML-clov}. 
The cutoff effect is substantial till much larger distances $\tau T \gtrsim 0.5$. 
Due to the overlap of the Wilson line with the leaves of the clover operator, the renormalization property of this operator is expected to be more complicated, and it is more difficult to obtain
continuum results from unsmeared operators. 

\section{Continuum extrapolations and discretizations} \label{app:cont_extra}
This appendix discusses different continuum extrapolations, how we extracted the systematic error, and the differences between using the CLO and 2PL discretization. 
We mainly focus on four different methods: Using all lattices in Table~\ref{tab:simulation_parameters} with a linear ansatz, labeled as $\mathrm{All}$; excluding the $N_\tau = 16$ lattice, labeled as $\mathrm{No\;16}$; excluding both the $N_\tau = 16$ and $N_\tau = 20$ lattices, labeled as $\mathrm{No\;16\;20}$; 
and using all lattices while adding a $1/N_\tau^4 = (aT)^4$ term, labeled as $\mathrm{All}\;\mathcal{O}(a^4)$.

\begin{figure*}
    \centering
    \includegraphics[width=0.425\linewidth]{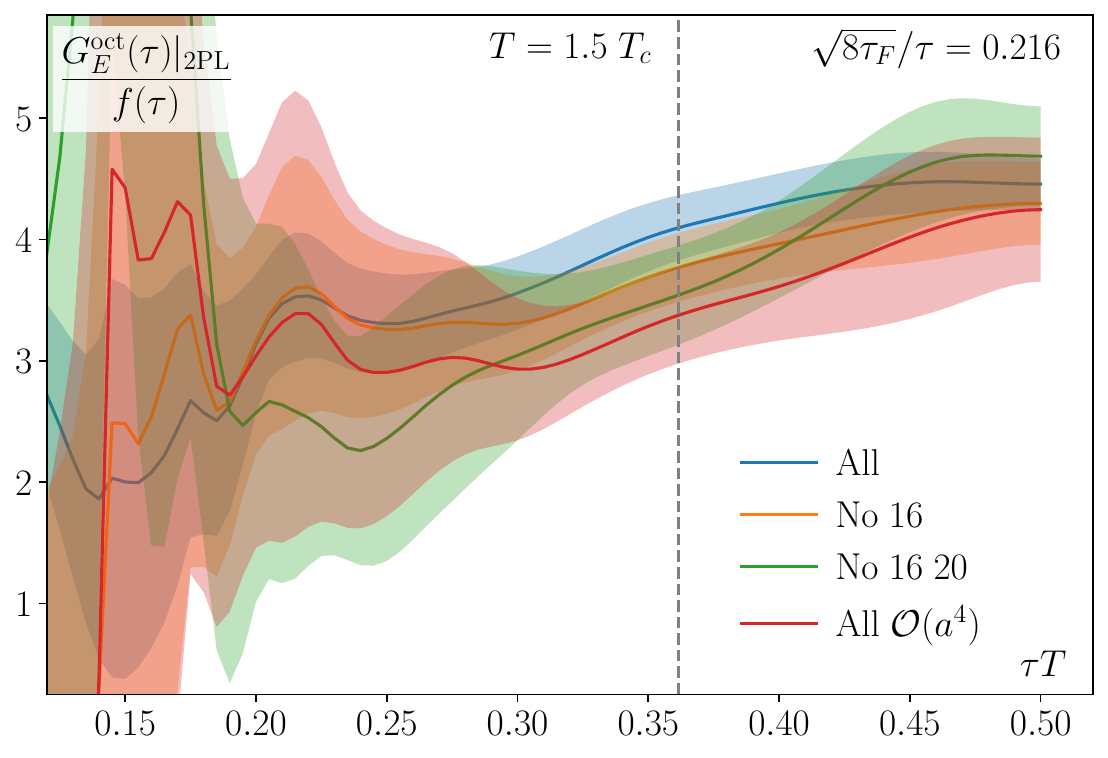} \includegraphics[width=0.425\linewidth]{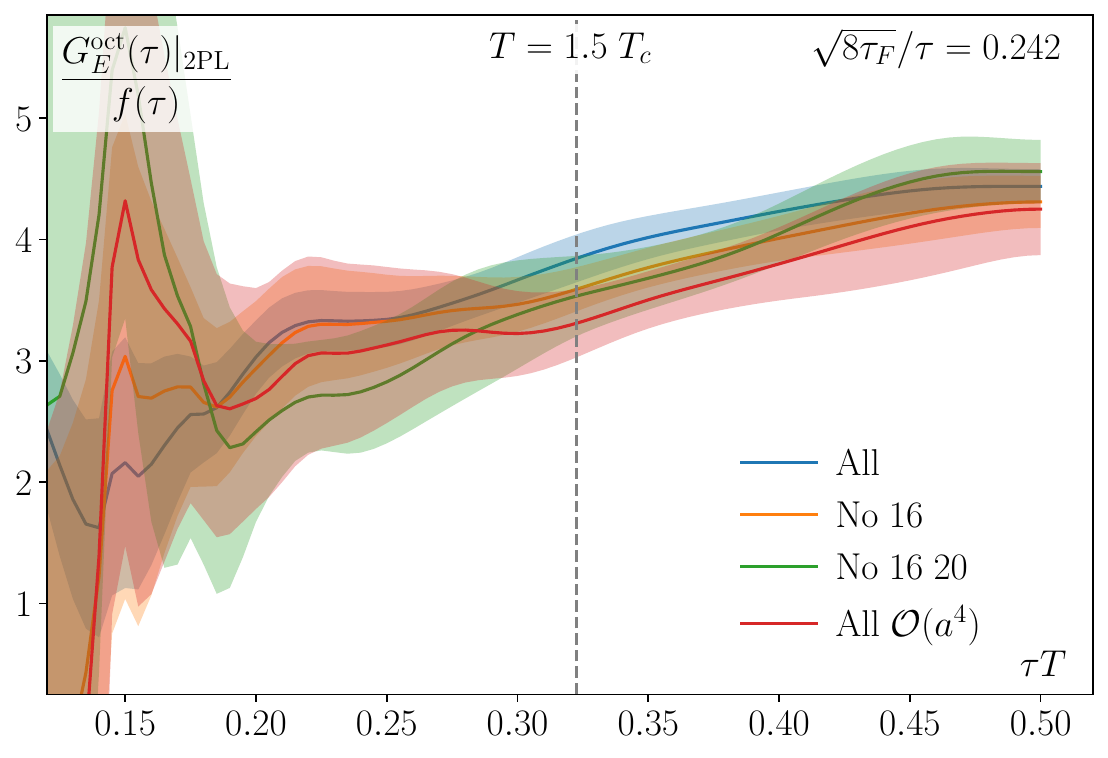} \\
    \includegraphics[width=0.425\linewidth]{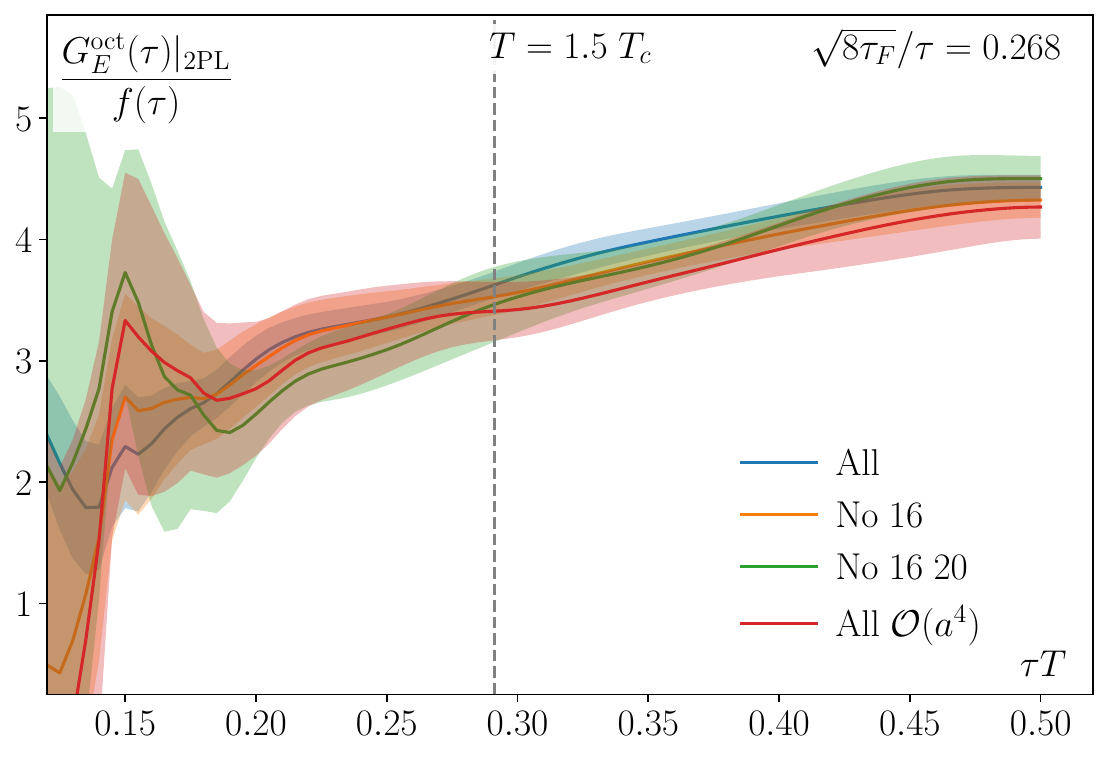} \includegraphics[width=0.425\linewidth]{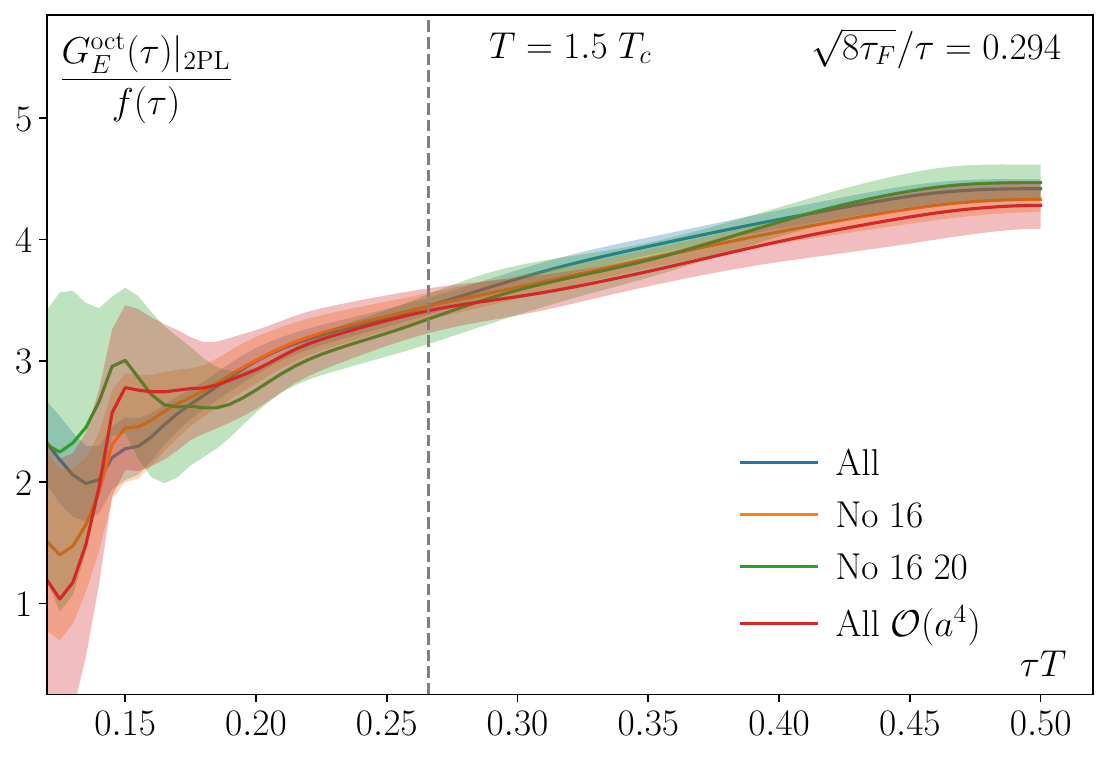}
    \caption{$G_E^{\mathrm{oct}}$ with different continuum extrapolations for different flow times at $T = 1.5\;T_c$.}
    \label{G_E_oct_comparison_1_5}
\end{figure*}

\begin{figure*}
    \centering
    \includegraphics[width=0.425\linewidth]{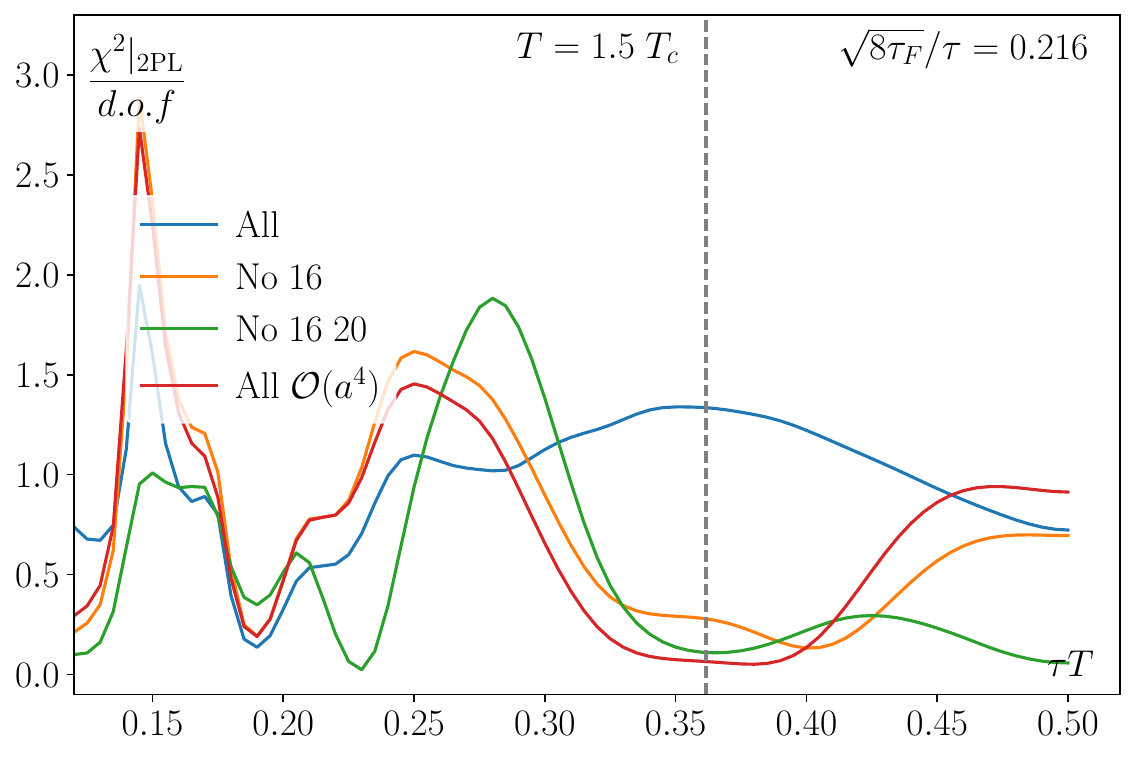} \includegraphics[width=0.425\linewidth]{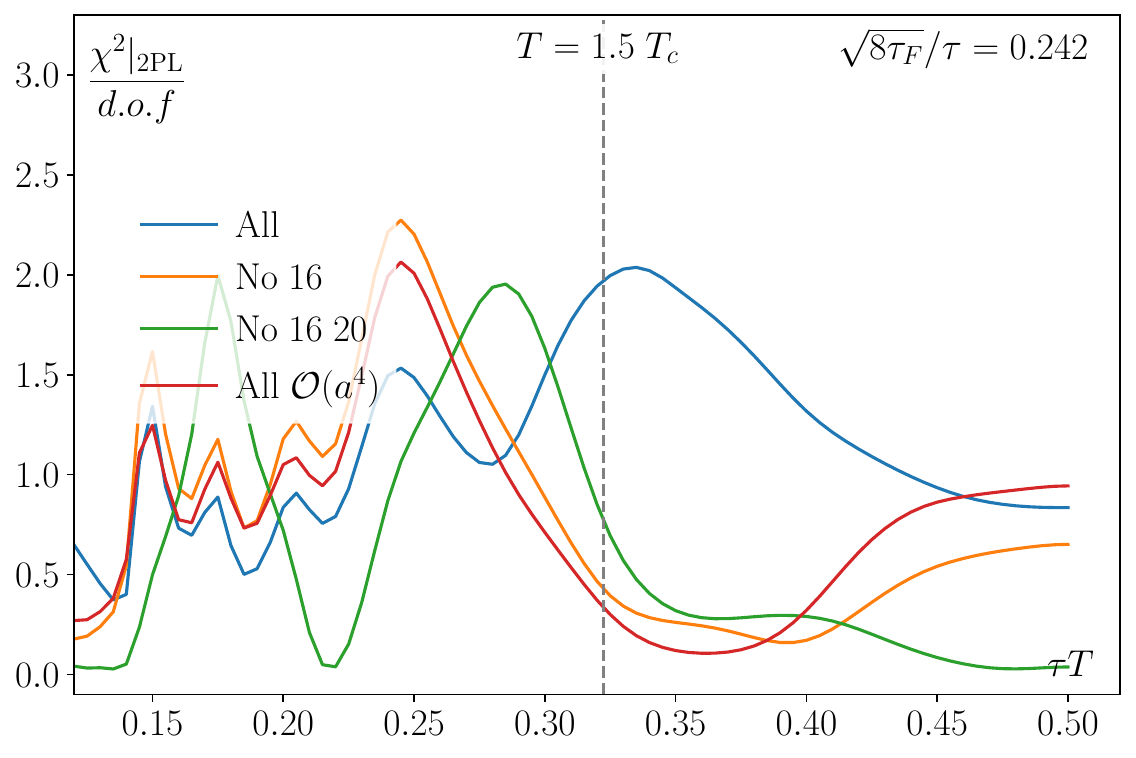} \\
    \includegraphics[width=0.425\linewidth]{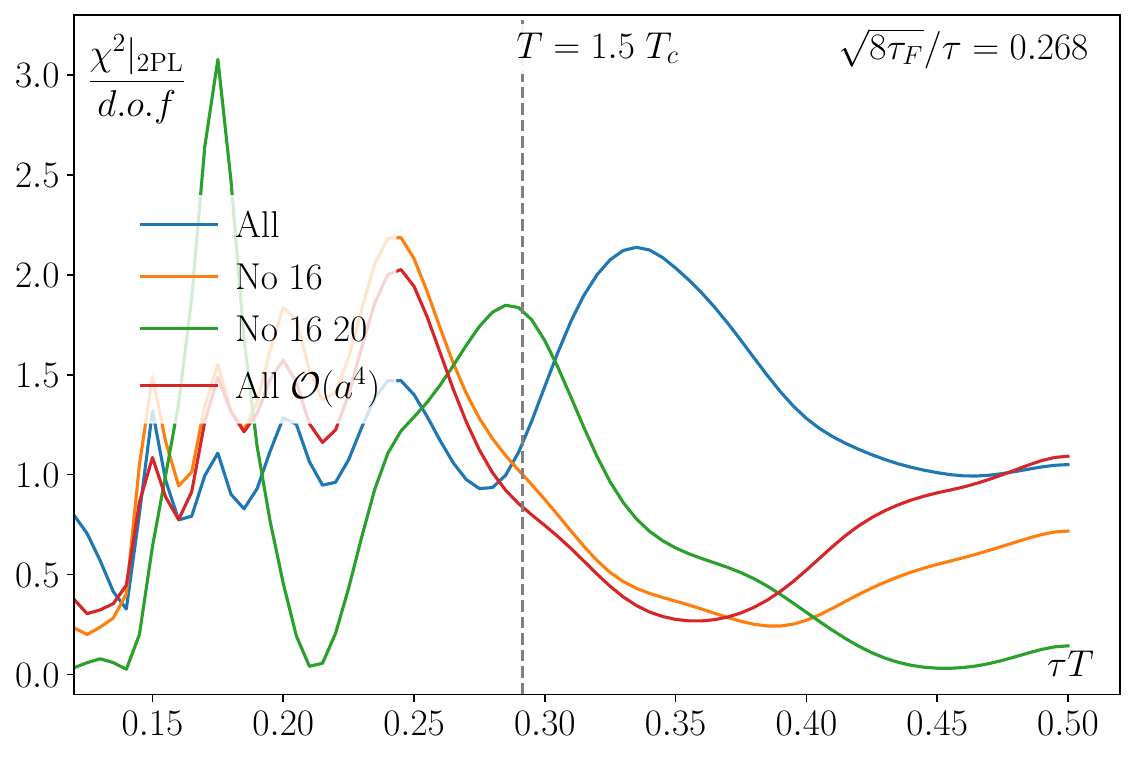} \includegraphics[width=0.425\linewidth]{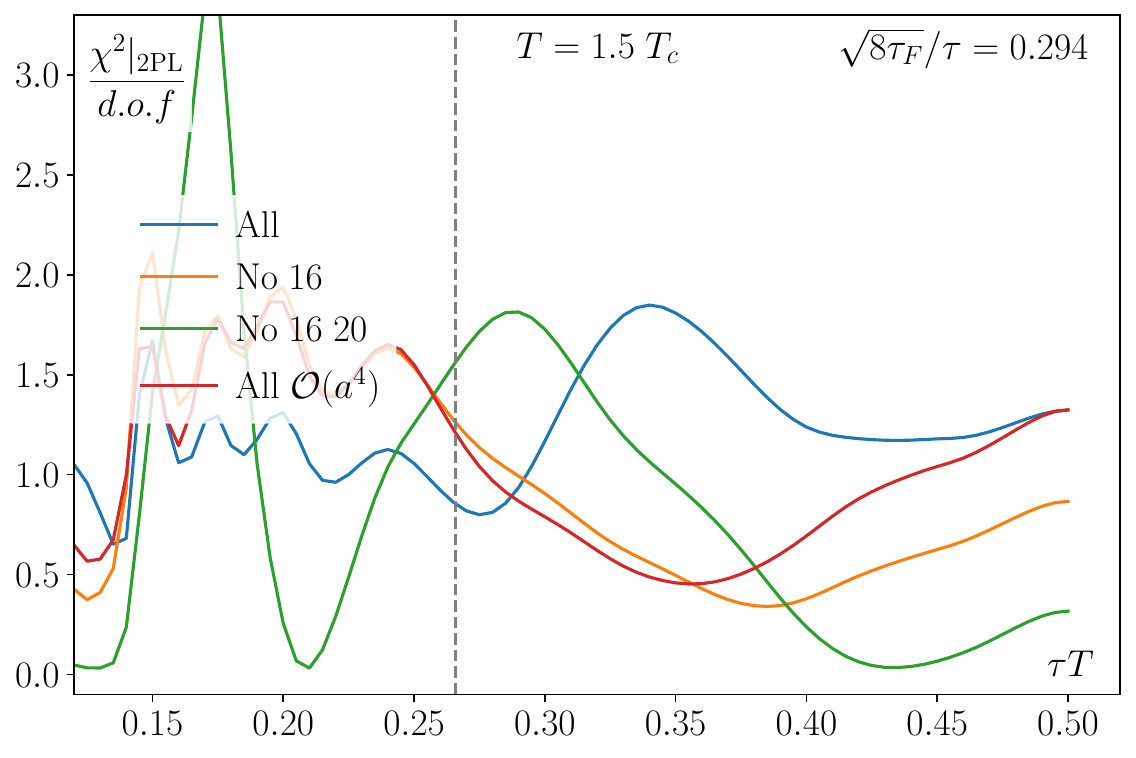}
    \caption{$\chi^2$ over degrees of freedom of the continuum extrapolations corresponding to Fig.~\ref{G_E_oct_comparison_1_5}.}
    \label{G_E_oct_chi2_comparison_1_5}
\end{figure*}

In Fig.~\ref{G_E_oct_comparison_1_5}, we show the continuum extrapolated $G_E^{\mathrm{oct}}$ at $T = 1.5\;T_c$ for different flow time values. 
As we increase the flow time, the different methods converge on the region where the condition Eq.~\eqref{flowtimecondition} is fulfilled. 
The corresponding values for the reduced $\chi^2$ values are shown in Fig.~\ref{G_E_oct_chi2_comparison_1_5}. There, we can see that using the $a^4$ term is better when including the $N_\tau=16$ lattice. 
The methods that are around and below one are $\mathrm{All}\;\mathcal{O}(a^4)$ and $\mathrm{No\;16}$. 
We estimate the systematic error by adding in quadrature the difference of these two measurements to our error budget.

\begin{figure*}
    \centering
    \includegraphics[width=0.425\linewidth]{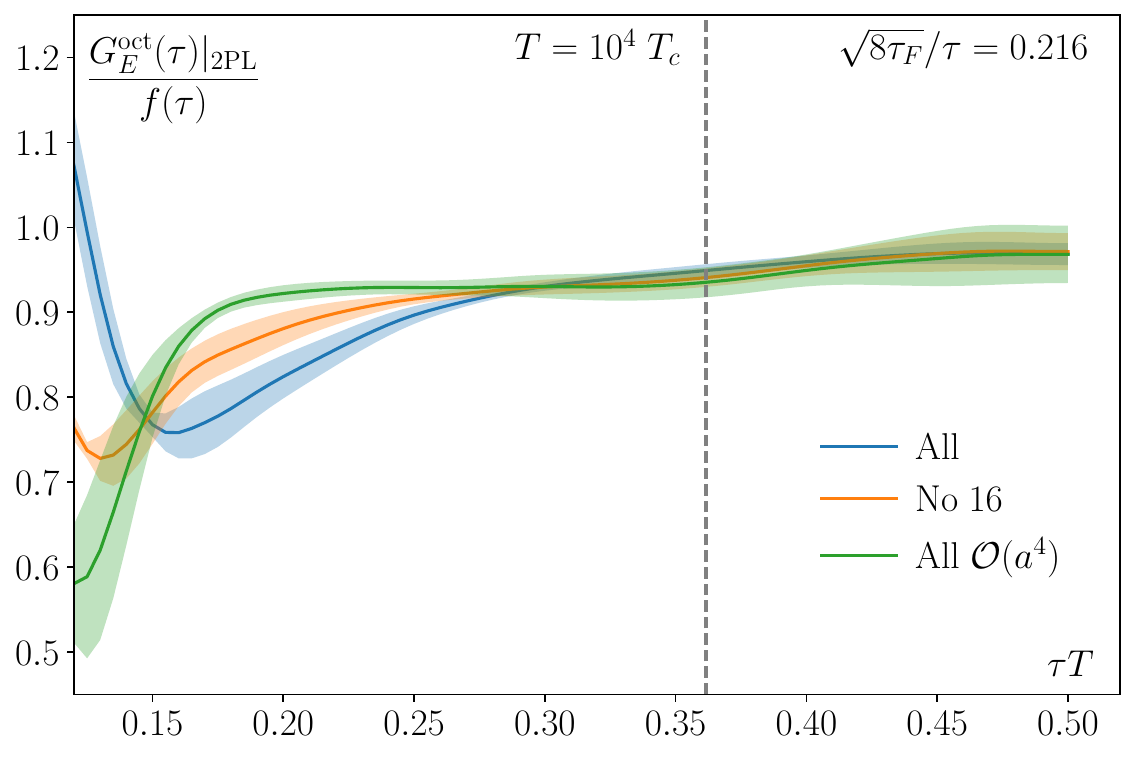} \includegraphics[width=0.425\linewidth]{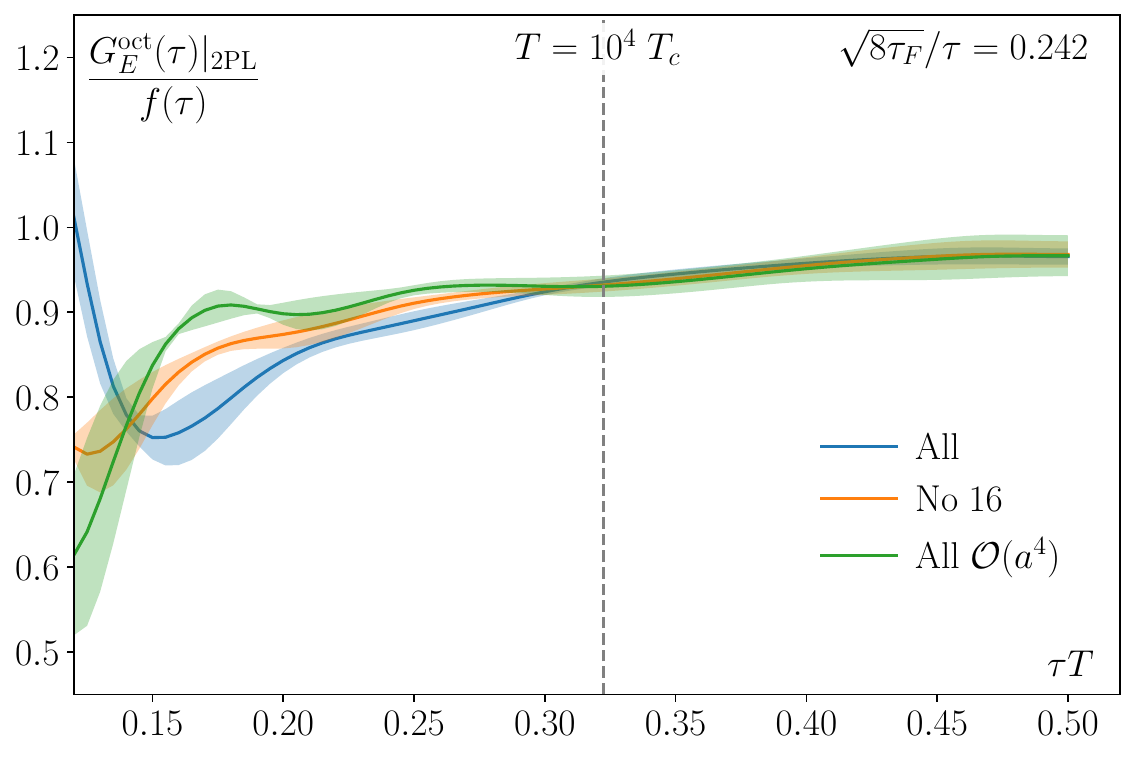} \\
    \includegraphics[width=0.425\linewidth]{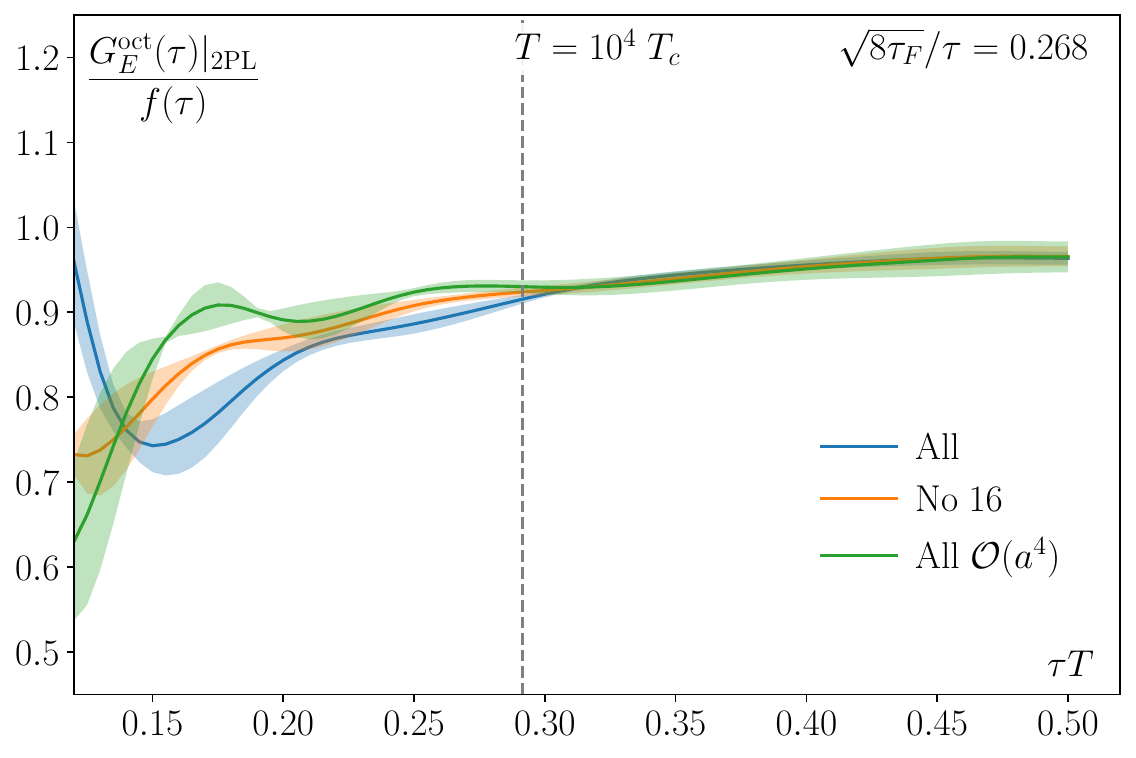} \includegraphics[width=0.425\linewidth]{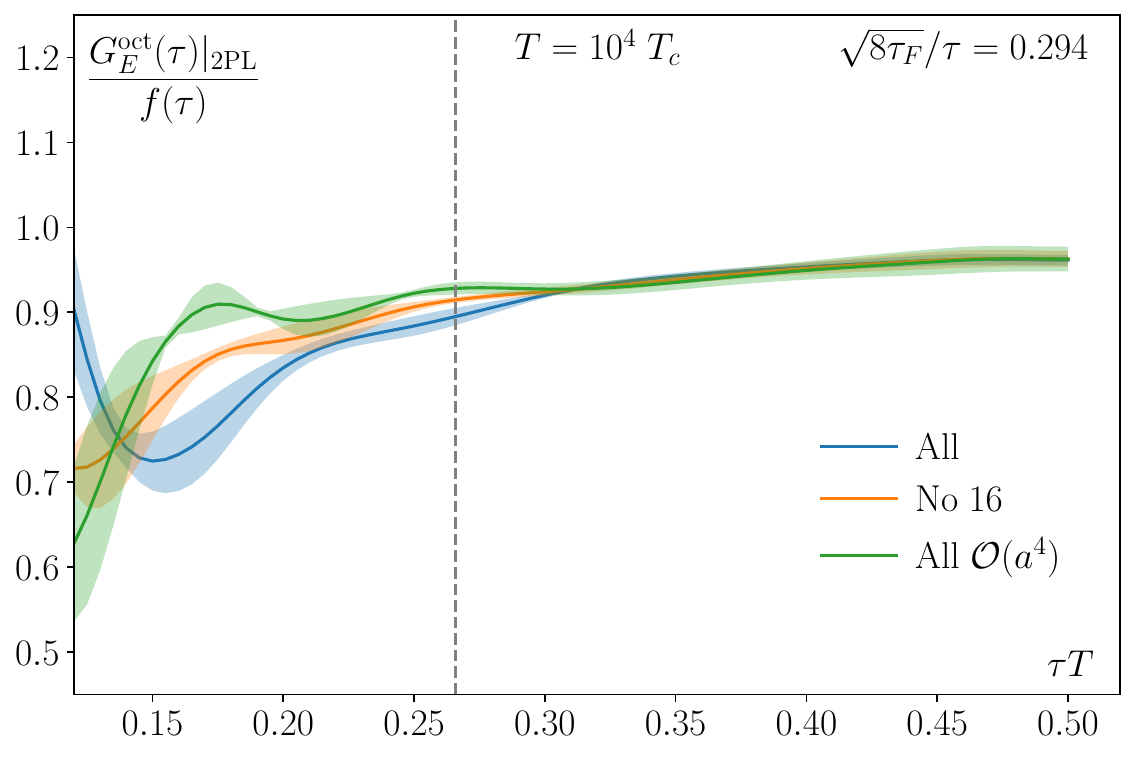}
    \caption{$G_E^{\mathrm{oct}}$ with different continuum extrapolations for different flow times at $T = 10^4\;T_c$.}
    \label{G_E_oct_comparison_10_4}
\end{figure*}

\begin{figure*}
    \centering
    \includegraphics[width=0.425\linewidth]{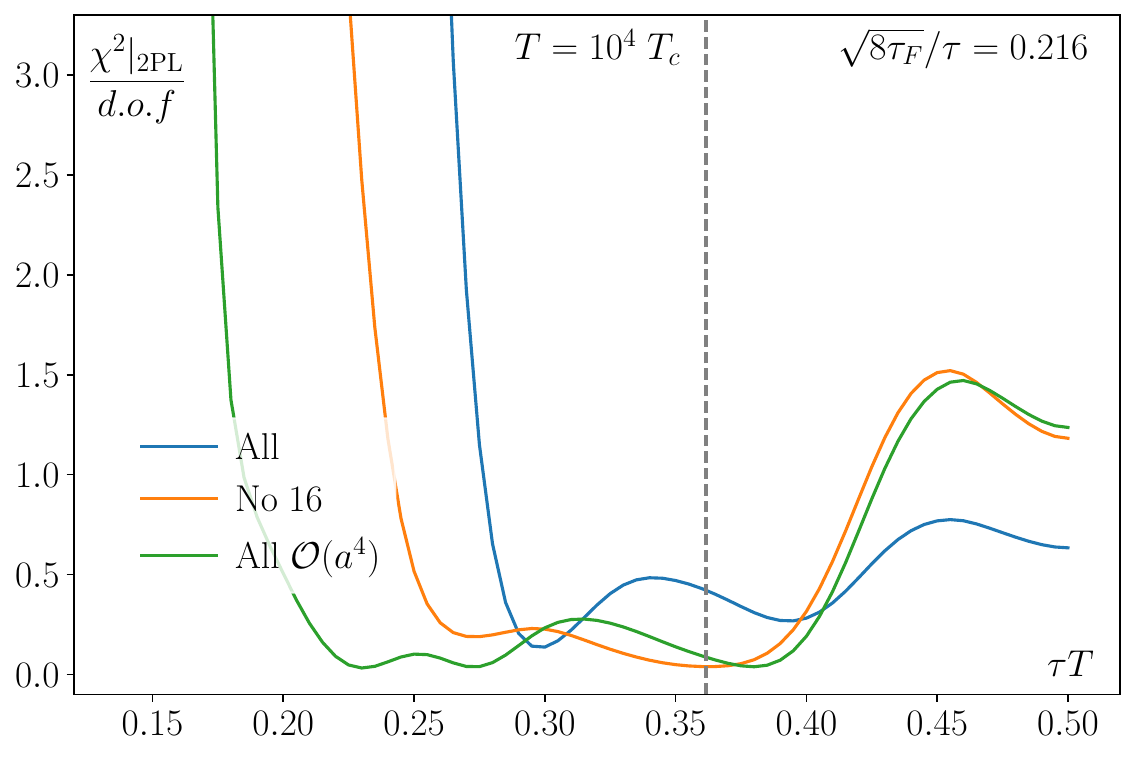} \includegraphics[width=0.425\linewidth]{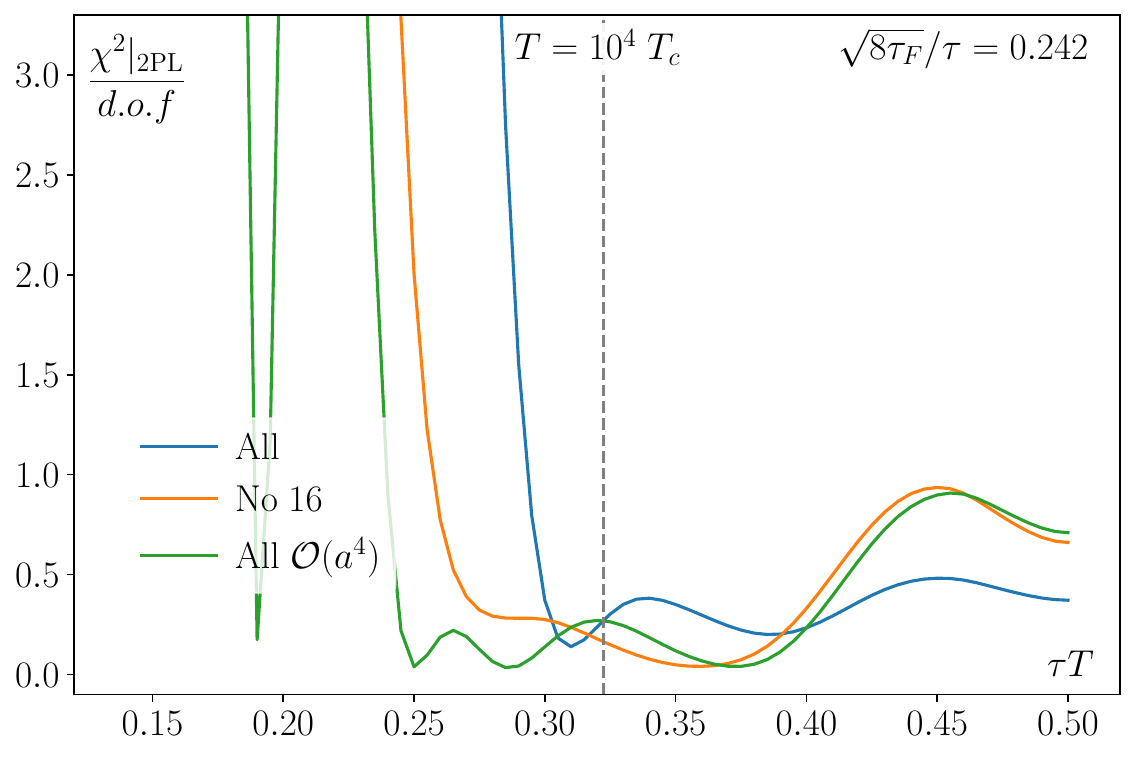} \\
    \includegraphics[width=0.425\linewidth]{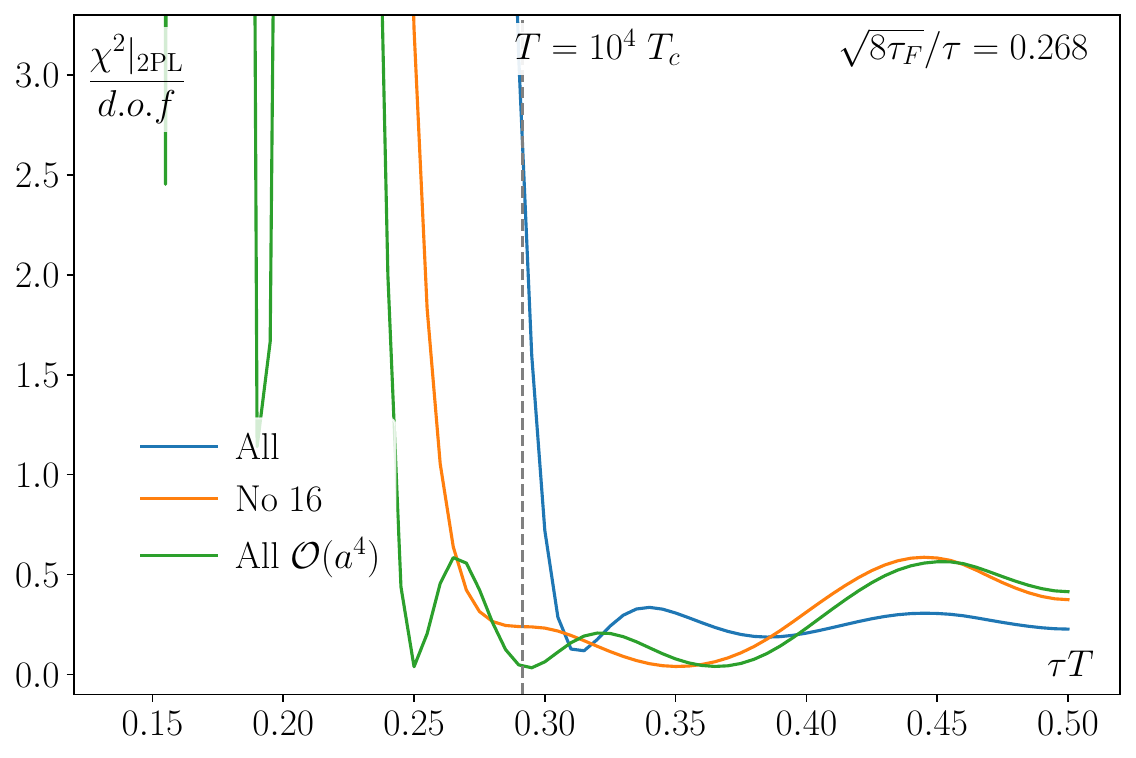} \includegraphics[width=0.425\linewidth]{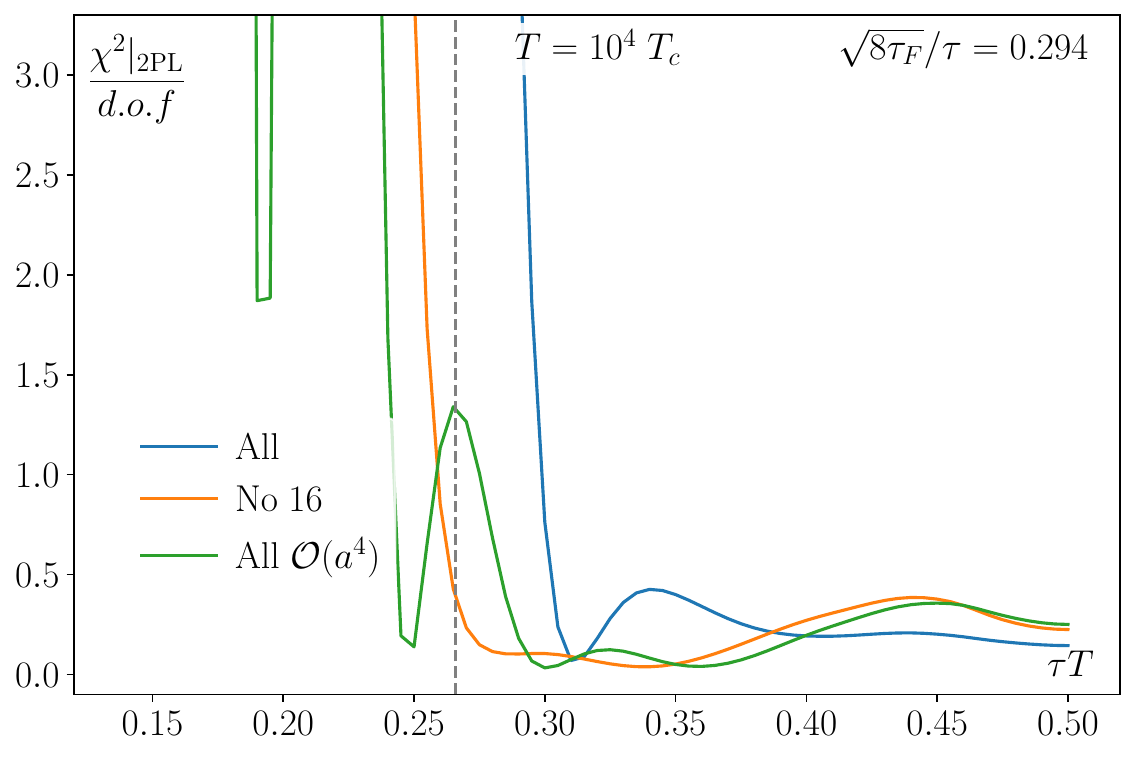}
    \caption{$\chi^2$ over degrees of freedom of the continuum extrapolations corresponding to Fig.~\ref{G_E_oct_comparison_10_4}.}
    \label{G_E_oct_chi2_comparison_10_4}
\end{figure*}

When going to $T=10^4\; T_c$ in Fig.~\ref{G_E_oct_comparison_10_4} with the corresponding values of $\chi^2$ in Fig.~\ref{G_E_oct_chi2_comparison_10_4}, we observe that the linear ansatz with all the lattices already has a good fit and including a $a^4$ term makes it worse. 
For this case, we estimate the systematic error by comparing $\mathrm{All}$ and $\mathrm{No\;16}$.

\begin{figure*}
    \centering
    \includegraphics[width=0.425\linewidth]{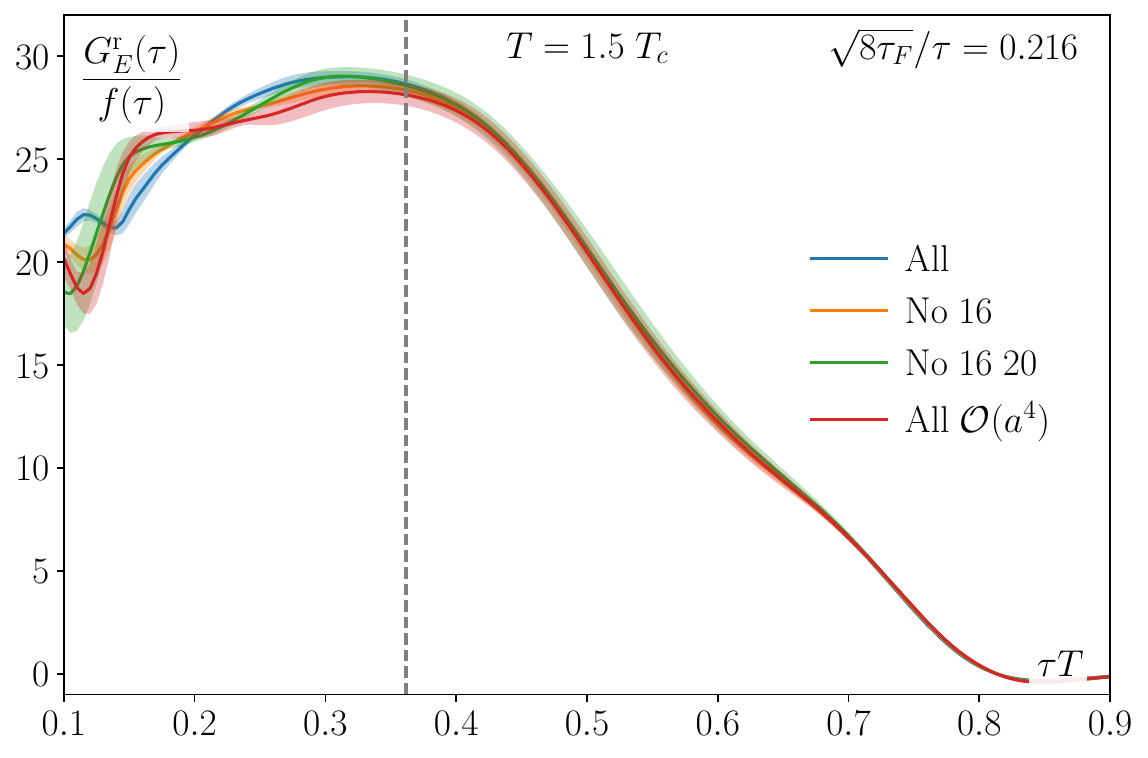} \includegraphics[width=0.425\linewidth]{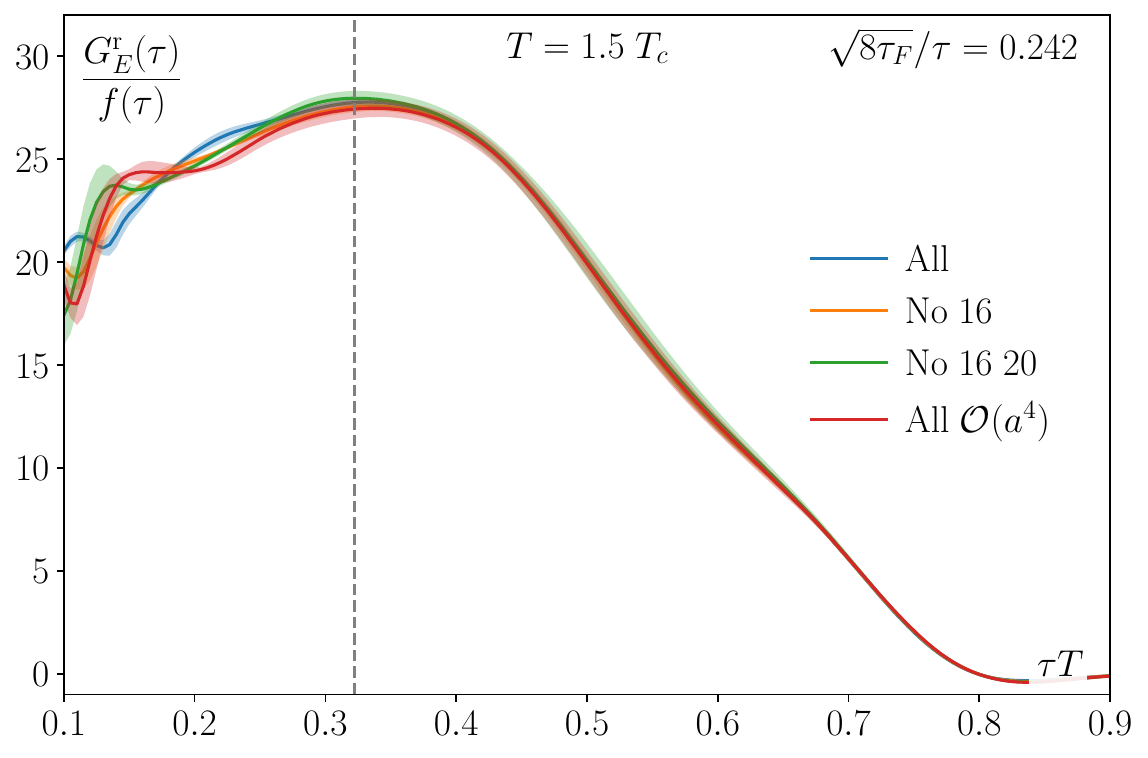} \\
    \includegraphics[width=0.425\linewidth]{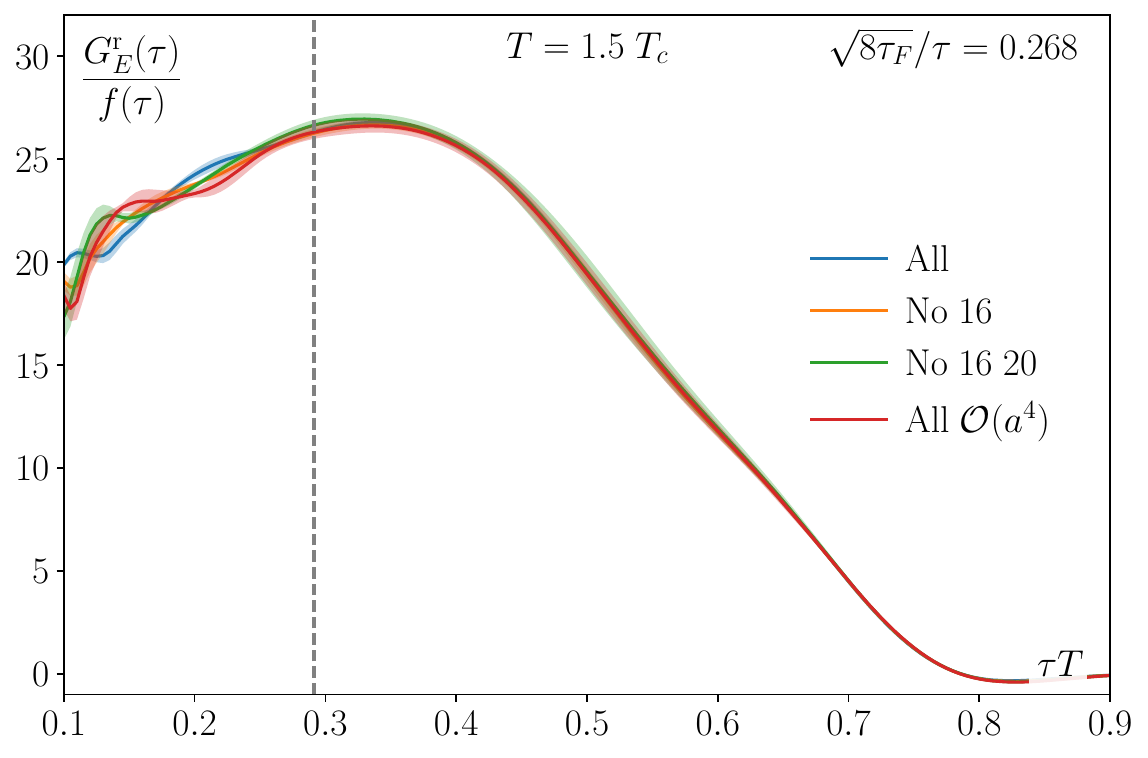} \includegraphics[width=0.425\linewidth]{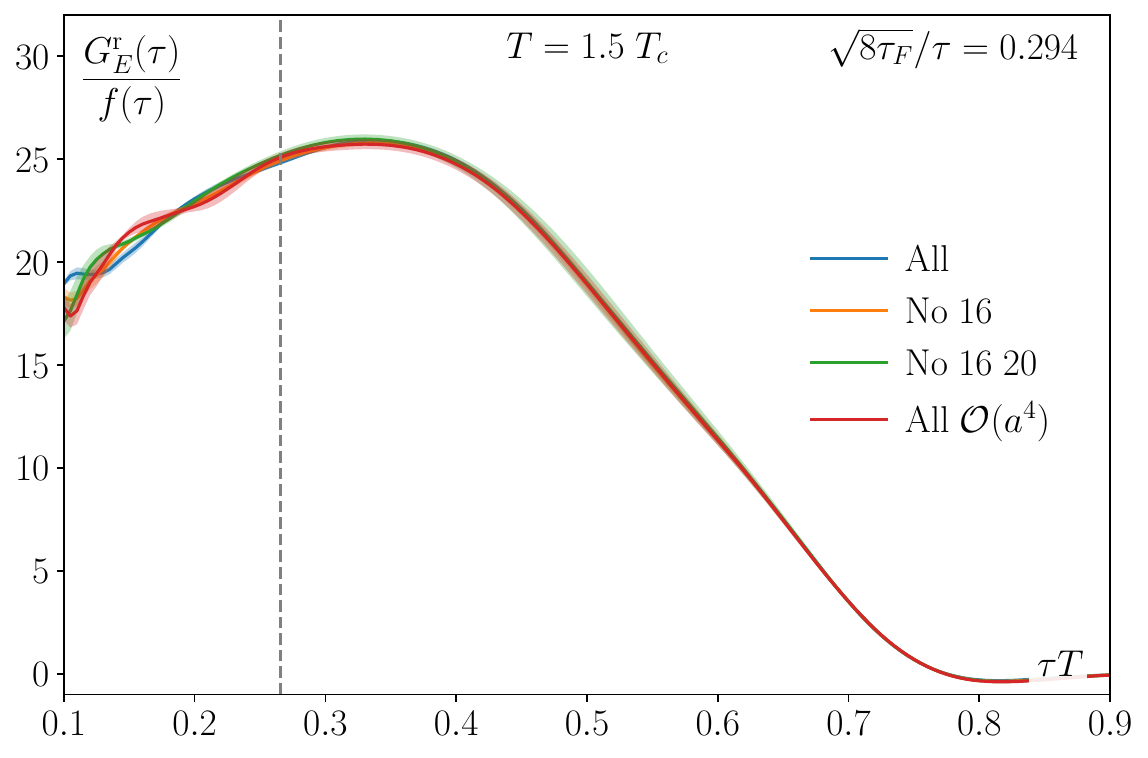}
    \caption{$G_E^{\mathrm{r}}$ with different continuum extrapolations for different flow times at $T = 1.5\;T_c$.}
    \label{G_E_r_comparison_1_5}
\end{figure*}

\begin{figure*}
    \centering
    \includegraphics[width=0.425\linewidth]{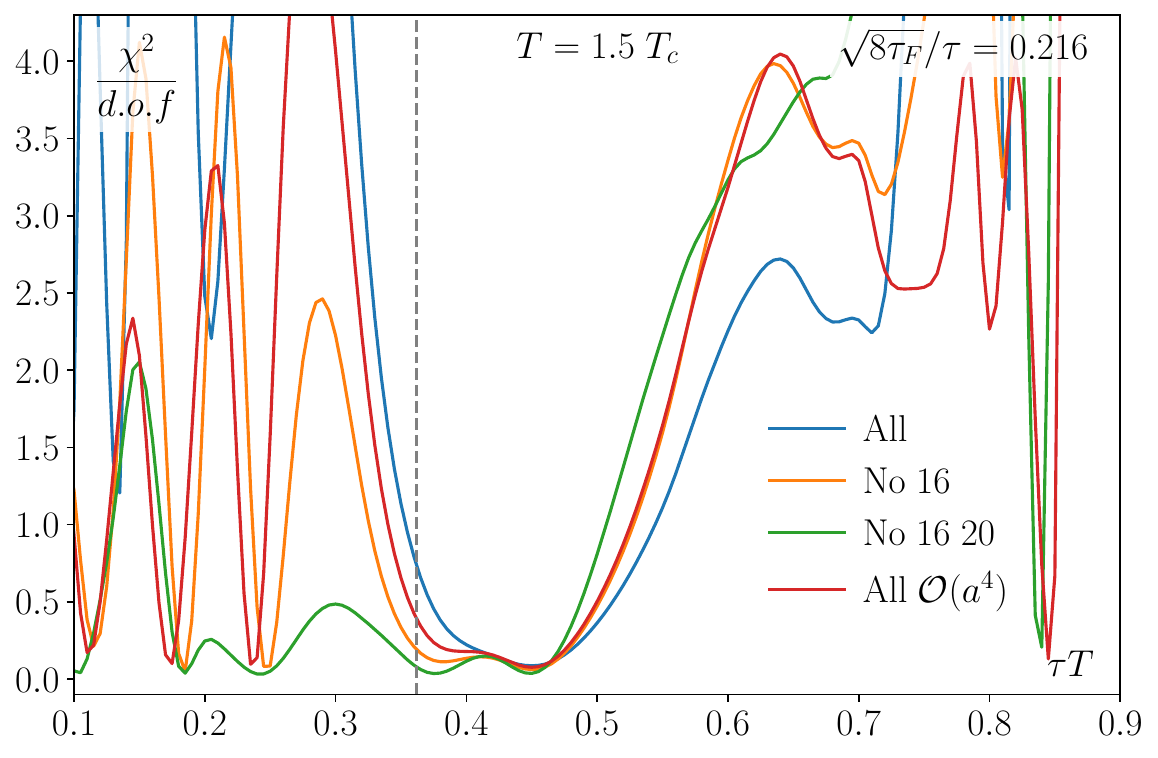} \includegraphics[width=0.425\linewidth]{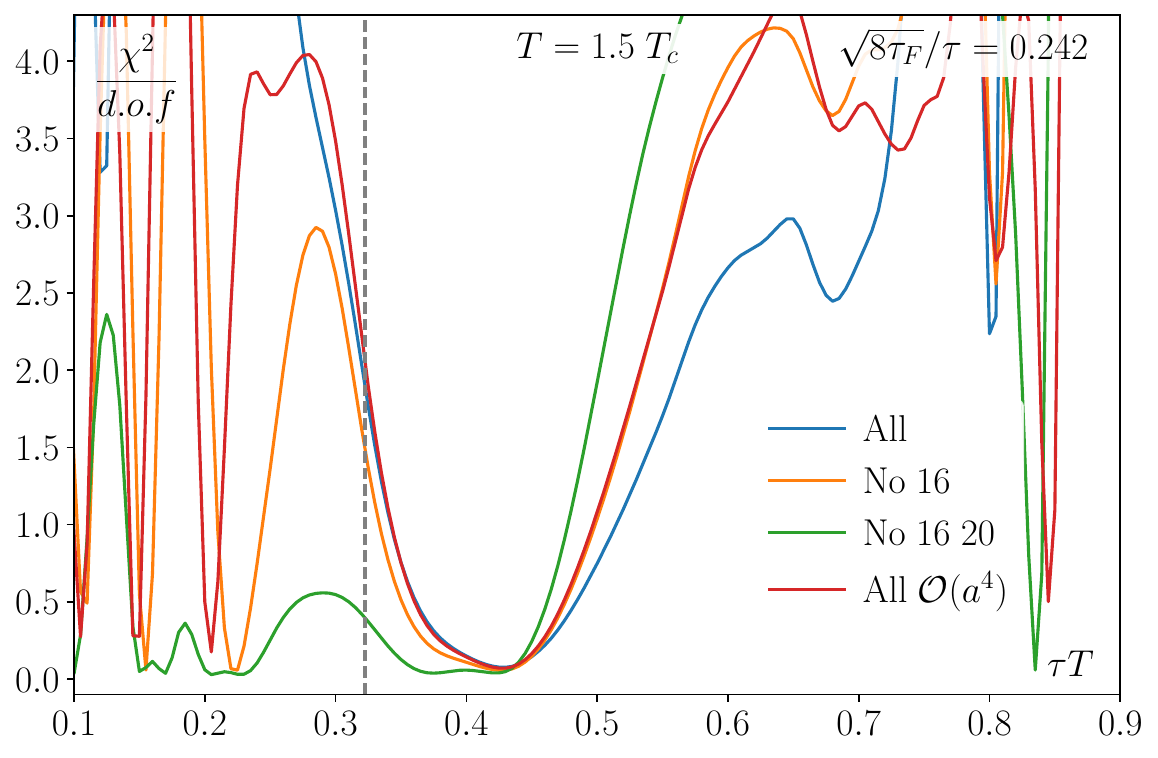} \\
    \includegraphics[width=0.425\linewidth]{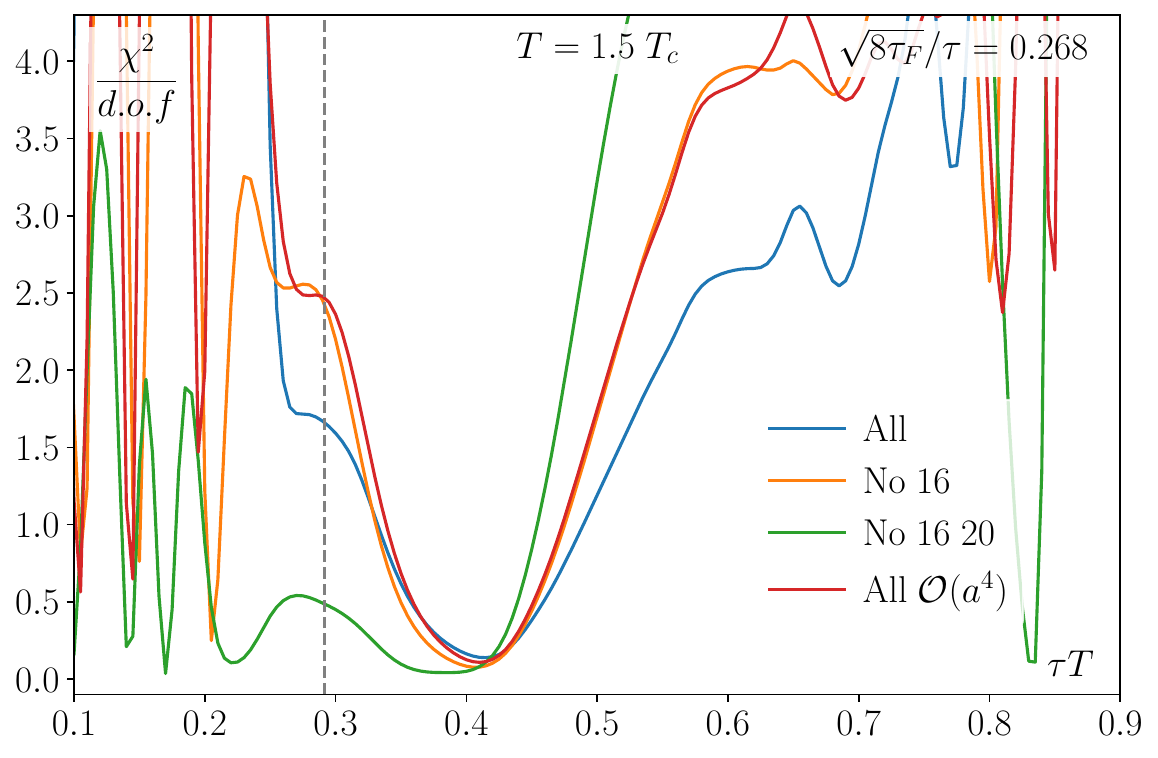} \includegraphics[width=0.425\linewidth]{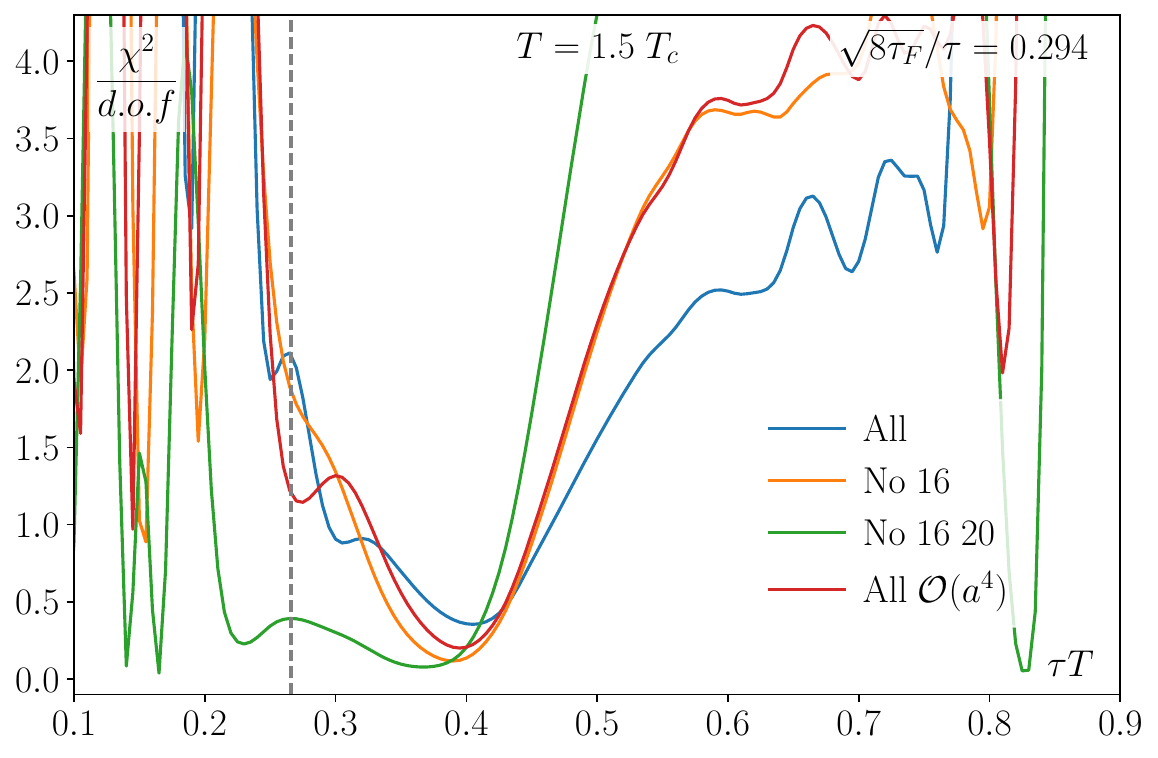}
    \caption{$\chi^2$ over degrees of freedom of the continuum extrapolations corresponding to Fig.~\ref{G_E_r_comparison_1_5}.}
    \label{G_E_r_chi2_comparison_1_5}
\end{figure*}

Next, we perform the same analysis for the renormalized non-symmetric correlator $G_E^{\mathrm{r}}$. 
In Fig.~\ref{G_E_r_comparison_1_5}, we note that all the methods give a similar result. Figure~\ref{G_E_r_chi2_comparison_1_5} shows the $\chi^2$ values, where at the extremes of the validity range, we reach values of 4. 
This increase with respect to $G_E^{\mathrm{oct}}$ can be due to having smaller error bars for $G_E^{\mathrm{r}}$. 
We do not see any improvement from adding the $a^4$ term, so we use again $\mathrm{All}$ and $\mathrm{No\;16}$ to estimate the systematics.

\begin{figure*}
    \centering
    \includegraphics[width=0.425\linewidth]{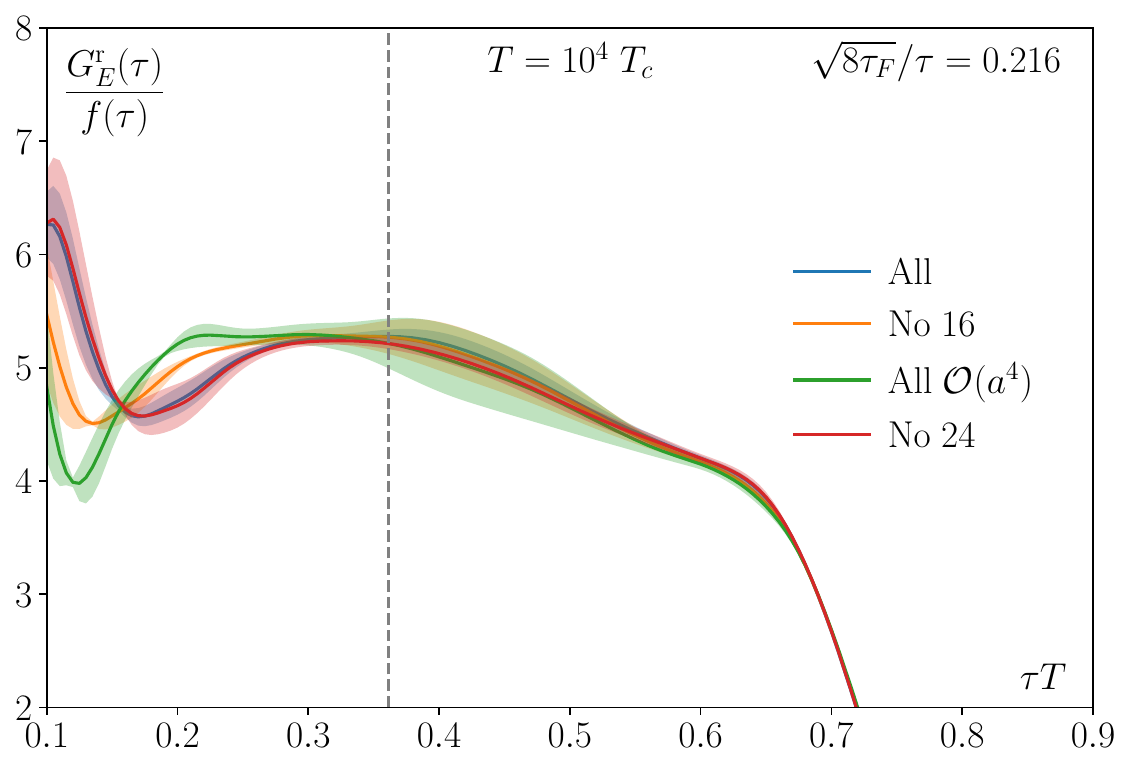} \includegraphics[width=0.425\linewidth]{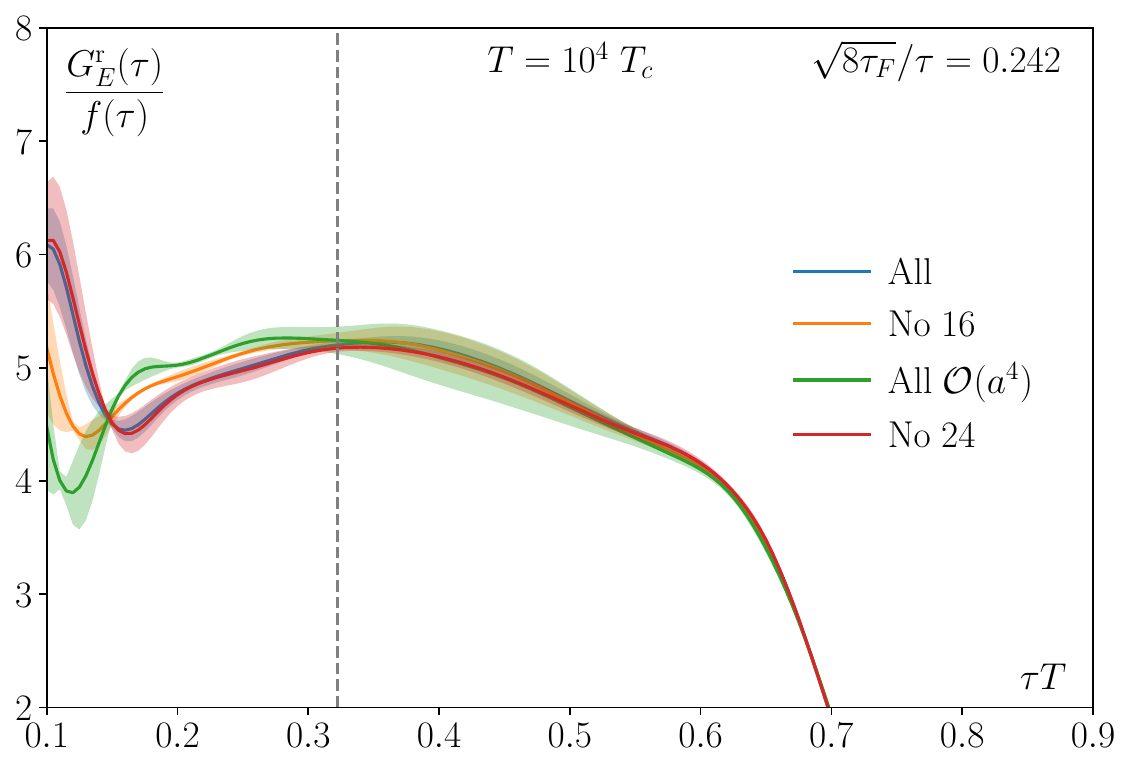} \\
    \includegraphics[width=0.425\linewidth]{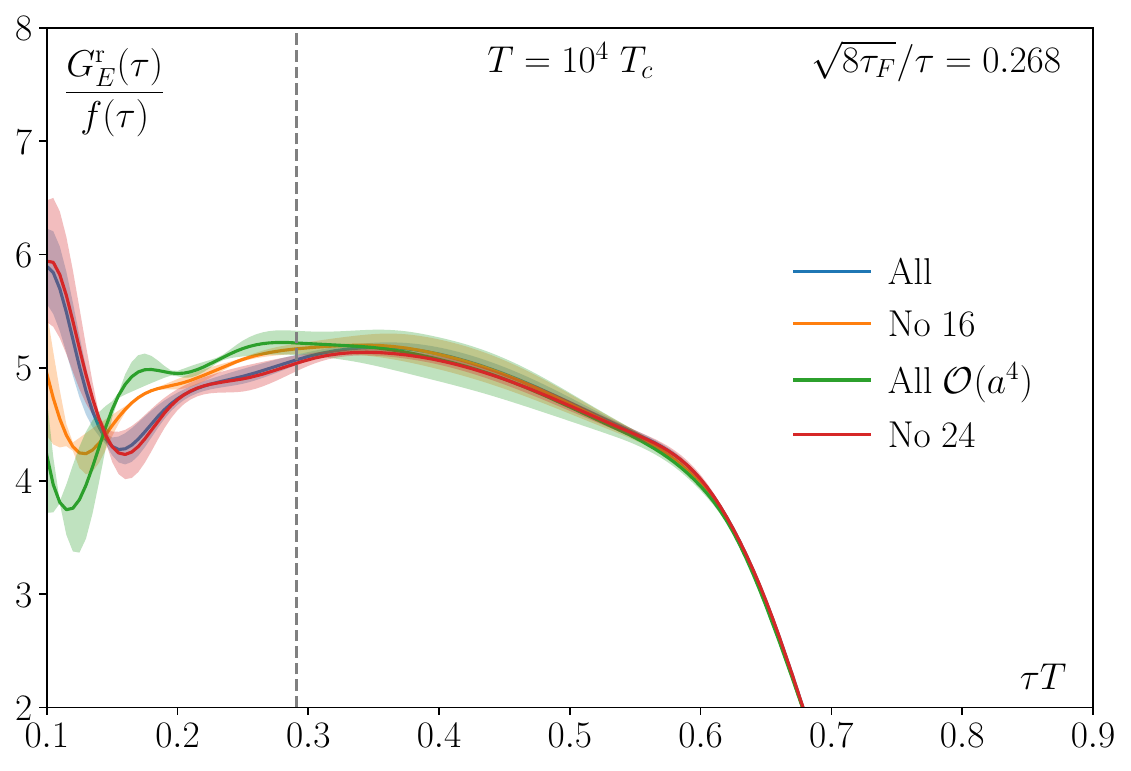} \includegraphics[width=0.425\linewidth]{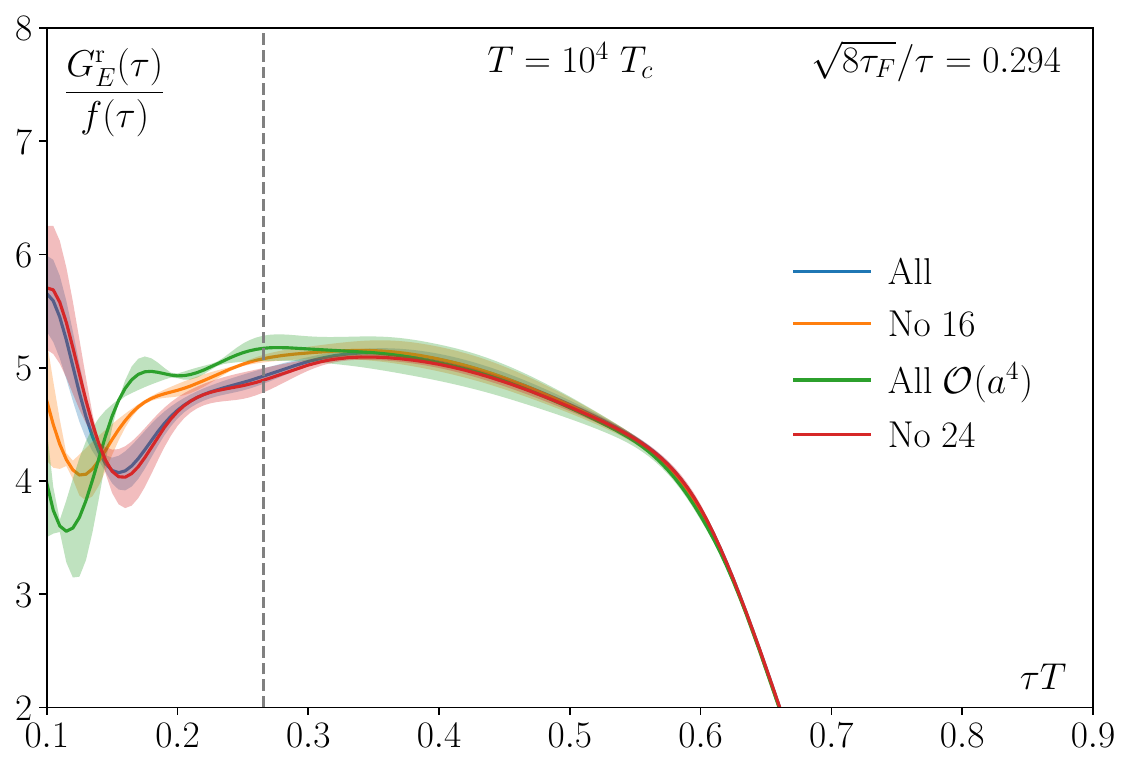}
    \caption{$G_E^{\mathrm{r}}$ with different continuum extrapolations for different flow times at $T = 10^4\;T_c$.}
    \label{G_E_r_comparison_10_4}
\end{figure*}

\begin{figure*}
    \centering
    \includegraphics[width=0.425\linewidth]{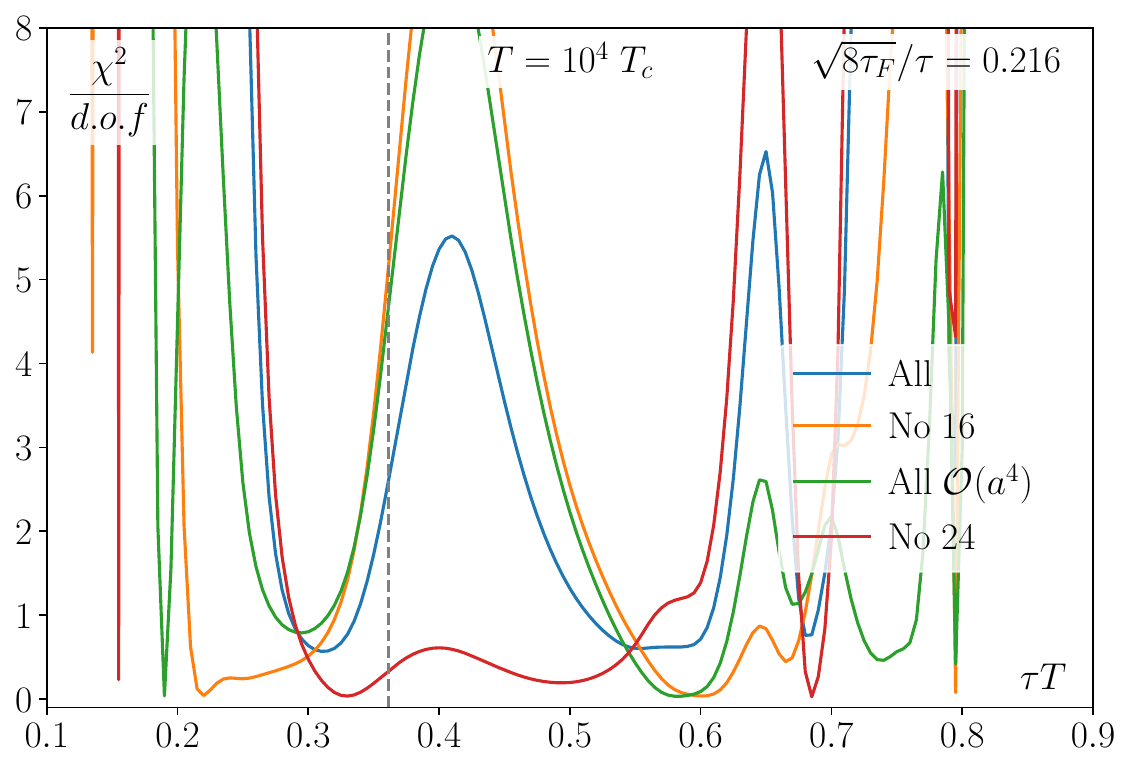} \includegraphics[width=0.425\linewidth]{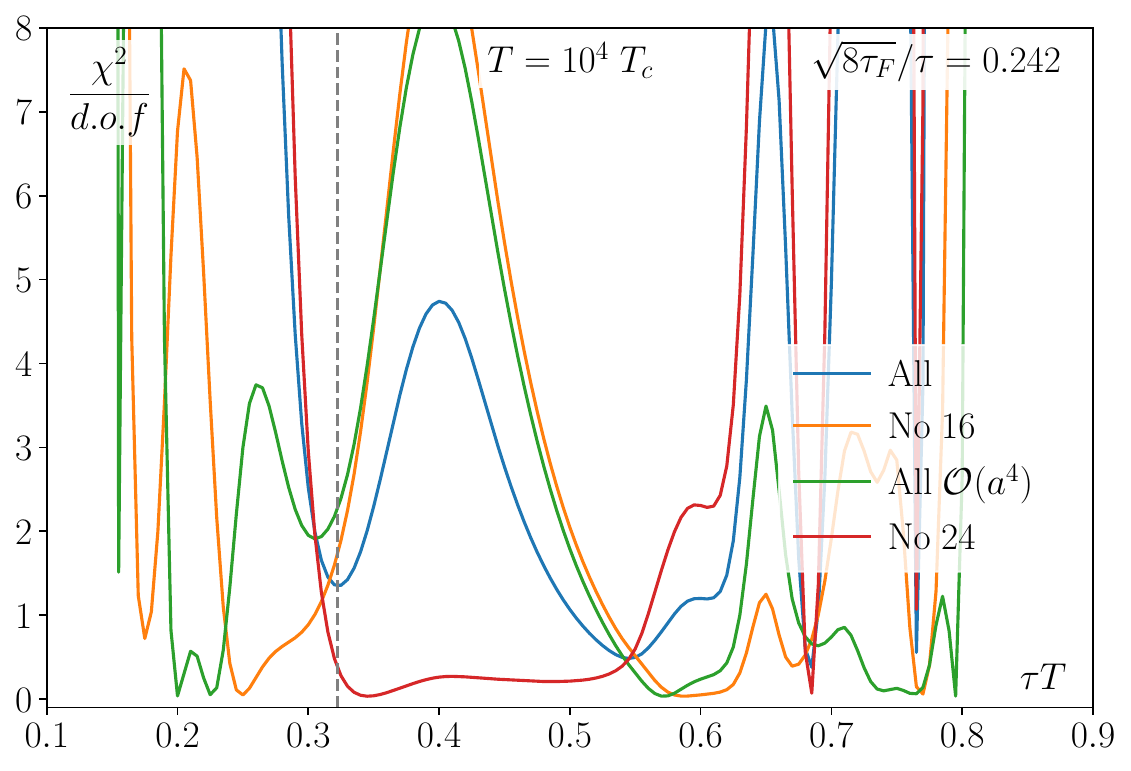} \\
    \includegraphics[width=0.425\linewidth]{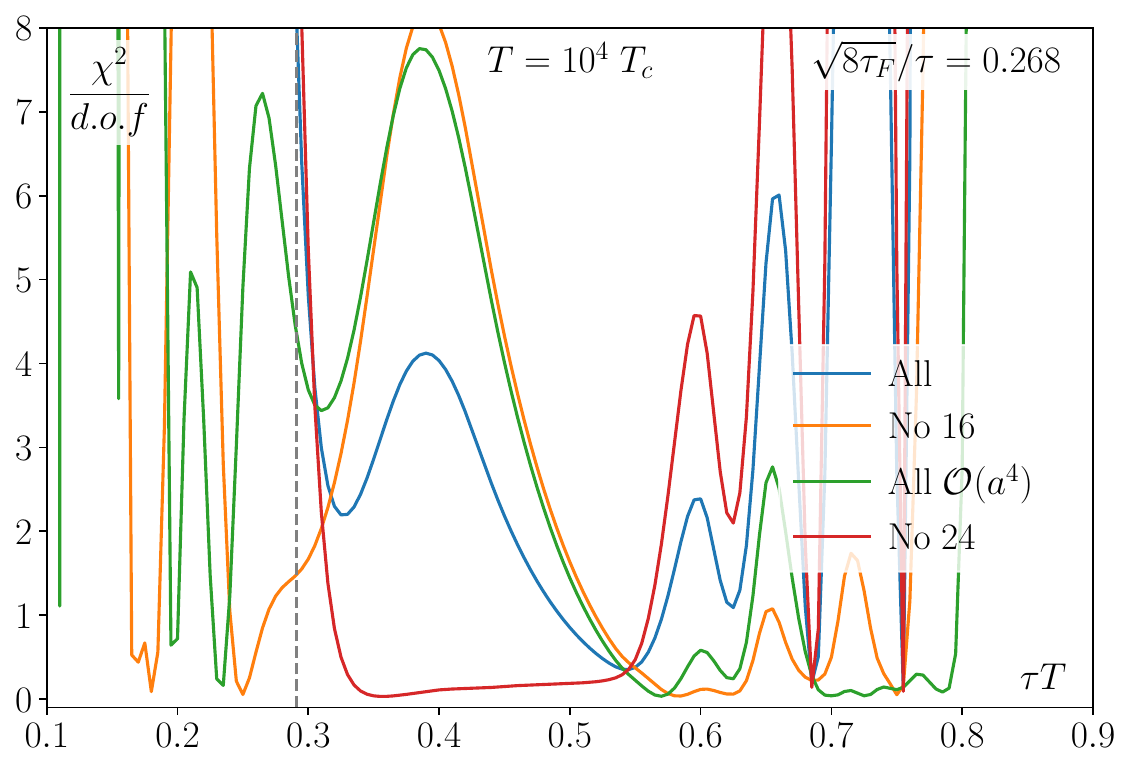} \includegraphics[width=0.425\linewidth]{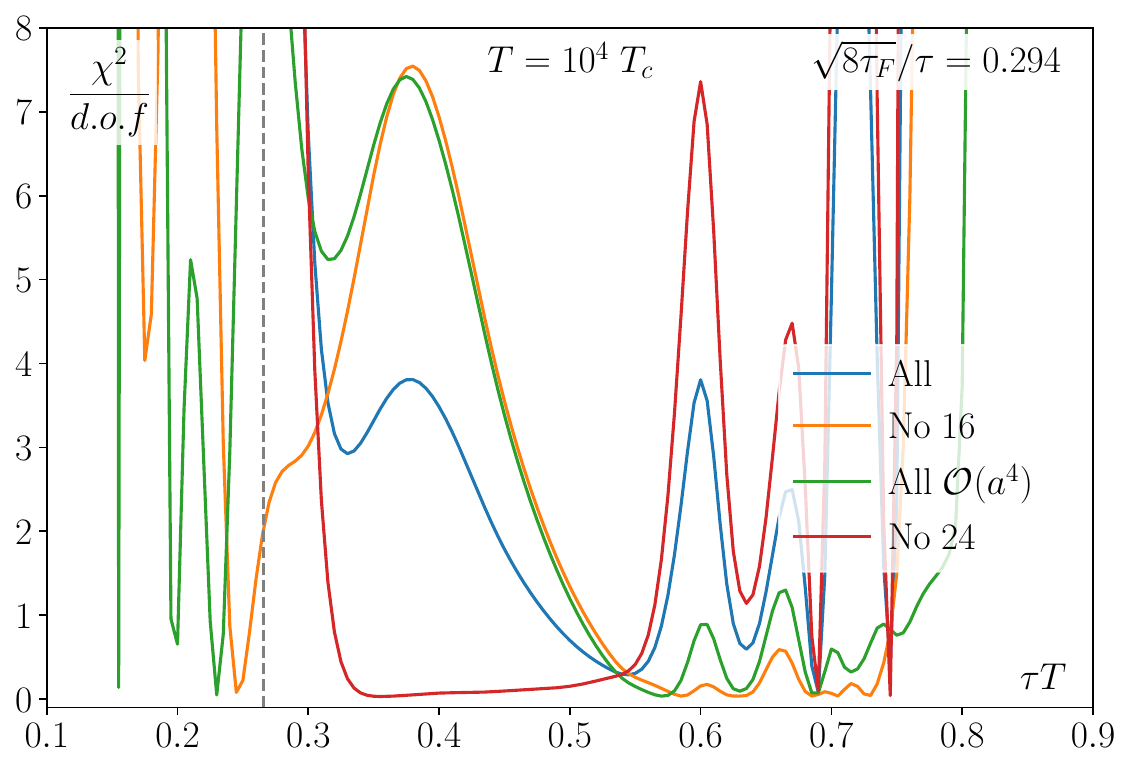}
    \caption{$\chi^2$ over degrees of freedom of the continuum extrapolations corresponding to Fig.~\ref{G_E_r_comparison_10_4}.}
    \label{G_E_r_chi2_comparison_10_4}
\end{figure*}

Lastly, in Fig.~\ref{G_E_r_comparison_10_4} and Fig.~\ref{G_E_r_chi2_comparison_10_4} at $T=10^4\;T_c$, we obtain even bigger values of $\chi^2$. 
This increase seems to be due to the $N_\tau = 24$ and $N_\tau = 34$ lattices not working well together, and adding a $a^4$ term does not help with the extrapolation. 
This can be solved by neglecting the $N_\tau = 24$ lattice, which gives us good $\chi^2$ values. Since the extrapolated results agree within error bars with the ones using all lattices, we continue using all lattices and do not worry about this increase in the $\chi^2$ value. 
We then estimate the systematics with $\mathrm{All}$ and $\mathrm{No\;16}$.

\begin{figure*}
    \centering
    \includegraphics[width=0.425\linewidth]{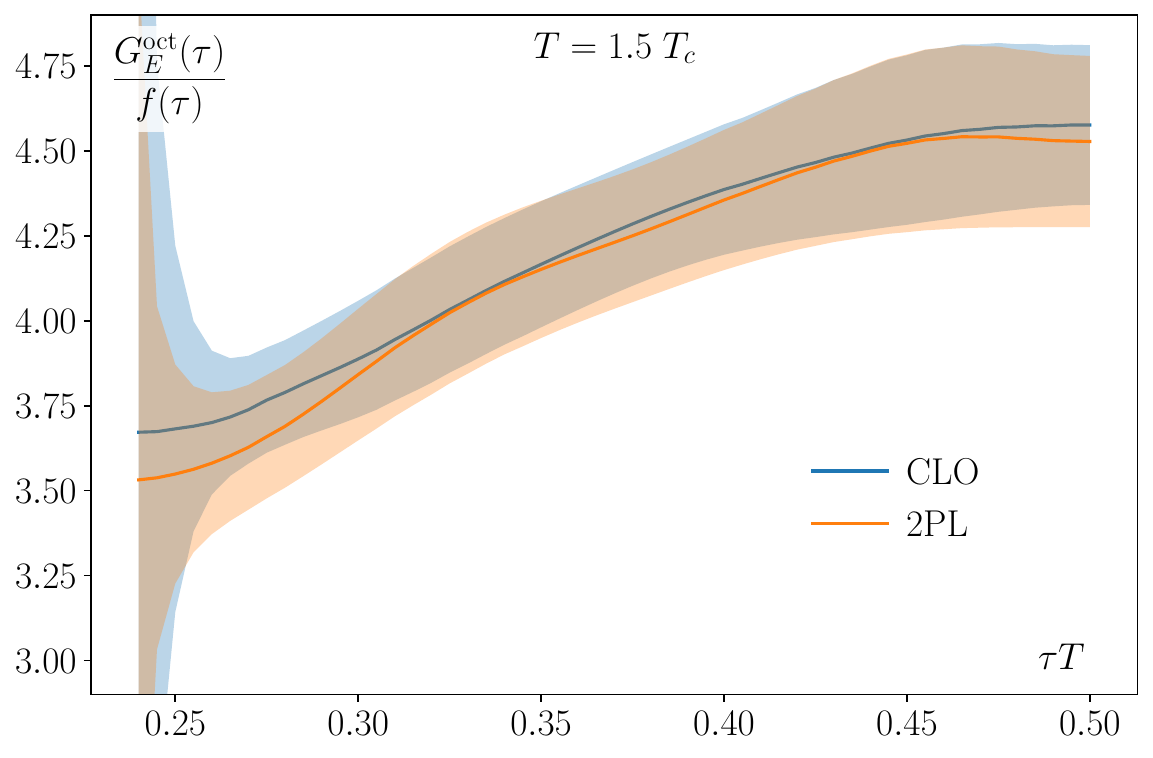} \includegraphics[width=0.425\linewidth]{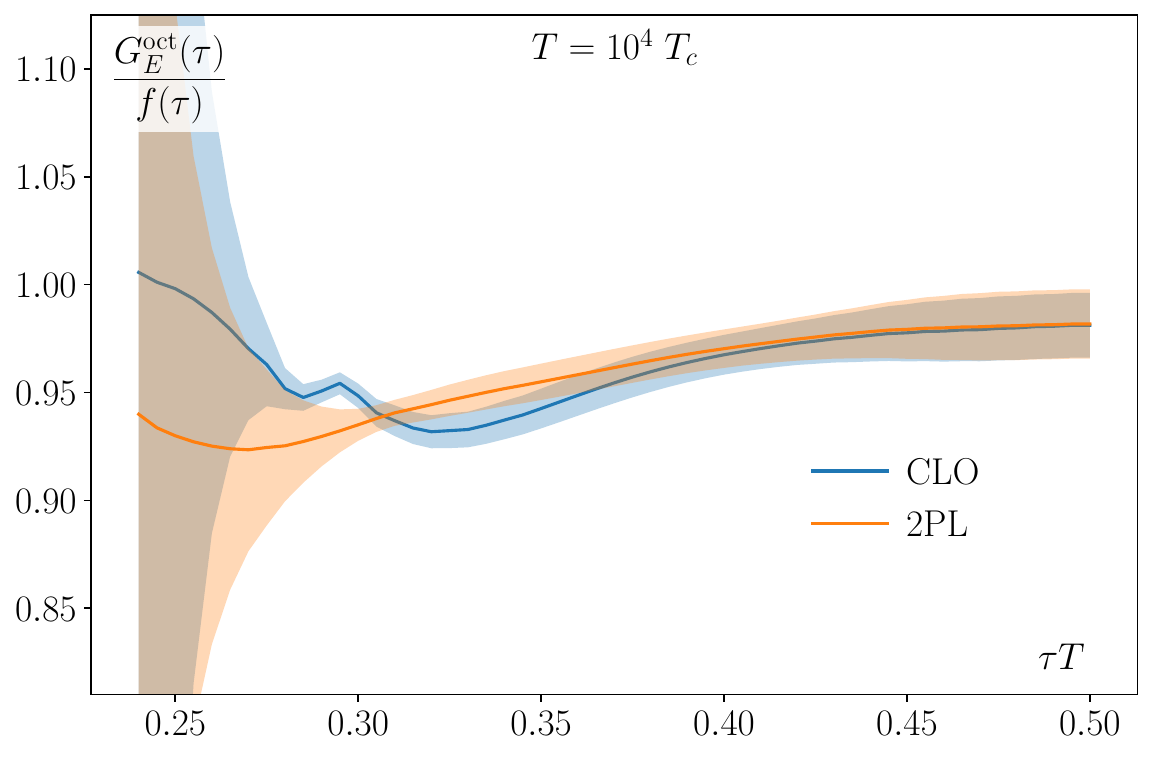} \\
    \includegraphics[width=0.425\linewidth]{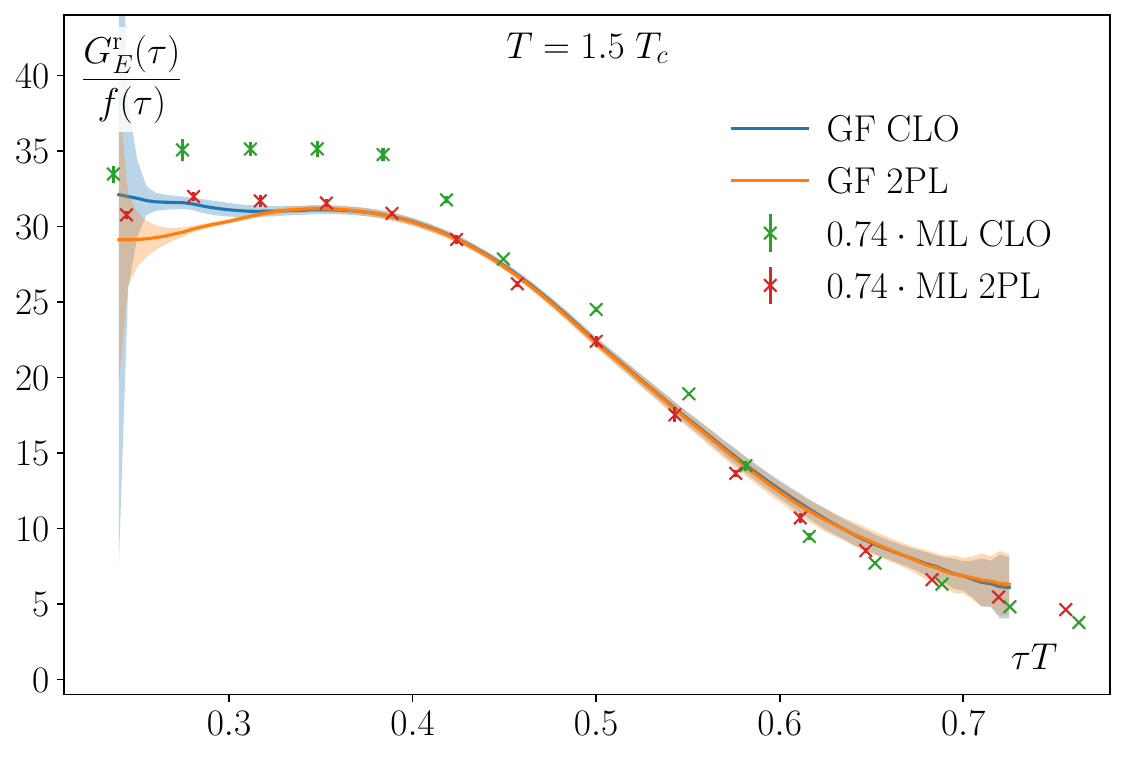} \includegraphics[width=0.425\linewidth]{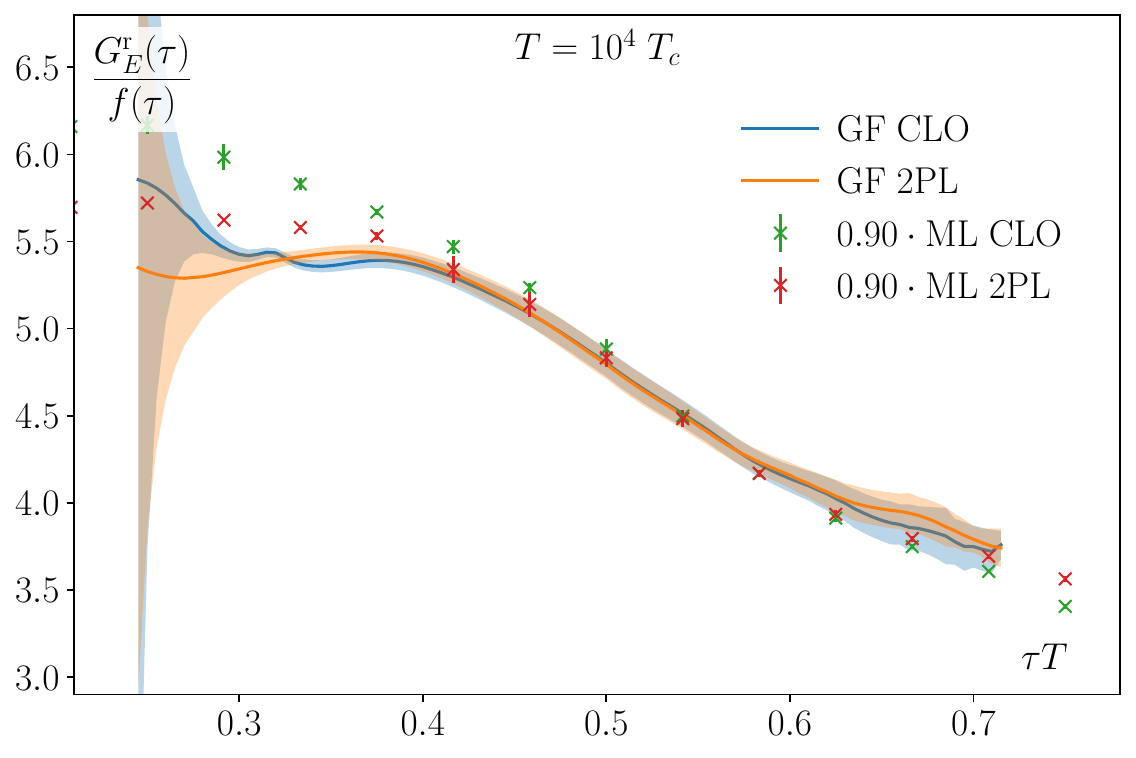}
    \caption{Comparison between the CLO and 2PL discretizations for $G_E^{\mathrm{oct}}$ (top) and $G_E^{\mathrm{r}}$ (bottom) at $T=1.5\;T_c$ (left) and $T=10^4\;T_c$ (right). For the non-symmetric correlator, we also compare between gradient flow (GF) and multilevel (ML).}
    \label{comparison_CLO_2PL}
\end{figure*}

Figure~\ref{comparison_CLO_2PL} shows the final results for $G_E^{\mathrm{oct}}$ and $G_E^{\mathrm{r}}$ after the continuum and zero-flow-time extrapolations for both, CLO and 2PL discretizations. 
We observe that $G_E^{\mathrm{r}}$ is slightly more dependent on the discretization and that the differences between them in multilevel are bigger than in gradient flow.

\FloatBarrier
\bibliography{kappa}{}

\end{document}